\documentclass[onecolumn,preprintnumbers,superscriptaddress,nofootinbib]{revtex4}

\usepackage[a4paper,left=2.5cm,right=2.5cm, top=1.5cm,bottom=1.5cm]{geometry}
\usepackage[english]{babel}
\usepackage[applemac]{inputenc}
\usepackage[colorlinks,citecolor=blue,linktoc=all,linkcolor=cyan,urlcolor=black]{hyperref}
\usepackage{amsfonts,amsmath,amssymb}
\usepackage{graphicx}
\usepackage{booktabs, multirow}
\usepackage[detect-all]{siunitx}
\usepackage{hyperxmp}
\usepackage{color}
\usepackage{simplewick}

\usepackage{relsize}
\newcommand{\babar}{\mbox{\slshape B\kern-0.1em{\smaller A}\kern-0.1em B\kern-0.1em{\smaller A\kern-0.2em R}}}
    
\newcommand{\dprime}{\prime\prime}
\newcommand{\ZV}{Z_{\mathrm{V}}}
\newcommand{\cV}{c_{\mathrm{V}}}
\newcommand{\bV}{b_{\rm V}}
\newcommand{\bbV}{\overline{b}_{\rm V}}
\newcommand{\bg}{b_{\rm g}}
\newcommand{\fV}{f_{\rm V}}
\newcommand{\bA}{b_{\rm A}}
\newcommand{\bbA}{\overline{b}_{\rm A}}
\newcommand{\mqav}{m_{\rm q}^{\rm av}}
\newcommand{\phys}{\mathrm{phys}}
\newcommand{\syst}{\mathrm{syst}}
\newcommand{\stat}{\mathrm{stat}}
\newcommand{\VMD}{\mathrm{VMD}}
\newcommand{\LMD}{\mathrm{LMD}}
\newcommand{\LMDV}{\mathrm{LMD+V}}
\newcommand{\dof}{\mathrm{d.o.f.}}
\newcommand{\fm}{\mathrm{fm}}
\newcommand{\MeV}{\mathrm{MeV}}
\newcommand{\GeV}{\mathrm{GeV}}

\newcommand{\psib}{\overline{\psi}}	
\newcommand{\FF}{{\cal F}_{\pi^0\gamma^*\gamma^*}}
\newcommand{\amu}{a_\mu^{\mathrm{HLbL}; \pi^0}}

\newcommand{\amuLMDV}{a_{\mu; \mathrm{LMD}+\mathrm{V}}^{\mathrm{HLbL}; \pi^0}}
\newcommand{\alphaQED}{\alpha_{\rm e}}

\makeatother

\begin{document}
%
\title{Lattice calculation of the pion transition form factor with $N_f=2+1$ Wilson quarks}
\author{Antoine G\'erardin}
\email{antoine.gerardin@desy.de}
\affiliation{Institut f\"ur Kernphysik \& Clusters of Excellence PRISMA and PRISMA$^+$,  Johannes Gutenberg-Universit\"at Mainz, D-55099 Mainz, Germany}
\affiliation{John von Neumann Institute for Computing, DESY, Platanenallee 6, D-15738 Zeuthen, Germany}
\author{Harvey B.\ Meyer} 
\email{meyerh@uni-mainz.de}
\affiliation{Institut f\"ur Kernphysik \& Clusters of Excellence PRISMA and PRISMA$^+$,  Johannes Gutenberg-Universit\"at Mainz, D-55099 Mainz, Germany}
\affiliation{Helmholtz Institut Mainz,  D-55099 Mainz, Germany}
\author{Andreas Nyf\/feler}
\email{nyffeler@uni-mainz.de}
\affiliation{Institut f\"ur Kernphysik \& Clusters of Excellence PRISMA and PRISMA$^+$,  Johannes Gutenberg-Universit\"at Mainz, D-55099 Mainz, Germany}
\date{\today}

\preprint{MITP/19-014}

\begin{abstract}
We present a lattice QCD calculation of the double-virtual neutral pion transition form factor, with the goal to cover the kinematic range relevant to hadronic light-by-light scattering in the muon $g-2$. Several improvements have been made compared to our previous work.  First, we take into account the effects of the strange quark by using the $N_f=2+1$ CLS gauge ensembles.  Second, we have  implemented the on-shell $\mathcal{O}(a)$-improvement of the vector current to reduce the discretization effects associated with Wilson
quarks. Finally, in order to have access to a wider range of photon virtualities, we have computed the transition form factor in a moving frame as well as in the pion rest frame.  After extrapolating the form factor to the continuum and to physical quark masses, we compare our results with phenomenology. We extract the normalization of the form factor with a precision of 3.5\% and confirm within our uncertainty previous somewhat conflicting estimates for a low-energy constant that appears in chiral perturbation theory for the decay $\pi^0 \to \gamma\gamma$ at NLO. With additional input from experiment and theory, we reproduce recent estimates for the decay width $\Gamma(\pi^0 \to \gamma\gamma)$. We also study the asymptotic large-$Q^2$ behavior of the transition form factor in the double-virtual case. Finally, we provide as our main result a more precise model-independent lattice estimate of the pion-pole contribution to hadronic light-by-light scattering in the muon $g-2$: $a_{\mu}^{\mathrm{HLbL}; \pi^0} = (59.7 \pm 3.6) \times 10^{-11}$. Using in addition the normalization of the form factor obtained by the PrimEx experiment, we get the lattice and data-driven estimate $a_{\mu}^{\mathrm{HLbL}; \pi^0} = (62.3 \pm 2.3) \times 10^{-11}$.
\end{abstract}
\maketitle

\section{Introduction \label{sec:introduction} }

There is a long-standing discrepancy between the Standard Model estimate of the muon anomalous magnetic moment and its experimental determination~\cite{pdg:2018}. Two new experiments, E989 at Fermilab~\cite{Venanzoni:2014ixa} and E34 at J-PARC~\cite{Otani:2015lra}, plan to reduce the experimental error by a factor of 4 in the near future. The theory error is completely dominated by hadronic contributions: the hadronic vacuum polarization (HVP), which enters at order $\alphaQED^2$ in the fine-structure constant $\alphaQED$, and the hadronic light-by-light (HLbL) scattering at order $\alphaQED^3$. The former is usually obtained using dispersive methods which rely on the $e^+e^- \to~$hadrons cross sections, accessible from experiments~\cite{Davier:2017zfy,Anastasi:2017eio}. 
Thus, a more accurate determination essentially relies on the availability of precise measurements. 
Lattice QCD is also a promising tool, and has made a lot of progress in recent years. Even if not yet competitive with the dispersive approach, it became a mature field where most sources of systematic uncertainties have now been addressed, see e.g. Ref.~\cite{Meyer:2018til} for a recent review.
As for the HLbL contribution, the situation is less favorable. Until recently, all estimates for the HLbL contribution were based on model calculations, leading to the Glasgow consensus~\cite{Prades:2009tw} (see also~\cite{Jegerlehner:2009ry}), for which errors are difficult to estimate. 
The single largest contribution to $a_{\mu}^{\rm HLbL}$ is given by the pion-pole contribution~\cite{Knecht:2001qf} with a prescription for its evaluation that has been confirmed in the recently proposed dispersive approaches to HLbL~\cite{HLbL_DR,Pauk:2014jza}. 
Its determination relies on the knowledge of the neutral pion transition form factor (TFF). 
Two groups have also started the direct calculation of the HLbL scattering contribution, from first principles, using lattice QCD~\cite{Blum:2014oka,Blum:2015gfa,Blum:2016lnc,Green:2015mva,Asmussen:2016lse,Asmussen:2017bup}. In the Mainz approach, the calculation involves the convolution of a QED kernel function, computed semianalytically in infinite, continuous coordinate space, and a QCD four-point correlation function, computed on the lattice. The long-distance contribution is expected to be dominated by the pion-pole contribution. However, this region suffers from large statistical errors and is also more affected by finite-size effects. The TFF is therefore a key ingredient to first reduce the statistical error by constraining the tail of the integrand at large distances and, second, to estimate and correct for the dominant finite-size effects due to pions.

The TFF is also interesting from a theoretical point of view. First, in the low energy region, it is directly related to the Adler-Bell-Jackiw (ABJ) chiral anomaly~\cite{Adler:1969gk,Bell:1969ts}. Secondly, for large virtualities, the asymptotic behavior of the TFF is predicted by the
Brodsky-Lepage analysis and the operator product expansion (OPE) in the single- and double-virtual case respectively~\cite{BL_3_papers,Nesterenko:1982dn,Novikov:1983jt}. Testing these predictions is a remarkable test of QCD over a large range of length scales.

The normalization of the TFF has been measured by the PrimEx experiment with a precision of 1.4~\%~\cite{Larin:2010kq} and this error should be reduced by a factor of 2 by the PrimeEx-II experiment~\cite{Gasparian:2016oyl,Primex2Mainz}. At finite virtualities, experimental data are only available in the single-virtual case and for rather large virtualities above $0.6~\GeV^2$~\cite{exp} but the BESIII experiment plans to have data in the region 0.3 - 3~$\GeV^2$, relevant for the muon $g-2$, in the near future~\cite{Denig:2014mma}. Finally, no experimental data exist to date in the regime of two virtual photons.

In our previous work~\cite{Gerardin:2016cqj}, we have shown that lattice QCD can provide a precise estimate of the TFF in the full spacelike region relevant for the pion-pole contribution to the muon $(g-2)$. 
This work is an update of our previous lattice calculation of the neutral pion TFF~\cite{Gerardin:2016cqj}, and 
includes several major improvements. First, it is based on gauge configurations with $N_f=2+1$ dynamical flavors, which have been generated as part of the CLS initiative. Second, on-shell $\mathcal{O}(a)$-improvement of correlation functions has been implemented to reduce discretization effects when approaching the continuum limit. Finally, in addition to the pion rest frame, a new frame where the pion carries one unit of momentum, typically in the range of 300 to 400\,MeV, is considered. This allows us to probe larger photon virtualities, especially in the single-virtual case, where virtualities as high as $1.5~\GeV^2$ are now available. In all cases, we probe a much wider range of virtualities than in our previous study and have increased statistics significantly. We also address many sources of systematic errors, including finite-size effects, hypercubic lattice artifacts due to the broken rotational invariance down to the isometry group H(3) on the lattice and disconnected contributions.

As a benchmark of our calculation, we reproduce the anomaly constraint in the continuum and at the physical pion mass with a precision of 3.5\%. From the pion mass dependence of the normalization of the form factor we extract the corresponding low-energy constant (LEC) appearing in the chiral Lagrangian needed for the next-to-leading order (NLO) calculation of the decay $\pi^0 \to \gamma\gamma$ in chiral perturbation theory. We confirm previous estimates of this and a related LEC and obtain with additional input from theory and experiment the value $\Gamma(\pi^0 \to \gamma\gamma) = 8.07(10)~\mathrm{eV}$ in perfect agreement with recent results in the literature~\cite{Kampf:2009tk}. The improvements of our lattice calculation allow for a model-independent determination of the pion-pole contribution to hadronic light-by-light scattering in the muon $(g-2)$, expected to be numerically dominant, along with the $\eta$ and $\eta^{\prime}$ pseudoscalar mesons. Our final lattice result reads $\amu = (59.7 \pm 3.6) \times 10^{-11}$ which corresponds to a precision of $6\%$. With the normalization of the form factor obtained from the measurement of the decay width $\pi^0 \to \gamma\gamma$ by the PrimEx experiment, we get the lattice and data-driven estimate $a_{\mu}^{\mathrm{HLbL}; \pi^0} = (62.3 \pm 2.3) \times 10^{-11}$ with a precision of 4\%.

This paper is organized as follows. In Sec.~\ref{sec:methodology}, we describe our methodology to extract the pion TFF. In particular, we explain how $\mathcal{O}(a)$-improvement is implemented. In Sec.~\ref{sec:results}, we present our results, extrapolated to the physical point, and discuss potential sources of systematic errors. Then, in Sec.~\ref{sec:pheno}, after comparing our data with various phenomenological models, we use our result to compute relevant phenomenological quantities, including the pion decay $\pi^0 \to \gamma\gamma$ and the pion-pole contribution in the hadronic light-by-light scattering contribution to the muon $(g-2)$. We conclude in Sec.~\ref{sec:ccl} with a summary of our work and present some possible improvements for future calculations.

\section{Methodology \label{sec:methodology} }

\subsection{Extraction of the transition form factor}

In this section, we use the same notations as in Ref.~\cite{Gerardin:2016cqj} and recall only the main equations.
In Minkowski spacetime, the TFF describing the interaction between a neutral pion with momentum $p$ and two off-shell photons with momenta $q_1$ and $q_2$ is defined via the following matrix element,
\begin{equation} 
M_{\mu\nu}(p,q_1)  = i \int \mathrm{d}^4 x \, e^{i q_1 \cdot x} \, \langle \Omega | T \{ J_{\mu}(x) J_{\nu}(0) \} | \pi^0(p) \rangle =
\epsilon_{\mu\nu\alpha\beta} \, q_1^{\alpha} \, q_2^{\beta} \, \FF(q_1^2, q_2^2) \,,
\label{eq:M}
\end{equation}
where $J_{\mu}$ is the hadronic component of the electromagnetic current, $p=q_1+q_2$ and $\epsilon_{\mu\nu\alpha\beta}$ is the fully antisymmetric tensor with\footnote{Latin indices run from 1 to 3 and Greek indices run from 0 to 3} $\epsilon^{0123} = 1$.
In Euclidean spacetime, the TFF is obtained after analytical continuation.
The latter is valid for $q_{1,2}^2 < s_0$, with $s_0$ the threshold for hadron production in the 
vector channel~\cite{Ji:2001wha,Ji:2001nf}\footnote{In the isovector case, the threshold $s_0$ is given by $4m_\pi^2$ or the square $\rho$ meson mass,
depending on how light the quarks are.}, and reads
\begin{equation}
M_{\mu\nu}  =  (i^{n_0}) M_{\mu\nu}^{\rm E}, \quad 
M_{\mu\nu}^{\rm E}  \equiv  - \int \mathrm{d} \tau \, e^{\omega_1 \tau}  \int \mathrm{d}^3 z \, e^{-i \vec{q}_1 \vec{z}} \, 
\langle 0 | T \left\{ J_{\mu}(\vec{z}, \tau) J_{\nu}(\vec{0}, 0) \right\} | \pi(p) \rangle  \,,
\label{eq:ME}
\end{equation}
where $n_0$ denotes the number of temporal indices carried by the two vector currents, $\omega_1$ is a real free parameter such that $q_1 = (\omega_1, \vec{q}_1)$ and the superscript E stands for Euclidean. It is convenient to write Eq.~(\ref{eq:ME}) as
\begin{equation}
 M_{\mu\nu}^{\rm E} = \frac{2 E_{\pi}}{ Z_{\pi} }  \int_{-\infty}^{\infty} \, \mathrm{d}\tau \, e^{\omega_1 \tau} \, \widetilde{A}_{\mu\nu}(\tau) \,, 
\label{eq:Mlat}
\end{equation}
where $E_{\pi}$ is the pion energy, $Z_{\pi}=\langle0|P(0)|\pi\rangle$ is the overlap of the pseudoscalar operator with the pion state\footnote{Fixing the phase of the pion state via the relation $\langle 0|A_\mu^a(x)|\pi^b,\vec p\rangle = i F_\pi p_\mu \delta^{ab} e^{-ip\cdot x}$, with $A_\mu^a = \bar\psi \gamma_\mu\gamma_5 \frac{\tau^a}{2}\psi$ the isovector axial current, the partially conserved axial current (PCAC) relation implies that the overlap $Z_\pi = -\sqrt{2}im_\pi^2 F_\pi/m$ of the operator $P = \bar\psi \gamma_5 \frac{\tau^3}{\sqrt{2}}\psi$ is purely imaginary. Here $F_\pi\simeq92\,$MeV is the pion decay constant and $m$ is the average up/down quark mass.}
and $\widetilde{A}_{\mu\nu}(\tau)$ is related to the three-point correlation function computed on the lattice,
\begin{gather}
\widetilde{A}_{\mu\nu}(\tau) \equiv \lim_{t_{\pi} \rightarrow + \infty} e^{E_\pi (t_f-t_0)} C^{(3)}_{\mu\nu}(\tau,t_{\pi}) \,.
\label{eq:Amunu}
\end{gather}
The three-point correlation function is defined as 
\begin{equation}
C^{(3)}_{\mu\nu}(\tau,t_{\pi}) \equiv a^6\sum_{\vec{x}, \vec{z}} \, \big\langle   J_{\mu}(\vec{z}, t_i) J_{\nu}(\vec{0}, t_f)  P^{\dag}(\vec{x},t_0) \big\rangle \, e^{i \vec{p}\, \vec{x}} \, e^{-i \vec{q}_1 \vec{z}} \,, 
\label{eq:C3} 
\end{equation}
where $\tau=t_i-t_f$ is the time separation between the two vector currents and  $t_{\pi}={\rm min}(t_f-t_0,t_i-t_0)$ is the minimal time separation between the pion interpolating operator and the two vector currents.
Finally, even if not explicitly written, the functions $\widetilde{A}_{\mu\nu}(\tau)$, as well as the three-point correlation functions, depend on the photon spatial momenta $\vec{q}_1$ and~$\vec{q}_2$. 

\subsection{Orbits and kinematical reach}
\label{sec:orbs}

\begin{figure}[t]
	\includegraphics*[width=0.49\linewidth]{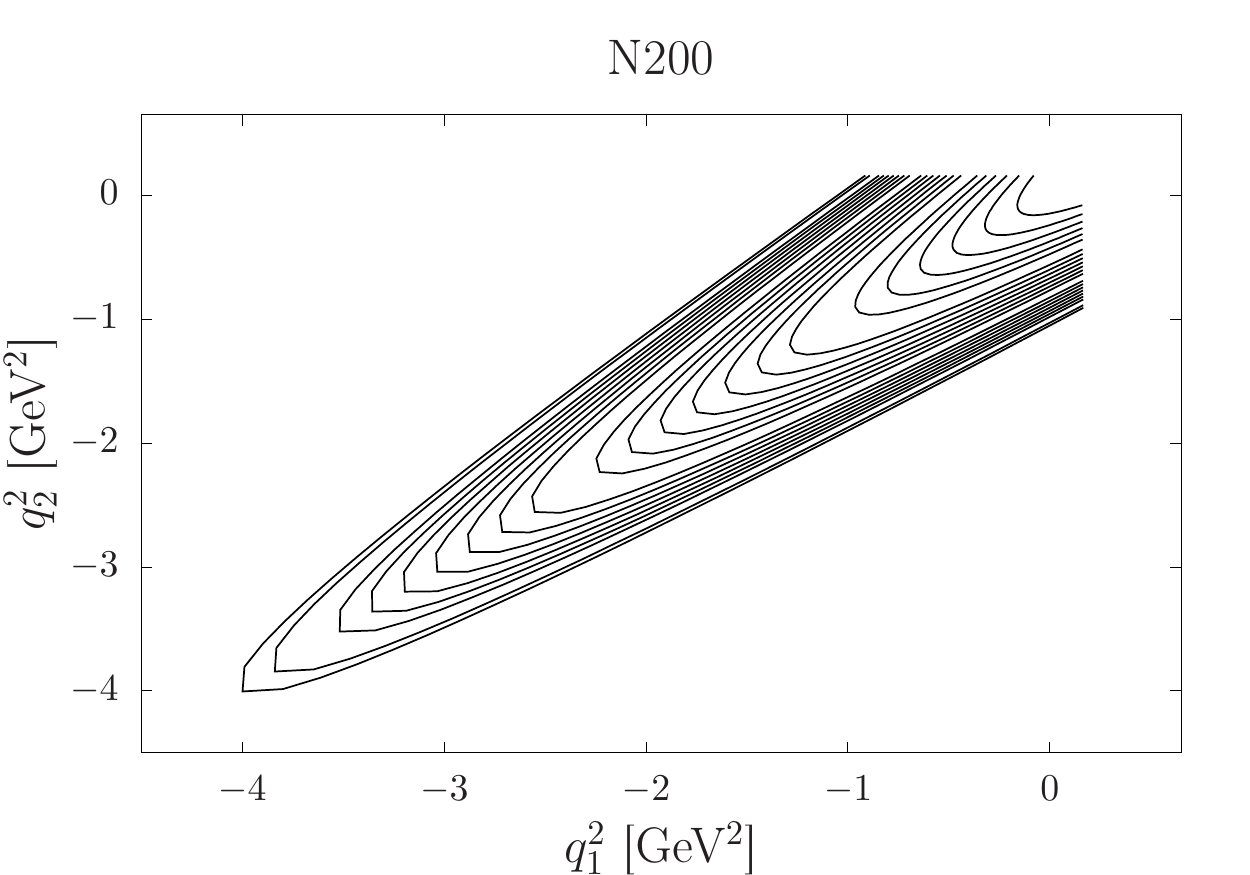}
	\includegraphics*[width=0.49\linewidth]{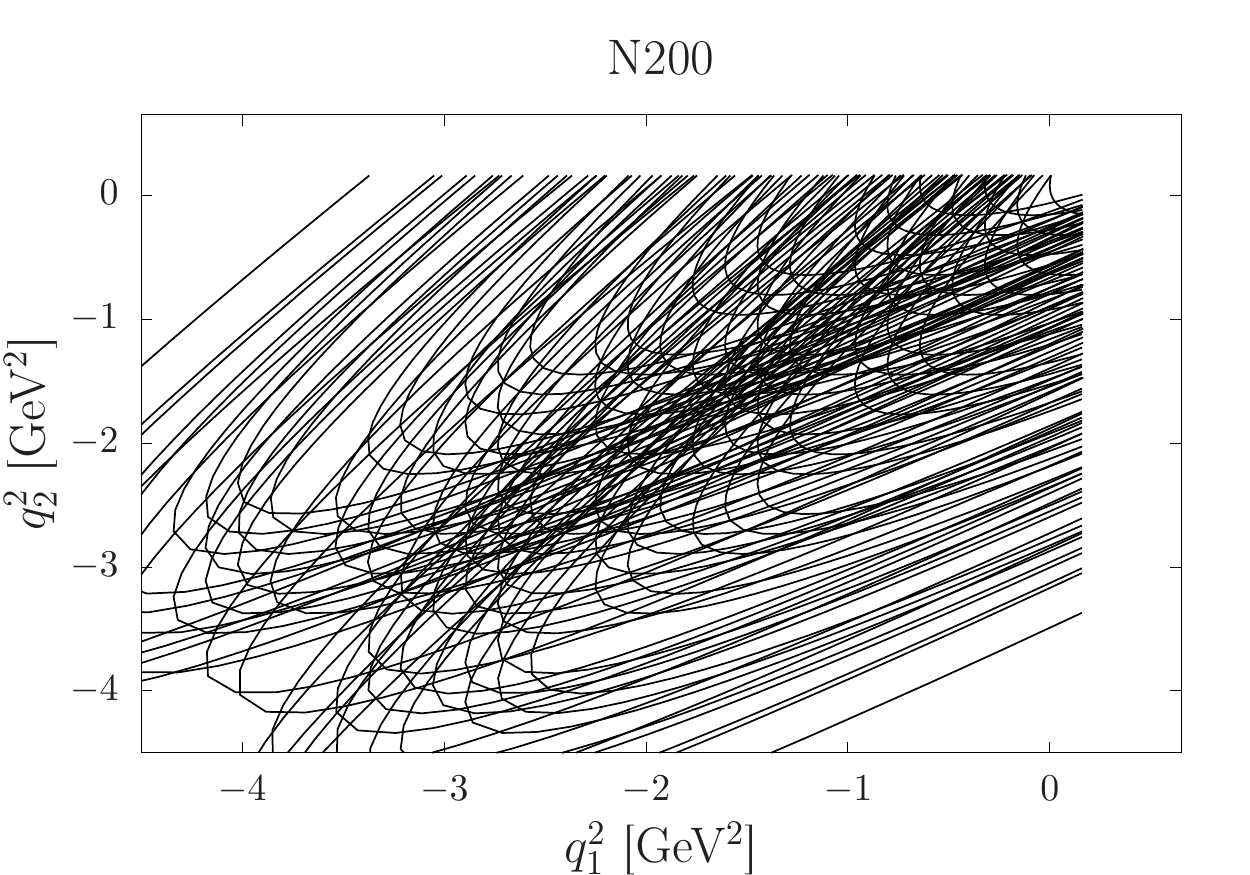}
	\caption{Kinematic reach in the photon virtualities ($q_1^2,q_2^2$) for the ensemble N200; see Table \ref{tab:sim} for its parameters. Left: Pion rest frame. Right: Moving frame where the pion has one unit of momentum in the $z$-direction $\vec{p} = (2\pi/L) \hat{z}$.}	
	\label{fig:kin}
\end{figure}

The TFF depends on the two virtualities $q_1^2$ and $q_2^2$.
Given that we use periodic boundary conditions in space, the kinematical range accessible on the lattice can be parametrized by 
\begin{align}
\begin{aligned}
q_1^2 &= \omega_1^2 - \vec{q}_1^{\, 2} \\
q_2^2 &= (E_{\pi} - \omega_1)^2 - (\vec{p}-\vec{q}_1)^2
\end{aligned}
 \quad \mathrm{with}  \quad \vec{q}_1 = \frac{2\pi}{L} \vec{n} \,, \quad \vec{n} \in \mathbb{Z}^3 \,.
\label{eq:kin}
\end{align}
Let the spatial momentum $\vec p$ of the pion  be fixed.  For a given $\vec q_1$, the second photon momentum $\vec q_2 = \vec p - \vec q_1$ is fully determined and the system (\ref{eq:kin}) describes a single curve in the $q_1^2,q_2^2$ plane, parametrized by $\omega_1$.  If we now consider a fixed value of $|\vec q_1|$, there is a finite number of realizations of $\vec q_1$, forming an orbit or a set of orbits\footnote{For instance, the vectors $\vec n=(3,0,0)$ and $\vec n=(2,2,1)$, which have the same norm, do not belong to the same lattice orbit.} on the reciprocal cubic lattice.  To this set of $\vec q_1$-values correspond in general several values of $|\vec q_2|^2$, each associated with a separate curve in the $(q_1^2,q_2^2)$ plane.  In our calculation, the three-point function evaluations obtained for a given set of $(|\vec q_1|,\,|\vec q_2|) $ are averaged over in order to increase the statistical precision. The presence of hypercubic artifacts, due to the breakdown of rotational symmetry on the lattice, is discussed in Sec.~\ref{sec:hypercubic}.

In Ref.~\cite{Gerardin:2016cqj}, we chose the pion to be at rest, and the corresponding orbits for the ensemble N200, whose parameters are given in Table \ref{tab:sim}, are shown in the left panel of Fig.~\ref{fig:kin}. This setup is well suited to probe large virtualities in the double-virtual case, the kinematical region where no experimental data are available. However, because of their large eccentricity, these orbits are limited to rather low virtualities in the single-virtual case. Here we therefore include a second frame where the pion has one unit of momentum in the $z$-direction. In this case, we can access larger virtualities for the single-virtual form factor, as can be seen in the right panel of Fig.~\ref{fig:kin}. In principle, even larger virtualities could be reached by increasing the pion momentum, but the signal-to-noise ratio would deteriorate rapidly. Moreover, since within our computational setup the pion interpolating operator is implemented using a sequential quark propagator, every new pion momentum $\vec{p}$ requires a new inversion of the Dirac operator, the most expensive part of the numerical
simulation.

\subsection{Decomposition of the integrand $\widetilde{A}_{\mu\nu}(\tau)$}

\begin{figure}[t]
	\includegraphics*[width=0.49\linewidth]{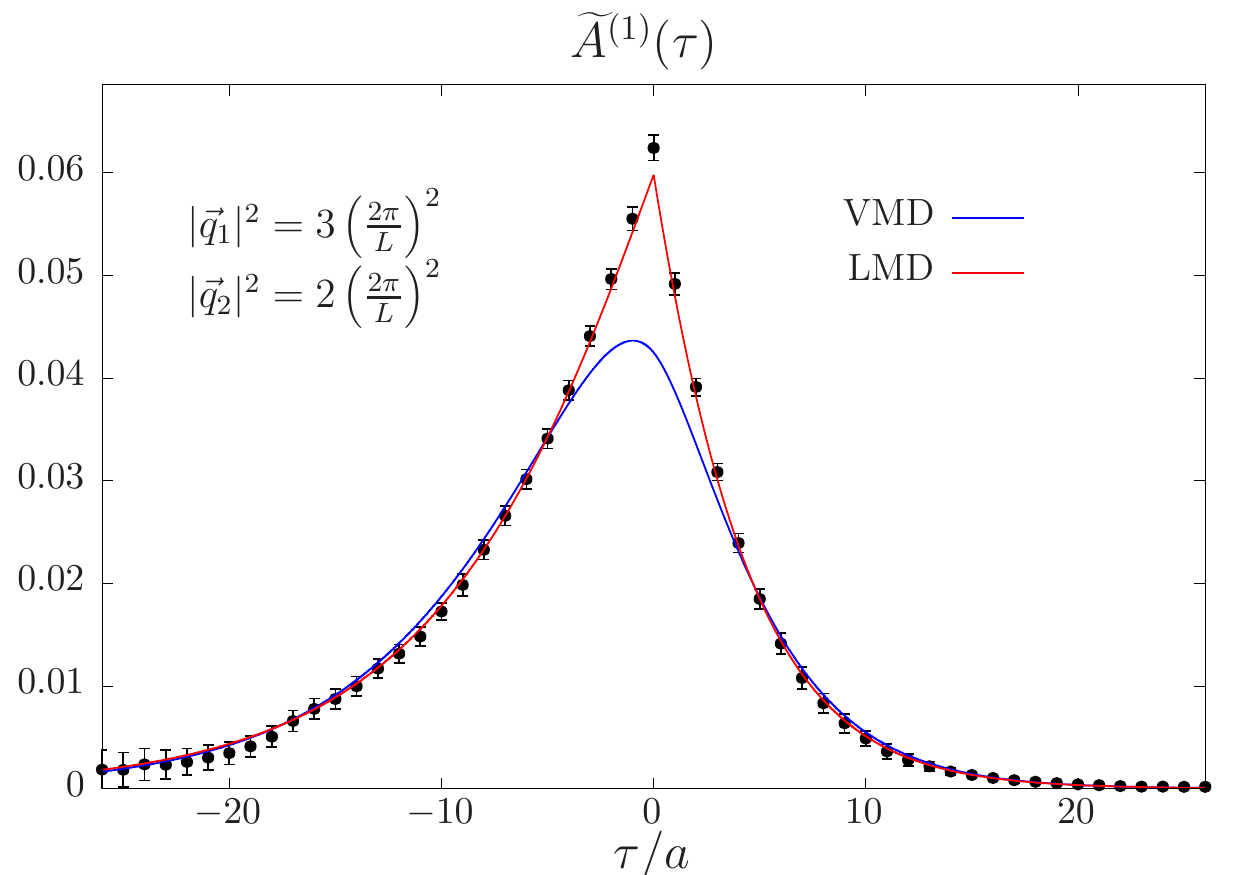}
	\includegraphics*[width=0.49\linewidth]{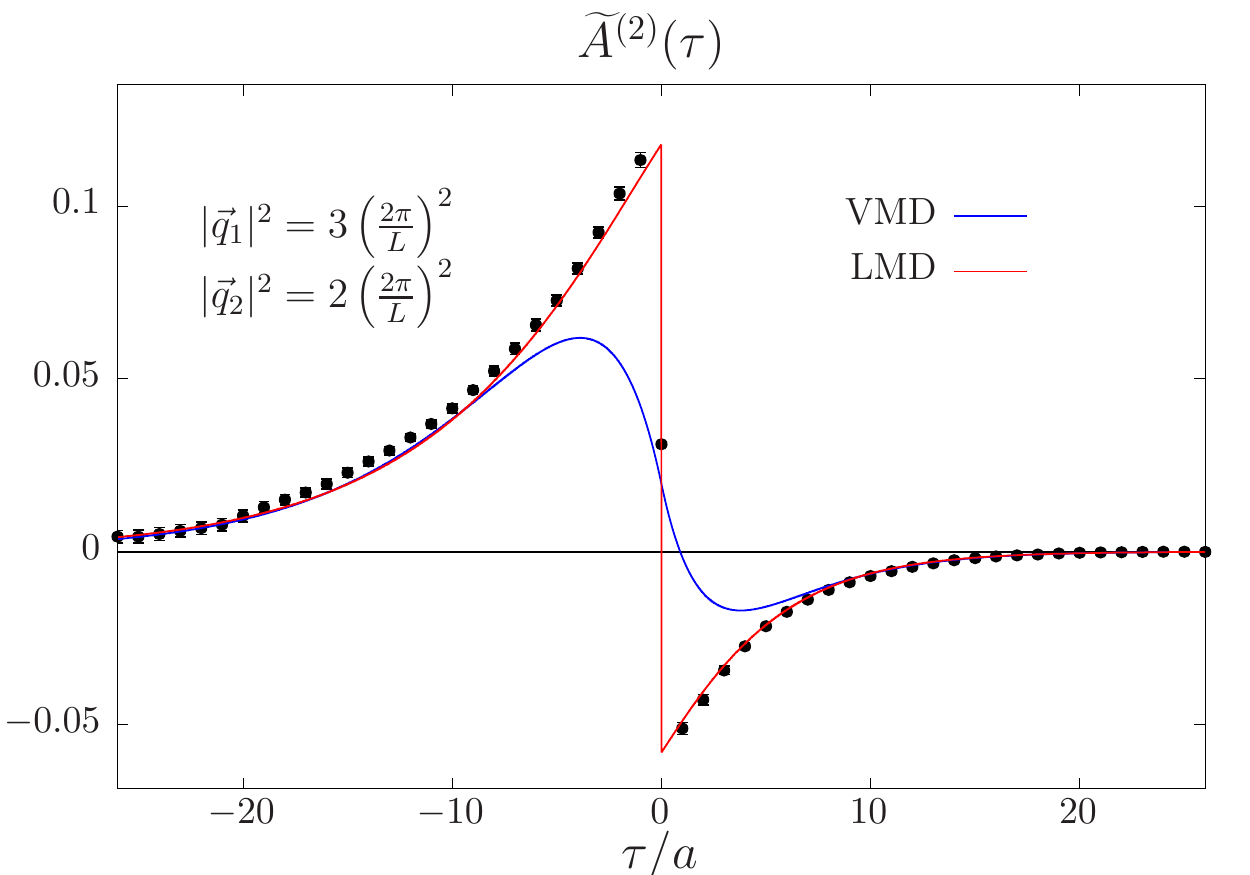}
	\caption{The functions $\widetilde{A}^{(1)}(\tau)$ and $\widetilde{A}^{(2)}(\tau)$ for two different orbits with $(|\vec{q}_1|^2 , |\vec{q}_2|^2)=\left(\frac{2\pi}{L}\right)^2(3,2)$ and $|\vec{p}| = 2\pi/L$ for ensemble D200, whose parameters are given in Table \ref{tab:sim}. Lattice data are in black. The blue and red lines correspond to a fit of the tail using the VMD and LMD models respectively, as explained in Sec.~\ref{sec:integration}.}	
	\label{fig:A}
\end{figure}

The main quantity of interest in the lattice calculation is the function $\widetilde{A}_{\mu\nu}(\tau)$ defined through Eq.~(\ref{eq:Amunu}) and directly related to the three-point correlation function. Starting from Eq.~(\ref{eq:Mlat}), which holds for all real values of $\omega_1$ such that $q_{1,2}^2<s_0$, we consider the analytic continuation for complex values of $\omega_1 = i \widetilde{\omega}$,
\begin{align}\label{eq:MunuAmunu}
M_{\mu\nu}^{\rm E}  = \frac{2 E_{\pi}}{Z_{\pi} } \int_{-\infty}^{\infty} \, \mathrm{d}\tau \, e^{i \widetilde{\omega} \tau} \, \widetilde{A}_{\mu\nu}(\tau)  \,.
\end{align}
Again, we do not write the dependence on the spatial momenta explicitly: $M_{\mu\nu}^{\rm E}$ is a function of $q_1$ and $q_2$ and   depends implicitly on $\widetilde{\omega}$. Relation (\ref{eq:MunuAmunu}) is valid as long as $\widetilde{A}_{\mu\nu}(\tau)$ falls off exponentially for $\tau\to-\infty$~\footnote{For $\tau\to+\infty$, an exponential falloff is guaranteed, irrespective of the kinematics.}. If $E_\pi<\sqrt{s_0}$, which holds in all cases considered in this paper, the relation is guaranteed to hold for all values of $(\vec q_1,\vec q_2)$. The inversion of the Fourier transform then yields
\begin{align}
\widetilde{A}_{\mu\nu}(\tau)  = \frac{ Z_{\pi}  }{ 4 \pi E_{\pi} } \int_{-\infty}^{\infty} \, \mathrm{d} \widetilde{\omega} \, 
M_{\mu\nu}^{\rm E} \, e^{- i \widetilde{\omega} \tau}  \,. 
\label{eq:A_tilde}
\end{align}
From Eq.~(\ref{eq:M}), we have $M_{\mu\nu} = q_{\mu\nu}\, \FF$, with
\begin{equation}
q_{\mu\nu} \equiv \epsilon_{\mu\nu\alpha\beta} q_1^{\alpha} q_2^{\beta}  = P_{\mu\nu} \omega_1 + Q_{\mu\nu}\,, 
\label{eq:qmunu}
\end{equation} 
where the coefficients $P_{\mu\nu}$ and $Q_{\mu\nu}$ do not depend on $\omega_1$. Thus, we can write
\begin{equation}
\widetilde{A}_{\mu\nu}(\tau)  = -i Q^{E}_{\mu\nu} \ \widetilde{A}^{(1)}(\tau) + P^{E}_{\mu\nu} \ \frac{ \mathrm{d} \widetilde{A}^{(1)} }{\mathrm{d}\tau}(\tau) \,,
\label{eq:decomp}
\end{equation}
with $P^{E}_{\mu\nu} = i P_{\mu\nu}$ and $Q^{E}_{\mu\nu} = (-i)^{n_0} Q_{\mu\nu}$ and where all the information is encoded into a single 
rotationally invariant function $\widetilde{A}^{(1)}$ defined through
\begin{equation}
\widetilde{A}^{(1)}(\tau) = \frac{iZ_{\pi} }{ 4 \pi E_{\pi} }  \int_{-\infty}^{\infty} \, \mathrm{d} \widetilde{\omega} \, \FF(q_1^2,q_2^2) e^{-i \widetilde{w} \tau} \,.
\label{eq:A1}
\end{equation}
In Eq.~(\ref{eq:A1}) it is understood that the arguments of the TFF are given by Eq.~(\ref{eq:kin}) with $\omega_1$ set to $i \widetilde{\omega}$.
In particular, the integral transform does not keep either virtuality fixed and $q_2^2$ even carries an imaginary part.

It is easiest to interpret the components of $\widetilde A_{\mu\nu}(\tau)$ in a spatially covariant notation using the Euclidean metric.
With $\vec\epsilon$ and $\vec\epsilon^{\,\prime}$ two unit vectors, we can write
\begin{subequations}
\begin{align} 
\label{eq:A0kmaster}
\widetilde{A}_{0k}(\tau) &= (\vec q_1\times \vec p\,)^k\, \widetilde{A}^{(1)}(\tau) \,, \\ 
\epsilon^{\prime k}\,\widetilde{A}_{kl}(\tau)\;\epsilon^{l} &= -i \big(\vec\epsilon^{\,\prime} \times \vec\epsilon \,\big)\cdot \Big( \vec q_1\, E_\pi  \, \widetilde{A}^{(1)}(\tau) +   \vec p\;\frac{d\widetilde{A}^{(1)}}{d\tau}\Big)\,. \label{eq:Aklmaster}
\end{align}
\end{subequations}
These are the master relations expressing $\widetilde A_{\mu\nu}(\tau)$ in terms of an integral transform of the pion transition form factor.
In the pion rest frame, $\widetilde A_{kl}(\tau)$ measures the transform $\widetilde A_1(\tau)$ of the transition form factor. At nonzero $\vec p$, $\widetilde A_{kl}(\tau)$ measures a linear combination of $\widetilde A_1(\tau)$ and its temporal derivative; in addition, the components $\widetilde{A}_{0k}(\tau)$ are proportional to $\widetilde A_1(\tau)$. For convenience, with a pion carrying one unit of momentum in the $z$-direction, we use the notation $\widetilde{A}^{(2)}(\tau)$ to write 
\begin{equation}
\widetilde{A}_{12}(\tau) \equiv -i E_{\pi} p_z \widetilde{A}^{(2)}(\tau), \qquad \vec{p}=(2\pi/L) \hat{z}. 
\end{equation}

We remark that the matrix elements $M_{\mu\nu}$, being proportional to the TFF, are real. Therefore, $M^E_{kl}$ are real and $M^E_{0k}$ are imaginary. From Eq.~(\ref{eq:Mlat}), we conclude that $\widetilde{A}_{kl}$ are imaginary and $\widetilde{A}_{0k}$ are 
real\footnote{Remember that the pion overlap factor $Z_{\pi}$ is imaginary.}.  Thus, the two scalar functions $\widetilde{A}^{(1)}(\tau)$ and $\widetilde{A}^{(2)}(\tau)$ are real. For the vector meson dominance (VMD) and the lowest meson dominance (LMD)~\cite{Moussallam:1994xp,Knecht:1999gb} models, explicit expressions for $\widetilde{A}_{\mu\nu}(\tau)$ are given in Appendix~\ref{app:A}.

Finally, from Eqs.~(\ref{eq:A_tilde}) and (\ref{eq:qmunu}), we deduce the symmetry 
\begin{equation}\label{eq:Bosym}
\widetilde{A}_{\mu\nu}(\tau; \vec{q}_1, \vec{q}_2) = \widetilde{A}_{\nu\mu}(-\tau; \vec{q}_2,\vec{q}_1) \, e^{-E_{\pi} \tau }, 
\end{equation}
which corresponds to the Bose symmetry of the two photons coupling to the pion. Here we have written explicitly the dependence of $\widetilde{A}_{\mu\nu}(\tau)$ on the spatial momenta. Equivalently, we have $\widetilde{A}^{(1)}(\tau;\vec{q}_1, \vec{q}_2)=\widetilde{A}^{(1)}(-\tau; \vec{q}_2, \vec{q}_1)e^{-E_{\pi} \tau }$. The relation (\ref{eq:Bosym}) can be checked explicitly in the case of the VMD and LMD models using the expressions provided in Appendix~\ref{app:A}. These symmetries are exploited in the analysis of the numerical data.

\subsection{Numerical integration \label{sec:integration}} 

\begin{figure}[t!]
	\includegraphics*[width=0.32\linewidth]{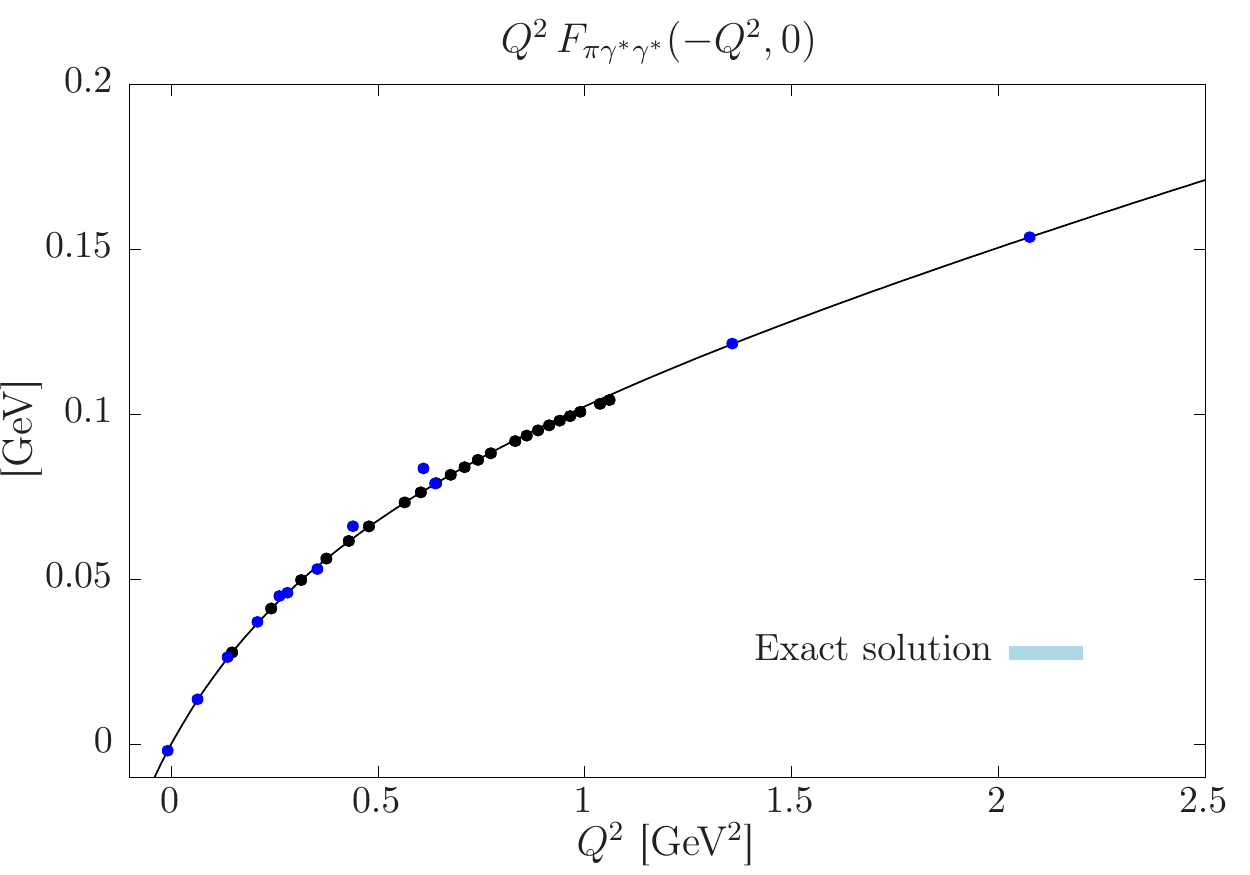} 
	\includegraphics*[width=0.32\linewidth]{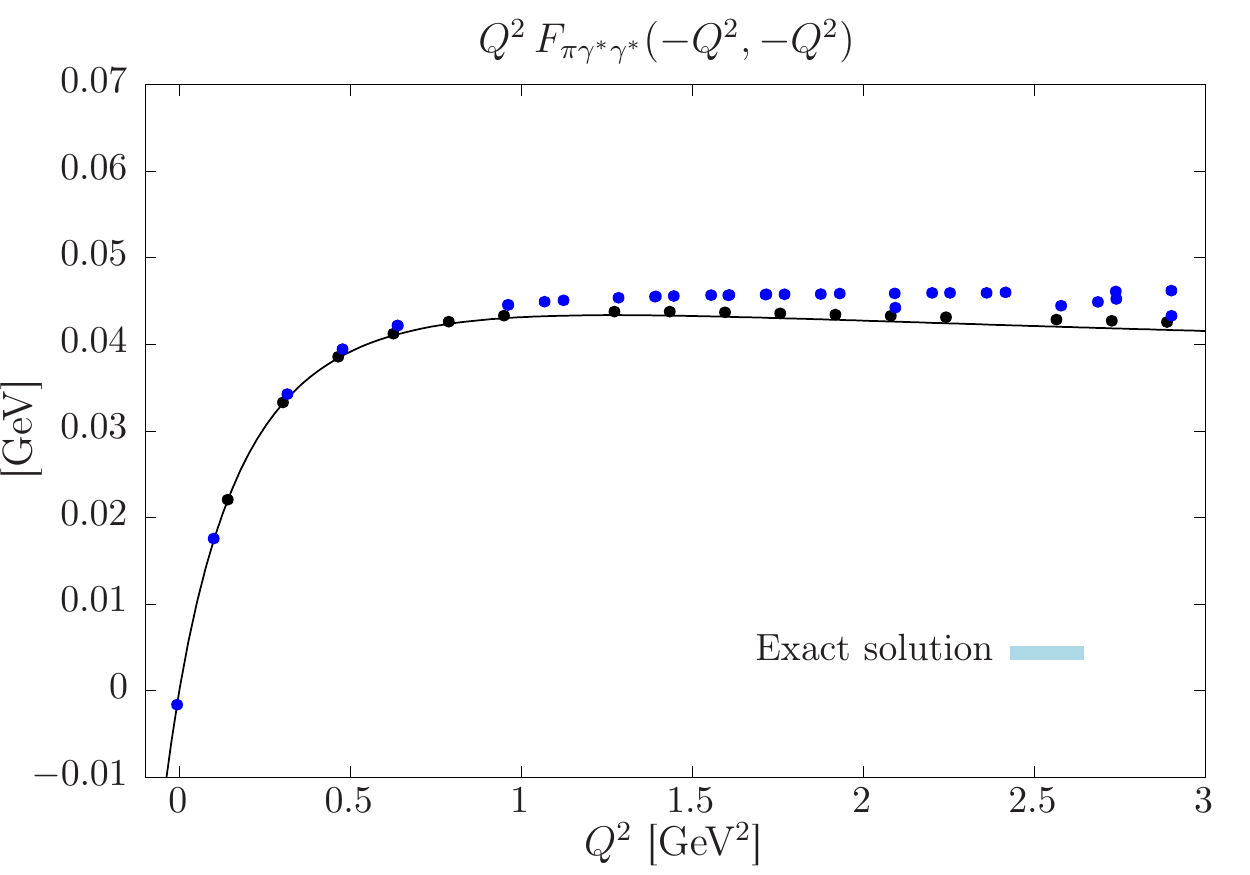}
	\includegraphics*[width=0.32\linewidth]{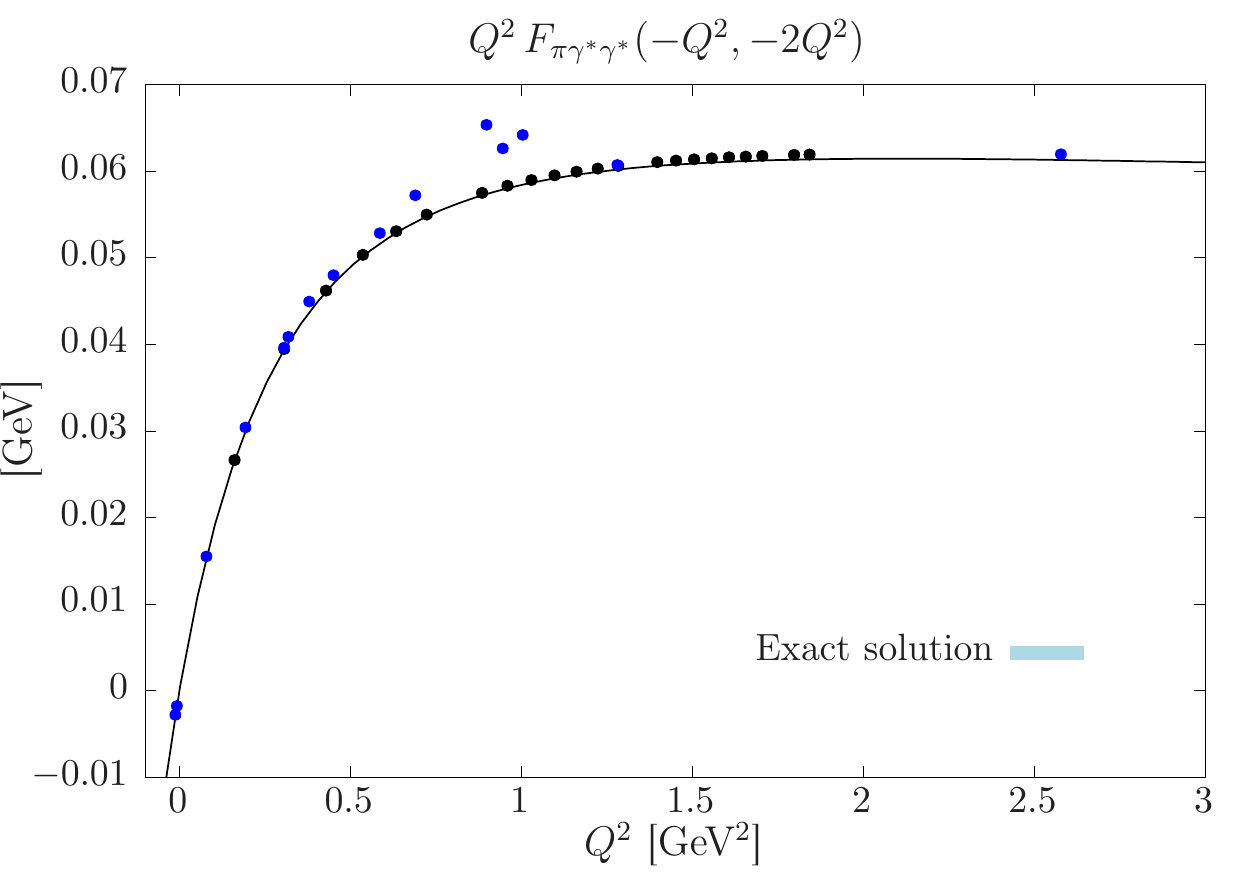}\\

	\includegraphics*[width=0.32\linewidth]{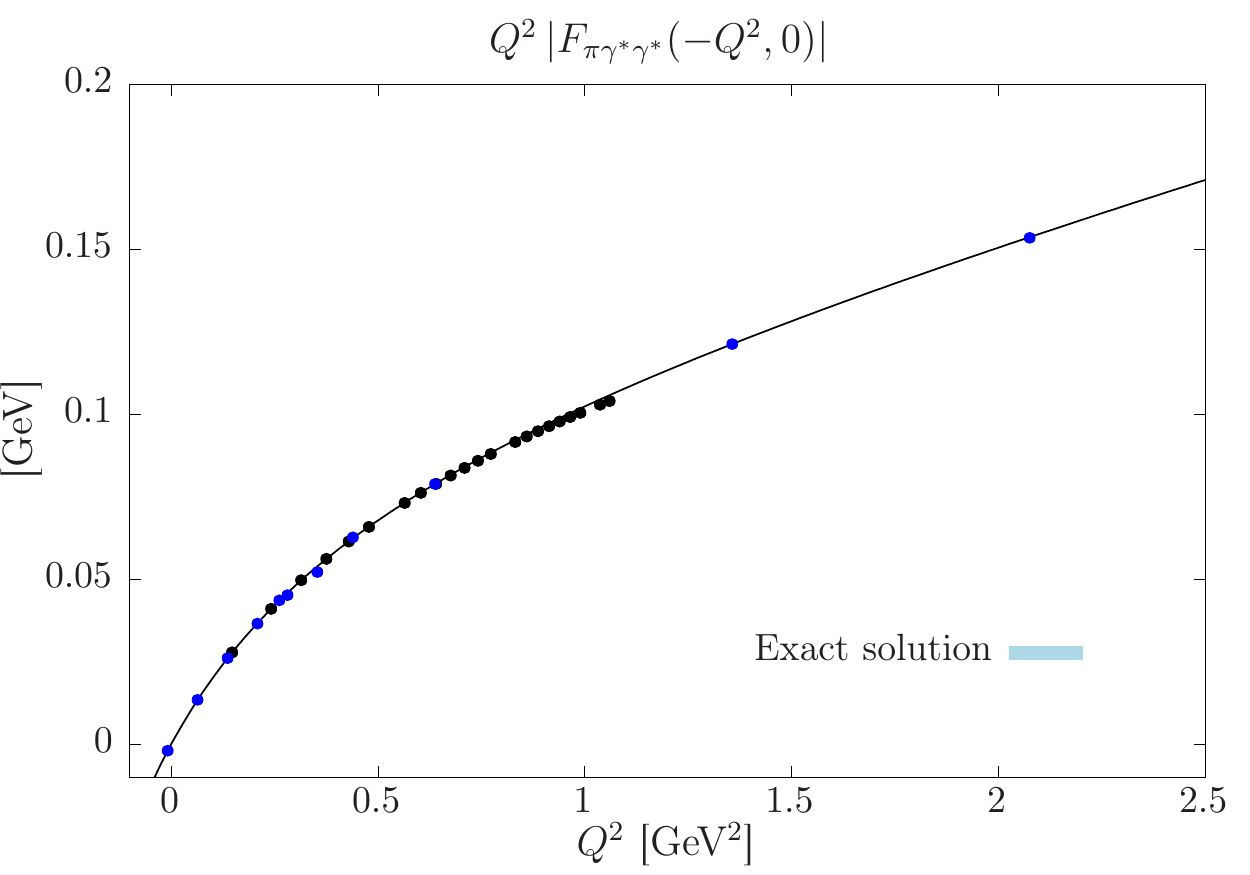}
	\includegraphics*[width=0.32\linewidth]{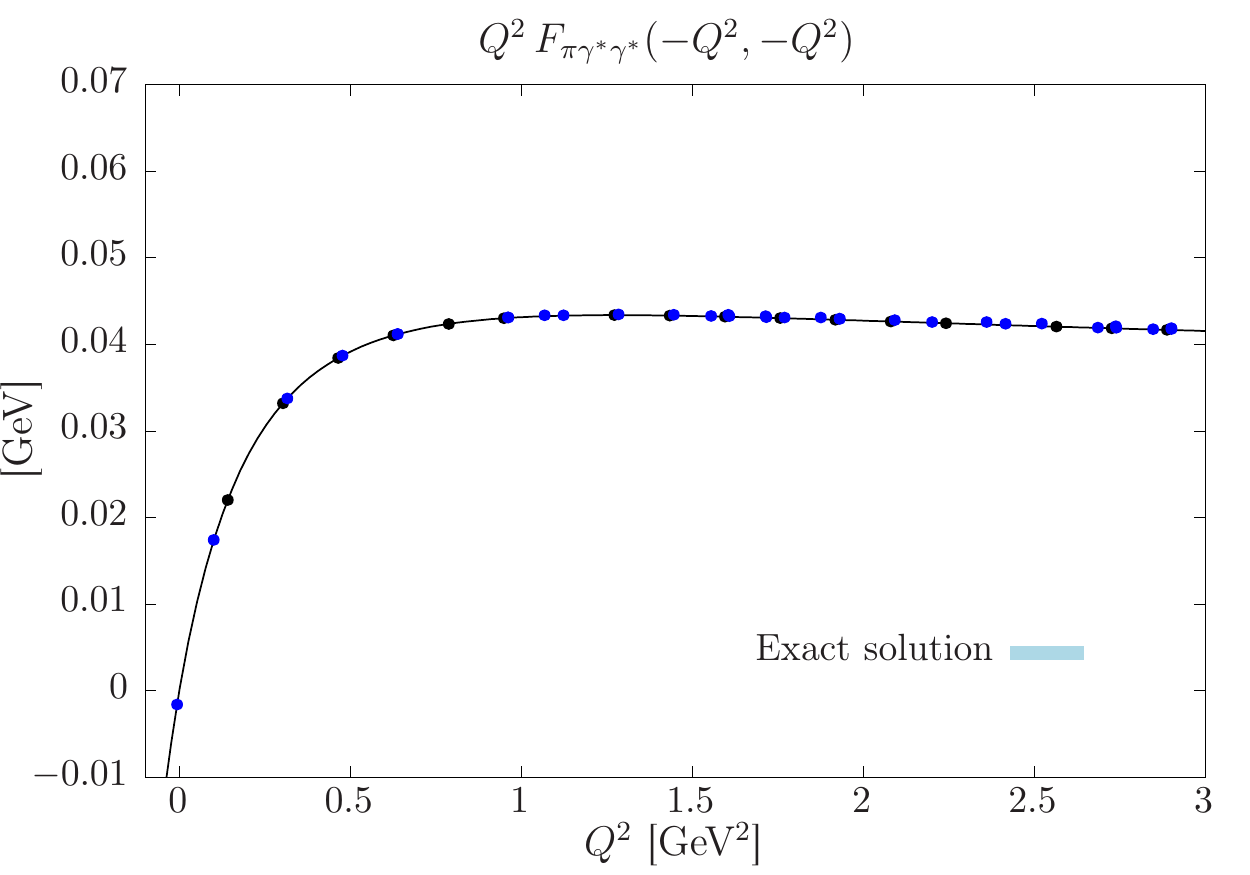}
	\includegraphics*[width=0.32\linewidth]{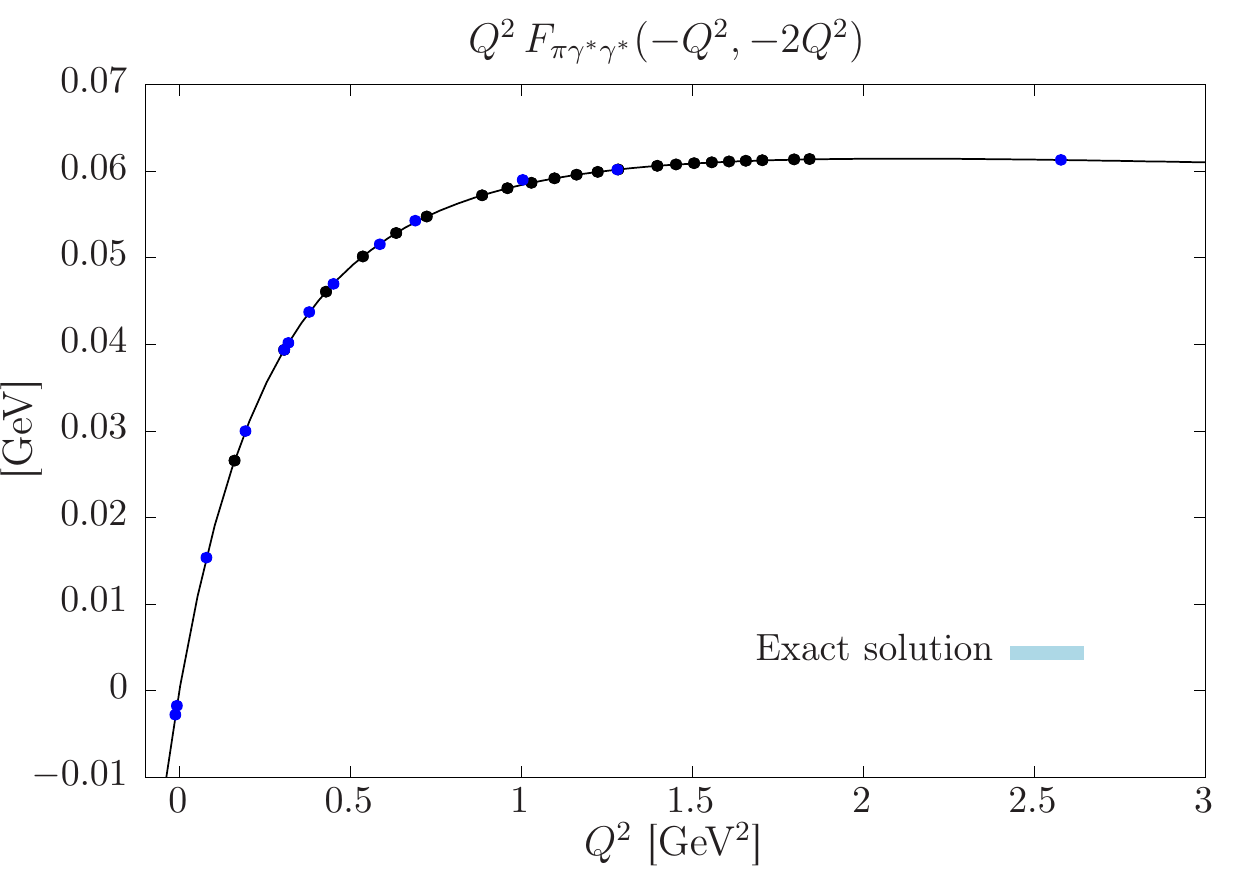}

	\caption{Numerical integration for data generated using a LMD model with lattice parameters close to N200. Blue (red) points correspond to the transition form factor $\FF(-Q_1^2,-Q_2^2)$ calculated using $\widetilde{A}^{(1)}(\tau)$ ($\widetilde{A}^{(2)}(\tau)$) respectively. The exact result is represented by the black line. Top: Using a trapezoidal integration rule. Bottom: Using a Simpson integration rule.} 	
	\label{fig:num_int}
\end{figure}

In the moving frame with $\vec{p} = (2\pi/L) \hat{z}$, both functions $\widetilde{A}^{(1)}(\tau)$ and $\widetilde{A}^{(2)}(\tau)$ are plotted in Fig.~\ref{fig:A}. The function $\widetilde{A}^{(1)}(\tau)$ is always positive with a cusp at $\tau=0$, related to the OPE at short distances, as discussed in Ref.~\cite{Gerardin:2016cqj}. On the contrary, the sign of $\widetilde{A}^{(2)}(\tau)$ changes at $\tau=0$ where the function is discontinuous. Therefore, the numerical integration is more challenging since large cancellations can occur in applying Eq.~(\ref{eq:Mlat}) to obtain the TFF. In particular, a naive replacement of the integral by a sum over discrete $\tau$ (trapezoidal rule) leads to noticeable numerical errors. A Simpson integration, however, reduces significantly this source of error as can be seen in Fig.~\ref{fig:num_int} : here, the results correspond to fictitious data where the three-point correlation function is generated assuming an LMD model with parameters obtained from a fit to the ensemble N200. When $\vec{p}=\vec{0}$, the function $\widetilde{A}^{(1)}(\tau)$ is always positive and we do not observe any significant difference between the two integration schemes.\\

Due to the finite time extent of the lattice, and because the signal deteriorates at large time separations $\tau$, the integration in Eq.~(\ref{eq:Mlat}) cannot be performed up to infinity. However, the VMD model is expected to give a good description of the data in a wide $\tau$ window, as shown in Appendix~\ref{app:A}.
Assuming a VMD parametrization, the functions $\widetilde{A}^{(1)}(\tau)$ and $\widetilde{A}^{(2)}(\tau)$, which depend on two parameters $\alpha$ and $M_V$, can be computed explicitly from Eqs.\ (\ref{eq:A1}, \ref{eq:A0kmaster}, \ref{eq:Aklmaster}) and the results are given in Appendix~\ref{app:A}. As in Ref.~\cite{Gerardin:2016cqj}, we start by fitting our data using the VMD parametrization at large $\tau$ where the normalization of the TFF, $\alpha$, is treated as a free fit parameter. The vector meson mass $M_V$ is set to the value extracted from the vector two-point correlation function evaluated on the same ensembles but with higher statistics. This fit is a global fit where all momenta and both $\widetilde{A}^{(1)}(\tau)$ and $\widetilde{A}^{(2)}(\tau)$ are fitted simultaneously. Then, we integrate over lattice data up to some cutoff $\tau_c$ and use the fit to evaluate the remaining part of the integral with $|\tau|>\tau_c$. The value of $\tau_c\gtrsim 1.5~\fm$ is chosen sufficiently large such that the two-point vector correlation function is dominated by the ground state. In practice, we only include points where more than $80\%$ of the integrand comes from lattice data.
We have also repeated the analysis with the LMD model and the difference between the VMD and LMD models is used to estimate the systematic uncertainty associated with the treatment of the tail. This systematic error is always smaller than the statistical error. Typical fits for our lightest ensemble, D200, are depicted in Fig.~\ref{fig:A}.

\subsection{Lattice ensembles, correlation functions and $\mathcal{O}(a)$-improvement}
\label{sec:ens}

This work is based on a subset of the $N_f=2+1$ Coordinated Lattice Simulations (CLS) ensembles~\cite{Bruno:2014jqa} generated using the openQCD suite \cite{Luscher:2012av}. They use $\mathcal{O}(a)$-improved Wilson fermions, with the nonperturbative coefficient $c_{\rm SW}$ determined in Ref.~\cite{Bulava:2013cta}, and tree-level $\mathcal{O}(a^2)$-improved Lüscher-Weisz action for the gauge field.
To avoid the freezing of the topological charge when approaching the continuum limit~\cite{Luscher:2012av,Luscher:2011kk}, CLS ensembles use periodic boundary conditions in space and open boundary conditions in time. The parameters of the simulations are summarized in Table~\ref{tab:sim}. We use four different lattice spacings in the range 0.050 - 0.086~fm and several pion masses down to 200~MeV to perform the continuum and chiral extrapolations. For the latter, the ensembles considered here have been generated keeping the average bare quark mass $m_{\rm q}^{\rm av} = (2 m_l + m_s)/3$ fixed and using two degenerate light quarks $m_u=m_d=m_l$. This chiral trajectory has the advantage of automatically keeping the $\mathcal{O}(a)$-improved bare coupling $\tilde g_0$ constant~\cite{Bietenholz:2010jr}. All our ensembles included in the final analysis satisfy $m_{\pi} L > 4$ such that finite-size effects are expected to be small. The scale setting was performed in Ref.~\cite{Bruno:2016plf} with a precision of $1\%$ using a linear combination of the pion and kaon decay constants. Statistical errors are estimated using the jackknife procedure and we propagate errors associated with the renormalization constant of the vector current, the scale setting, the pion mass $m_\pi$ and decay constant $f_\pi$ in the chiral extrapolations.\\

\begin{table}[t]
\renewcommand{\arraystretch}{1.1}
\caption{Parameters of the simulations: the bare coupling $\beta = 6/g_0^2$, the lattice resolution, the lattice spacing $a$ in physical units extracted from \cite{Bruno:2016plf}, the light and strange hopping parameters $\kappa_l$ and $\kappa_s$, the pion mass $m_{\pi}$, the ground state vector mass, the number of gauge configurations and the number of sources per configuration for the three-point correlation function. Ensembles with an asterisk are not included in the final analysis but used to control finite size-effects.}
\vskip 0.1in
\begin{tabular}{lcl@{\hskip 01em}c@{\hskip 02em}l@{\hskip 01em}l@{\hskip 01em}c@{\hskip 01em}c@{\hskip 01em}c@{\hskip 01em}l@{\hskip 01em}l}
\hline
Id	&	$\quad\beta\quad$	&	$L^3\times T$ 	&	$a~[\fm]$	&	$\kappa_l$		&	$\kappa_s$	&	$m_{\pi}~[\MeV]$	&	$m_{V}~[\MeV]$	& $m_{\pi}L$	&	$\#$confs	&	$\#$src \\
\hline
H101&	$3.40$	&	$32^3\times96$	& 	0.08636	& 0.136760 & 0.13675962	& 416(6) & 844(11) & 5.8 & 1000 & 10 \\
H102&			&	$32^3\times96$	& 			& 0.136865 & 0.13654934	& 354(5) & 832(09) & 5.0 & 1900 & 10 \\  
H105$^*$	&		&	$32^3\times96$	& 			& 0.136970 & 0.13634079	& 281(4) & 759(18) & 3.9 & 2800 & 10 \\  	
N101&			&	$48^3\times128$	& 			& 0.136970 & 0.13634079	& 280(4) & 774(08) & 5.9 & 1600 & 10 \\  	
C101&			&	$48^3\times96$	&			& 0.137030 & 0.13622204	& 224(3) & 741(10) & 4.7 & 2200 & 15 \\
\hline
S400	&	$3.46$	&	$32^3\times128$	& 	0.07634	& 0.136984 & 0.13670239	& 349(5)  & 821(09) & 4.3 & 1700 & 20 \\  	
N401&			&	$48^3\times128$	& 			& 0.137062 & 0.13654808	& 286(4)  & 793(09) & 5.3 & 950 & 10 \\  	
\hline
H200$^*$&$3.55$	&	$32^3\times96$	& 	0.06426	& 0.137000 & 0.137000	& 419(6)  & 875(17) & 4.4 & 2000 & 10\\  
N202&			&	$48^3\times128$	& 			& 0.137000 & 0.137000	& 411(5)  & 859(10) & 6.4 & 900 & 5 \\  
N203&			&	$48^3\times128$	& 			& 0.137080 & 0.13684028	& 346(5)  & 830(09) & 5.4 & 1500 & 10 \\ 
N200&			&	$48^3\times128$	& 			& 0.137140 & 0.13672086	& 284(3)  & 805(13) & 4.4 & 1700 & 10 \\ 
D200&			&	$64^3\times128$	& 			& 0.137200 & 0.13660175	& 200(3)  & 740(14) & 4.2 & 1100 & 30 \\  
\hline
N300&	$3.70$	&	$48^3\times128$	&	0.04981	& 0.137000 & 0.137000	& 422(5)  & 897(12) & 5.1	& 1200 & 5 \\
N302&			&	$48^3\times128$	&			& 0.137064 & 0.13687218	& 343(5)  & 856(16) & 4.2	& 1100 & 10 \\
J303	&			&	$64^3\times192$	&			& 0.137123 & 0.13675466	& 258(3)  & 796(09) & 4.2	& 650 & 10 \\
\hline
 \end{tabular} 
\label{tab:sim}
\end{table}

To reduce discretization effects and obtain a shorter continuum extrapolation than in~\cite{Gerardin:2016cqj}, on-shell $\mathcal{O}(a)$-improvement has been implemented. In addition to the improvement of the action, it requires the implementation of the renormalized $\mathcal{O}(a)$-improved vector current. Two discretizations are used, the local $(l)$ and the point-split $(c)$ lattice vector currents, 
\begin{subequations}
\begin{align}
V_{\mu}^{l,a}(x) &= \psib(x) \gamma_{\mu} \frac{\lambda^a}{2} \psi(x) \,,\\
V_{\mu}^{c,a}(x) &= \frac{1}{2} \left( \psib(x+a\hat{\mu})(1+\gamma_{\mu}) U^{\dag}_{\mu}(x) \frac{\lambda^a}{2} \psi(x) - \psib(x) (1-\gamma_{\mu} ) U_{\mu}(x) \frac{\lambda^a}{2} \psi(x+a\hat{\mu}) \right) \,,
\end{align}  
\end{subequations}
where $\lambda^a$ are the eight Gell-Mann matrices. With the tensor current defined as $\Sigma^{a}_{\mu\nu}(x) = -\frac{1}{2}\, \overline{\psi}(x) [\gamma_{\mu}, \gamma_{\nu}] \frac{\lambda^a}{2} \psi(x)$, the improved vector current is given by 
\begin{equation}
\label{eq:Vimp}
V^{I}_{\mu}(x) = V_{\mu}(x) + a\cV(g_0) \, \partial_{\nu} \Sigma_{\mu\nu}(x) 
\end{equation}
where the coefficient $\cV$ differs for the local and point-split vector currents and has been determined nonperturbatively in Ref.~\cite{Gerardin:2018kpy}.
The point-split vector current is conserved on the lattice and does not need to be further renormalized. For the local vector current, the renormalization pattern reads~\cite{Bhattacharya:2005rb}
\begin{multline}
\label{eq:Jimp}
\mathrm{tr}( \lambda V_{\mu} )_R = \ZV(\widetilde{g}_0) \Big[ \left(1+ 3 \, \bbV(g_0) \, a \mqav \right) \mathrm{tr}( \lambda V_{\mu}^{I} )  + \frac{1}{2} \bV(g_0) \, \mathrm{tr}( \{ \lambda, aM_{\rm q} \} V_{\mu}^{I} ) \\ +   \fV(g_0) \, \mathrm{tr}( \lambda\, aM_{\rm q} ) \, \mathrm{tr}(V_{\mu}^{I}) \Big] 
\end{multline}
where $\ZV$, $\bV$ and $\bbV$ have been evaluated nonperturbatively in Ref.~\cite{Gerardin:2018kpy} and $M_{\rm q}=\mathrm{diag}(m_l,m_l,m_s)$ is the bare (subtracted) quark mass matrix. The renormalized coupling is given by $\widetilde{g}_0^2 = g_0^2 (1 + \bg a \mqav)$. The coefficient $\fV$ starts at order $\mathcal{O}(g^6)$ in perturbation theory and is neglected here. For the electromagnetic current $J_{\mu}$ with up, down and strange quarks, it is convenient to use the isospin decomposition
\begin{equation}
J_\mu = \hat{V}_\mu^3 + \frac{1}{\sqrt{3}} \hat{V}_\mu^8 \,,
\label{eq:isodec}
\end{equation}
where $V_\mu^a = \bar\psi \gamma_\mu \frac{\lambda^a}{2}\psi$ is the octet of vector currents and the hat means that the current is both improved and renormalized. In particular, Eq.~(\ref{eq:Jimp}) reduces to
\begin{subequations}
\begin{align}
\hat{V}_\mu^3 &= Z_3 V^{3,I}_\mu \,, \\
\hat{V}_\mu^8 &= Z_8 V^{8,I}_\mu + Z_{80} V^{0,I}_\mu  \,,
\end{align}
\end{subequations}
with $V^{0,I}_\mu = \frac{1}{2} \bar\psi \gamma_\mu \psi$ the flavor-singlet current and
\begin{subequations}
\begin{align}
Z_3 &= \ZV \left[ 1 + 3\bbV a\mqav + \bV am_{l} \right] \,, \\
Z_8 &= \ZV \left[ 1 + 3\bbV a\mqav + \frac{\bV}{3} a( m_{l} + 2m_s) \right] \,, \\
Z_{80} &= \ZV \left( \frac{1}{3} \bV + \fV \right) \frac{2}{\sqrt{3}} a(m_l-m_s)  \,.
\end{align}
\end{subequations}
We use $P(x) = (1/\sqrt{2}) (\overline{u}(x) \gamma_{5} u(x) - \overline{d}(x) \gamma_{5} d(x))$ as an interpolating operator for the neutral pion. The renormalized and $\mathcal{O}(a)$-improved three-point correlation function appearing in Eq.~(\ref{eq:C3}), 
obtained with two local vector currents, reads  
\begin{align}
&\nonumber \langle J^{l}_{\mu}(z) J^{l}_{\nu}(y) P^{\dag}(x) \rangle = \frac{ \sqrt{2} }{3} Z_3 \left( Z_8 + \sqrt{3} Z_{80} \right) \mathrm{Re} \; \mathrm{Tr} \left[ G_l(x,z) \gamma_{\mu} G_l(z,y) \gamma_{\nu} G_l(y,x) \gamma_5 \right]  \\
\nonumber &- \frac{\sqrt{2}}{6} Z_3   \left[   \left( Z_8 + \sqrt{3} Z_{80} \right) \mathrm{Tr} \left[ G_l(y,y) \gamma_{\nu} \right]  -  \left( Z_8 - \frac{ \sqrt{3} }{2} Z_{80} \right) \mathrm{Tr} \left[ G_s(y,y) \gamma_{\nu} \right]     \right]     \mathrm{Tr} \left[ G_l(z,x) \gamma_{5} G_l(x,z) \gamma_{\mu} \right] \; \\
&- \frac{\sqrt{2}}{6} Z_3   \left[   \left( Z_8 + \sqrt{3} Z_{80} \right) \mathrm{Tr} \left[ G_l(z,z) \gamma_{\mu} \right]  -  \left( Z_8 - \frac{ \sqrt{3} }{2} Z_{80} \right) \mathrm{Tr} \left[ G_s(z,z) \gamma_{\mu} \right]     \right]        \mathrm{Tr} \left[ G_l(y,x) \gamma_{5} G_l(x,y) \gamma_{\nu} \right],
\end{align}
where the first line corresponds to the connected part and the second and third lines to the disconnected parts. We did not explicitly write the improvement term proportional to $\cV$. Here, $G_l$ and $G_s$ denote the light and strange quark propagators respectively. If $\fV$ is neglected, the connected part renormalizes proportionally to $Z_3^2$, since $Z_8 + \sqrt{3} Z_{80} = Z_3 - 2 \ZV \fV\, a (m_s-m_l)$.
Similarly, with one local and one conserved vector currents, one finds
\begin{align}
\nonumber &\langle J^{l}_{\mu}(z) J^{c}_{\nu}(y) P^{\dag}(x) \rangle = \frac{ \sqrt{2} }{6} \left( Z_3 + Z_8 + \sqrt{3} Z_{80} \right) \mathrm{Re} \; \mathrm{Tr} \left[ G_l(x,z) \gamma_{\mu} G_l(z,y^{\prime}) \mathcal{V}^{(y)}_{\nu}(y^{\prime},y^{\dprime}) G_l(y^{\dprime},x) \gamma_5 \right]  \\
\nonumber &- \frac{\sqrt{2}}{6} \left[   \left( Z_8 + \sqrt{3} Z_{80} \right) \mathrm{Tr} \left[ G_l(z,z) \gamma_{\mu} \right]  -  \left( Z_8 - \frac{ \sqrt{3} }{2} Z_{80} \right) \mathrm{Tr} \left[ G_s(z,z) \gamma_{\mu} \right]     \right]  \\
\nonumber & \times   \mathrm{Tr} \left[ G_l(y^{\dprime},x) \gamma_{5} G_l(x,y^{\prime}) \mathcal{V}^{(y)}_{\nu}(y^{\prime},y^{\dprime}) \right] \; \\
&- \frac{\sqrt{2}}{6} Z_3   \Big[  \mathrm{Tr} \left[ G_l(y^{\dprime},y^{\prime})  \mathcal{V}^{(y)}_{\nu}(y^{\prime},y^{\dprime}) \right]  - \mathrm{Tr} \left[ G_s(y^{\dprime},y^{\prime}) \mathcal{V}^{(y)}_{\nu}(y^{\prime},y^{\dprime})   \right]     \Big]        \mathrm{Tr} \left[ G_l(z,x) \gamma_{5} G_l(x,z) \gamma_{\mu} \right] \,,
\end{align}
where a summation over $y^{\prime}$ and $y^{\dprime}$ is understood and where 
\begin{equation}
\mathcal{V}^{(y)}_{\mu}(y^{\prime},y^{\dprime}) G(y^{\dprime},x) 
= \frac{1}{2} \left( \delta_{y^{\prime},y+a\hat{\mu}} [1+\gamma_{\mu}] U_{\mu}^{\dag}(y) G(y,x) - \delta_{y^{\prime},y} [1-\gamma_{\mu}] U_{\mu}(y) G(y+a\hat{\mu},x) \right) \,.
\end{equation} 
The lattice quark propagator is $\gamma_5$-Hermitian, $G^{\dag} = \gamma_5 G \gamma_5$, and the linear operator $\mathcal{V}^{(y)}_{\mu}$ is anti $\gamma_5$-Hermitian, $\mathcal{V}^{(y) \dag}_{\mu} = - \gamma_5 \mathcal{V}^{(y)}_{\mu} \gamma_5$.\\

\begin{figure}[t]
	\includegraphics*[width=0.49\linewidth]{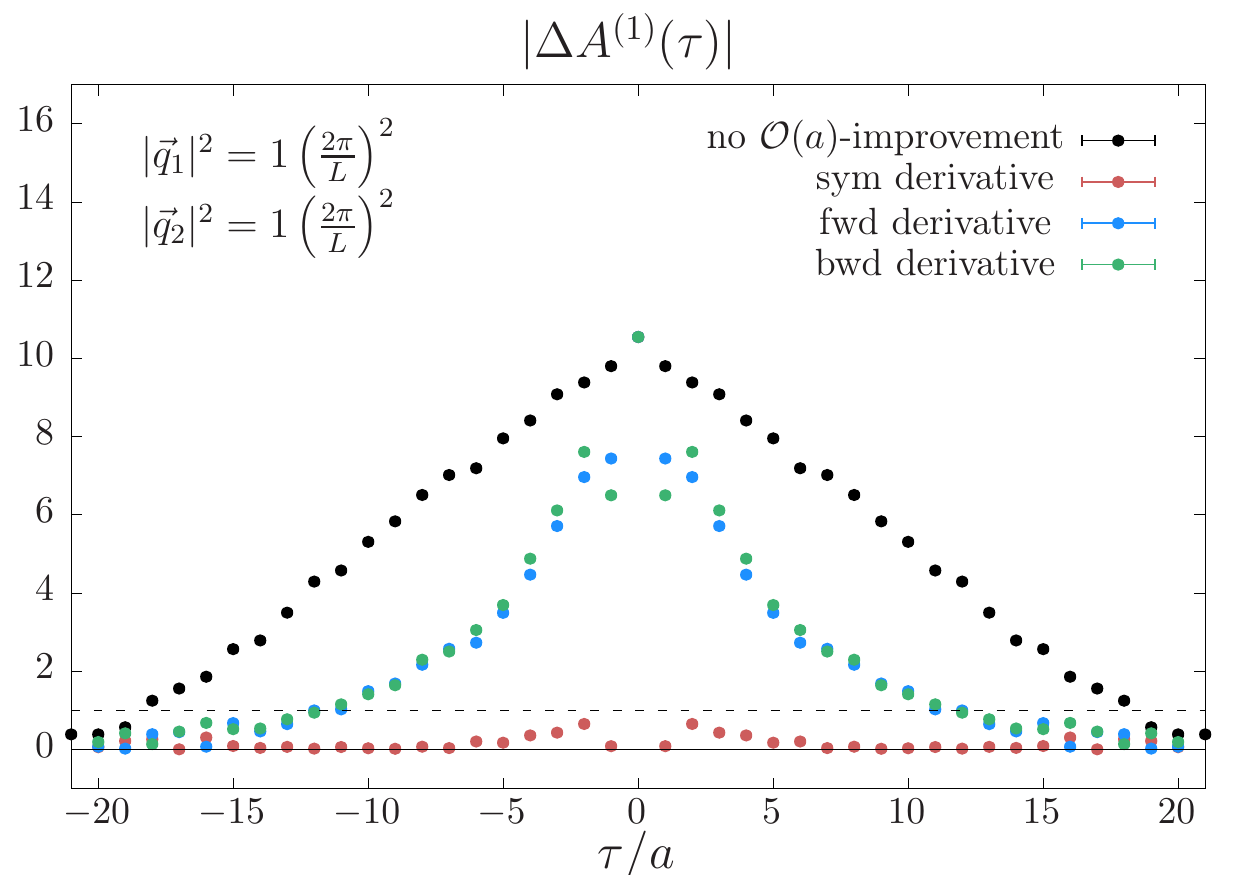}
	\includegraphics*[width=0.49\linewidth]{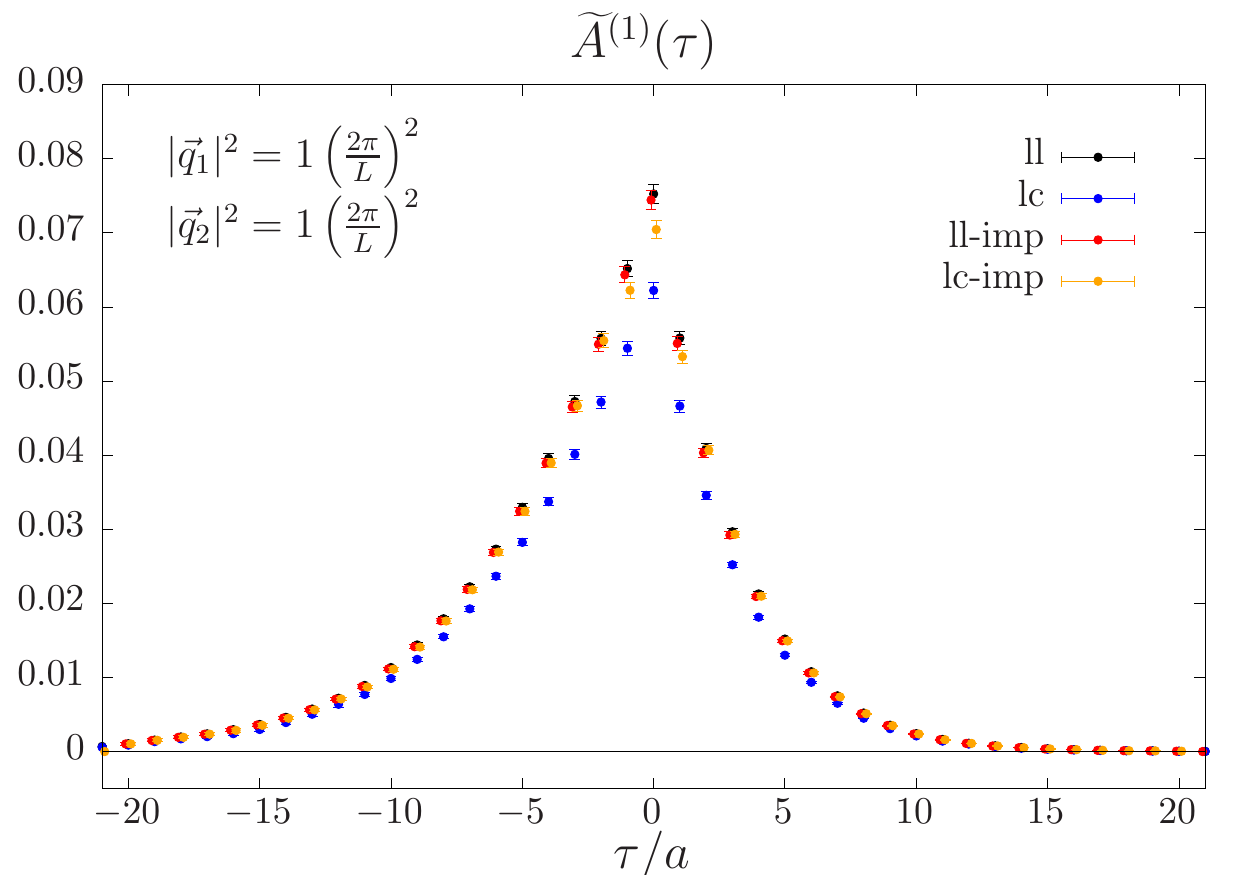}
	
	\caption{ Left: Influence of the discretization of the lattice derivative used to implement $\mathcal{O}(a)$-improvement on the function $A^{(1)}(\tau)$. Black points correspond to $\Delta \widetilde{A}^{(1)}(\tau)$ before improvement. Red, blue and green points correspond to $\Delta \widetilde{A}^{(1)}(\tau)$ after improvement using respectively the symmetric, forward or backward derivative at the sink. Right: The function $A^{(1)}(\tau)$ for different discretizations of the vector current. Black and blue (red and orange) points correspond to local-local and local-conserved correlation functions without (with) improvement of the vector currents using the symmetric lattice derivative at the sink. Results correspond to the ensemble H102.}	
	\label{fig:improvement}
\end{figure}
The three-point correlation function is computed using the same technique as in Ref.~\cite{Gerardin:2016cqj}. A point source is created on the time slice $y_0$ with a random value of $\vec{y}$ and a sequential propagator is computed at $x_0$, for two values of the pion spatial momenta $\vec{p} = \vec{0}$ and $\vec{p} = (2\pi/L) \hat{z}$. With our computational setup we obtain all values of the photon momenta $\vec{q}_1$ and $\vec{q}_2$ 
such that $\vec q_1+\vec q_2=\vec p$ without any new inversion of the Dirac operator. The number of sources per gauge configuration is given in Table~\ref{tab:sim} and the location of $y_0$ is discussed in Sec.~\ref{sec:2pt}. For all ensembles, we are able to probe photon virtualities up to $3~\GeV^2$  in the double-virtual case and up to $1.5~\GeV^2$ in the single-virtual case. This is a major improvement compared to our previous study, where we were limited to virtualities below $0.5~\GeV^2$ in the single-virtual case.\\

In Eq.~(\ref{eq:Vimp}), we have some freedom in the choice of the lattice derivative: different definitions differ by $\mathcal{O}(a)$ effects and therefore contribute only to order $a^2$ in correlation functions. 
We have considered the symmetric, the backward and the forward derivatives, respectively, given by
\begin{equation}
\partial_{\mu}^{s} f(x) = \frac{ f(x+a)-f(x-a) }{2a} \,, \quad \partial_{\mu}^{b} f(x) = \frac{ f(x)-f(x-a) }{a} \,, \quad \partial_{\mu}^{f} f(x) = \frac{ f(x+a)-f(x) }{a} \,.
\end{equation}
For the vector current located at the source, we choose the forward derivative $\partial_{\mu}^{f} f(x)$, since a symmetric derivative would require more inversions of the Dirac operator.
For the vector current located at the sink, we have compared the three different discretizations. We define the ratio 
\begin{equation}
\Delta \widetilde{A}^{(1)}(\tau) = \frac{| \widetilde{A}^{(1), {\rm lc}}(\tau) - \widetilde{A}^{(1), {\rm ll}}(\tau) | }{ \delta \widetilde{A}^{(1), {\rm lc}}(\tau) } \,,
\end{equation}
where $\widetilde{A}^{(1), {\rm lc}}(\tau)$ and $\widetilde{A}^{(1), {\rm ll}}(\tau)$, defined in Eq.~(\ref{eq:Mlat}), are, respectively, computed using one or two local vector currents  and $\delta \widetilde{A}^{(1), {\rm lc}}(\tau)$ is the statistical error associated with $\widetilde{A}^{(1), {\rm lc}}(\tau)$. In the left panel of Fig.~\ref{fig:improvement}, this ratio is plotted with and without improvement, using different discretizations of the lattice derivative. After $\mathcal{O}(a)$-improvement, the discrepancy between both discretizations is reduced as expected and the symmetric derivative turns out to be the best choice.

Finally, we note that on-shell improvement of correlation functions is not sufficient for $\tau={\rm O}(a)$. 
For $\tau = \pm a$, we do not use the symmetric derivative but the forward and backward derivatives, respectively, to avoid using the correlation function at $\tau = 0$ where $A^{(2)}(\tau)$ is discontinuous (see right panel of Fig.~\ref{fig:A})). In the right panel of Fig.~\ref{fig:improvement} we plot $\widetilde{A}(\tau)$ using both discretizations ($\widetilde{A}^{ll}(\tau)$ and $\widetilde{A}^{lc}(\tau)$) with and without improvement of the vector currents using the symmetric derivative. We observe that after improvement, with the exception of $\tau=0$, the two discretizations are in excellent agreement with each other.

\section{Results \label{sec:results} }

\subsection{Two-point pseudoscalar correlation function}
\label{sec:2pt}

\begin{figure}[t!]
	\includegraphics*[width=0.49\linewidth]{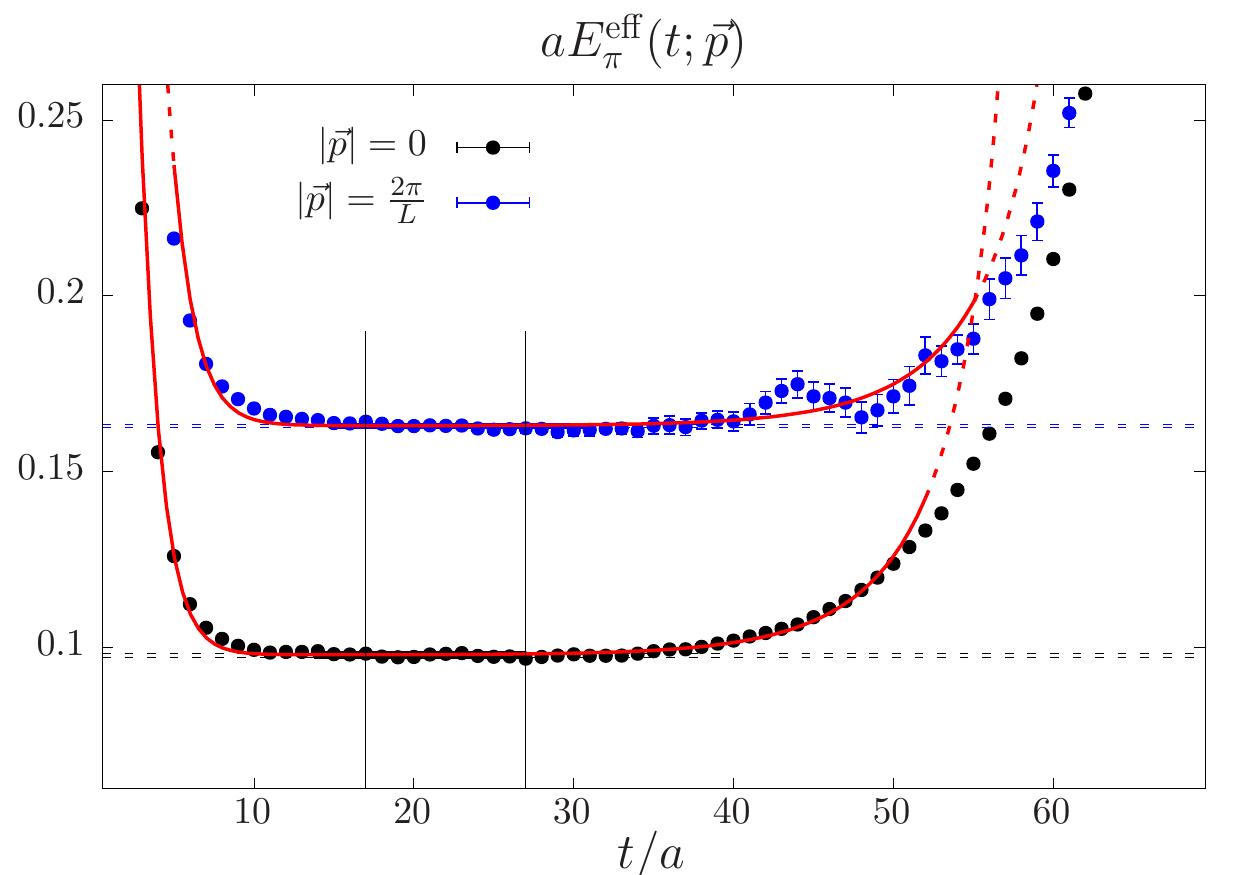}
	\includegraphics*[width=0.49\linewidth]{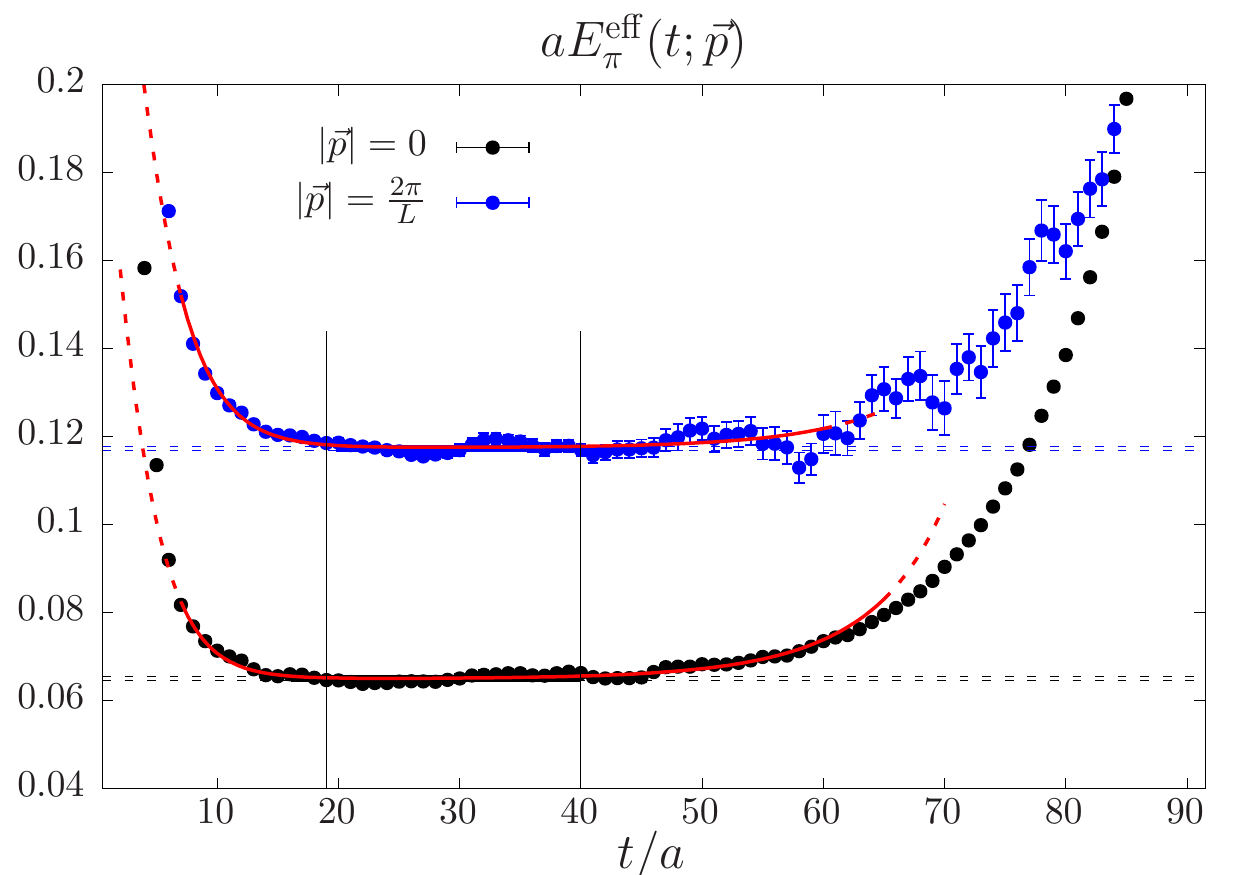}

	\caption{Effective masses given by Eq.~(\ref{eq:Eeff}) for the ensembles C101 (left panel) and D200 (right panel) with $t=x_0-y_0$. Black and blue points correspond respectively to $|\vec{p}| = 0$ and $|\vec{p}| = 2\pi/L$. The red lines correspond to the effective masses obtained from the fit using Eq.~(\ref{eq:cor_fit}). The vertical lines indicate the plateau region and the dashed horizontal lines the extracted pion energy and its error.}
	\label{fig:mass_fit}
\end{figure}  

The pion energy $E_{\pi}(\vec{p})$ and its overlap $Z_{\pi}$ with our interpolating operator are extracted from the pseudoscalar two-point correlation function. 
In the limit of large source-sink time separation, and with both source and sink far from the boundary, the correlation function behaves as 
\begin{equation}
C_{PP}(x_0,y_0; \vec{p}) =  a^3 \sum_{\vec{x}} \langle P(\vec{x},x_0) P(\vec{0},y_0) \rangle \, e^{-i \vec{p} \vec{x}}     \xrightarrow[ |x_0-y_0| \to \infty]{}     \frac{Z^2_{\pi}}{2E_{\pi}(\vec{p})} \, e^{-E_{\pi}(\vec{p}) (x_0-y_0)} \,.
\end{equation}
We want to extract $E_{\pi}(\vec{p})$ and $Z_{\pi}$ in a region where the excited-state contribution and boundary effects are both small. We therefore place the source far from the boundary located at $x_0=0$. In Fig.~\ref{fig:mass_fit}, we show the result for the effective mass defined by
\begin{equation}
aE_{\pi}^{\rm eff}(t;\vec{p}) = -\log\left( \frac{ C_{PP}(x_0+a,y_0;\vec{p}) }{ C_{PP}(x_0,y_0;\vec{p}) } \right) \,.
\label{eq:Eeff}
\end{equation}
For large source-sink time separation, we observe a plateau up to some time where the sink starts to be close to the boundary located at $x_0=T-a$. 
In the case of open boundary conditions in time, and including only the first excited state, we first fit our data using the \textit{Ansatz}
\begin{multline}
C_{PP}(x_0,y_0; \vec{p}) =  \frac{Z_{\pi}^2}{2E_{\pi}(\vec{p})} \, e^{-E_{\pi}(\vec{p}) (x_0-y_0)} + \frac{Z_{\pi}^{\prime 2}}{2E_{\pi}^{\prime}(\vec{p})} \, e^{-E_{\pi}^{\prime}(\vec{p}) (x_0-y_0)}  \\
+ A(\vec{p})  \, e^{-E_{\pi}(\vec{p}) (x_0-y_0)} \left( e^{-E_{2\pi} y_0} + e^{-E_{2\pi} (T-x_0)} \right)\,,
\label{eq:cor_fit}
\end{multline}
where $E_{\pi}^{\prime}(\vec{p})$ and $Z_{\pi}^{\prime}$ are the energy and the overlap of the first excited state with momentum $\vec{p}$ in the pseudoscalar channel and the third term includes the first boundary excited state: a finite volume two-pion state with vanishing momentum. In the fits, we assume that $E_{2\pi} = 2 E_{\pi}(\vec{0})$. Explicitly, denoting $|B\rangle$ the boundary state, the prefactor $A(\vec{p})$ is given by
\begin{equation}
A(\vec{p}) = \frac{ \langle 2\pi | B \rangle }{ \langle 0 | B \rangle } \frac{ \langle 0 | P | \pi \rangle \langle \pi | P | 2\pi \rangle }{ 2E_{2\pi} \, 2E_{\pi} } \,.
\end{equation}
To reduce the number of fit parameters, both momenta are fitted simultaneously. In a second step, the data are fitted using a single-exponential ansatz in the region where excited state contributions are small compared to the statistical precision, and the results are given in Table~\ref{tab:spectrum}. We note that the pion energies are compatible with the relativistic dispersion relation. 
Finally, we have checked that the source is sufficiently far from the boundary such that the overlaps are correctly extracted: the last term in Eq.~(\ref{eq:cor_fit}) contributes to $Z_{\pi}$ as less than half its statistical error. 
The same value of $y_0$ is used to compute the three-point correlation function. The statistical error on the ratio $E_{\pi}/Z_{\pi}$ in Eq.~(\ref{eq:Mlat}) lies between 0.5\% and 0.8\%. \\

\renewcommand{\arraystretch}{1.1}
\begin{table}[t]
\caption{Ground state energies $E_{\pi}(\vec{p})$ for the two values of the pion momentum and overlap factors $Z_{\pi}$ for each lattice ensemble. We also give the PCAC mass and the pion decay constant in lattice units without finite-size effect correction.} 
\begin{center}

\begin{tabular}{l @{\hskip 1.5em} S[table-format=0.4(2)] @{\hskip 1.5em} S[table-format=0.4(2)] @{\hskip 1.5em} S[table-format=0.4(2)] @{\hskip 1.5em} S[table-format=0.6(2),group-digits = false] @{\hskip 1.5em} S[table-format=0.6(2),group-digits = false] }
\hline
Id	  & 	${iZ_{\pi}}$	&	${aE_{\pi}(\vec{0})}$		&	${aE_{\pi}(\vec{p})}$ 	&	${am_{\rm PCAC}}$& ${a\sqrt2f_{\pi}}$ \\ 
\hline 
H101  & 	0.2139(13)	&	0.1821(07)	& 	0.2671(07)	&	0.009177(45)	&	0.06458(29) 	\\ 
H102  & 	0.2086(19)	&	0.1550(09)	& 	0.2492(09)	&	0.006499(51)	&	0.06151(27)	\\ 
H105  & 	0.2004(23) 	&	0.1230(12) 	& 	0.2316(11)	&	0.003985(57)	&	0.05802(39)	\\ 
N101  &	0.2011(15)	&	0.1224(05)	&	0.1792(05) 	&	0.003990(32)	&	0.05832(30)	\\ 
C101  & 	0.1947(11) 	&	0.0982(06)	& 	0.1629(05)	&	0.002435(29)	&	0.05534(37)	\\ 
\hline 
S400  & 	0.1606(16) 	&	0.1352(08)	& 	0.2387(09)	&	0.005684(27)	&	0.05463(21)	\\ 
N401  & 	0.1578(10)	&	0.1106(05)	& 	0.1714(05)	&	0.003770(28)	&	0.05327(17)	\\ 
\hline 
H200  & 	0.1160(12) 	&	0.1365(07)	& 	0.2382(10)	&	0.006863(24)	&	0.04805(27)	\\ 
N202  & 	0.1155(10)	&	0.1337(05)	& 	0.1870(06)	&	0.006866(15)	&	0.04884(18) 	\\ 
N203  & 	0.1135(08)	&	0.1127(04)	&	0.1728(05)	&	0.004749(17)	&	0.04699(16)	\\ 
N200  & 	0.1094(08)	&	0.0924(05) 	& 	0.1599(05)	&	0.003145(15)	&	0.04454(18)	\\ 
D200  & 	0.1044(08)	&	0.0650(04)	&	0.1173(05)	&	0.001554(12)	&	0.04259(17)	\\ 
\hline 
N300  & 	0.0700(07)	&	0.1066(04)	& 	0.1685(06)	&	0.005507(09)	&	0.03811(14)	\\ 
N302  & 	0.0646(08)	&	0.0867(05)	& 	0.1556(09)	&	0.003725(10)	&	0.03570(22)	\\ 
J303  & 	0.0632(05)	&	0.0652(02)	&	0.1177(04)	&	0.002056(07)	&	0.03412(15)	\\ 
\hline 
\end{tabular}
\label{tab:spectrum}
\end{center}
\end{table}

The chiral extrapolation of the TFF is done using the dimensionless parameter $\widetilde{y} = m_{\pi}^2 / (16 \pi^2 f_{\pi}^2)$. In order to propagate the error on $m_{\pi}$ and $f_{\pi}$, we also compute the pion decay constant on the same gauge configurations for each ensemble. The bare matrix element of interest is extracted from 
\begin{equation}
R(x_0,y_0) = \frac{2}{iZ_{\pi}} \ C_{AP}^I(x_0,y_0) \ e^{ E_{\pi}(\vec{0}) (x_0-y_0) } \,,
\end{equation}
where the axial-pseudoscalar two-point correlation function is given by
\begin{equation}
C_{AP}^I(x_0,y_0) = a^3 \sum_{\vec{x}} \langle A^I_0(\vec{x},x_0) P(\vec{0},y_0) \rangle \,.
\end{equation}
The $\mathcal{O}(a)$-improved axial current reads $A^I_0(x) = A_0(x) + c_{\rm A}\, \partial^{(s)}_0 P(x)$ where the improvement parameter $c_{\rm A}$ has been determined nonperturbatively in Ref.~\cite{Bulava:2015bxa}. 
The ratio $R(x_0,y_0)$ is fitted using the \textit{Ansatz} 
\begin{equation}
R(x_0,y_0) = R_{\rm plateau} + R_1\, e^{-E_{\pi}^{\prime}(\vec{0})(x_0-y_0) } + R_2(y_0)\, e^{ -(E_{2\pi} - E_{\pi}(\vec{0})) (T-x_0) } \,.
\end{equation}
From this bare matrix element, the renormalized and $\mathcal{O}(a)$-improved pseudoscalar decay constant is given by\footnote{Within our conventions, the physical value of $f_{\pi}$ is $F_{\pi} = 92.4~\MeV$. We distinguish between $F$ the pion decay constant in the chiral limit, $F_{\pi}$ the pion decay constant at physical pion mass and $f_{\pi}$ the pion decay constant at unphysical pion mass.}
\begin{equation}
f_{\pi}(a,\tilde{y}) = Z_{\rm A}(\widetilde{g}_0) \left( 1 + 3 \bbA a \mqav + \bA am_l  \right) \, R_{\rm plateau} \,,
\end{equation}
where $Z_{\rm A}$ is the renormalization factor of the axial current and $b_{\rm A}$, $\overline{b}_{\rm A}$ are improvement parameters~\cite{DallaBrida:2018tpn,Korcyl:2016ugy}. The results are summarized in Table~\ref{tab:spectrum}. In practice, we correct these results for finite-size effects (FSE) using
chiral perturbation theory ($\chi$PT) as described in Ref.~\cite{Colangelo:2005gd}. These corrections are small and we find that they correctly account for FSE on the ensembles H105/N101, which were generated using the same action parameters but different lattice volumes. However the FSE correction fails on ensemble H200, which corresponds to our smallest volume ($L \approx 2$~fm) and was not used to extract the pion TFF.
  
As a consistency check, we have performed an extrapolation to the physical point assuming the functional form    
\begin{equation}
f_{\pi}(a,\tilde{y}) =f_{\pi}(0,\tilde{y}_{\phys}) \left( 1 + \bar\ell_4 ( \tilde{y} - \tilde{y}_{\phys} ) + 2 \tilde{y} \ln \frac{m_{\pi}^{\exp}}{m_{\pi}}  + \epsilon_{f_\pi} ( \tilde{y}^2 - \tilde{y}_{\phys}^2 ) \right) + \delta_{f_{\pi}} \left(\frac{a}{a_{\beta=3.55}}\right)^2 \,,
\end{equation}
inspired from NLO $\chi PT$~\cite{Gasser:1983yg} and where we allow for a quadratic dependence in $\tilde{y}$. The four fit parameters are $f_{\pi}(0,\tilde{y}_{\phys})$, $\bar\ell_4$, $\epsilon_{f_\pi}$ and $\delta_{f_{\pi}}$. Our result at the physical point $f_{\pi} = 92.1(1.8)~\MeV$ and the value $\bar\ell_4 = 3.56(44)_{\stat}(80)_{\syst}$ turns out to be in good agreement with the Flavour Lattice Averaging Group (FLAG) average~\cite{Aoki:2019cca}.
  
\subsection{The transition form factor} 

\begin{figure}[t!]
	\includegraphics*[width=0.49\linewidth]{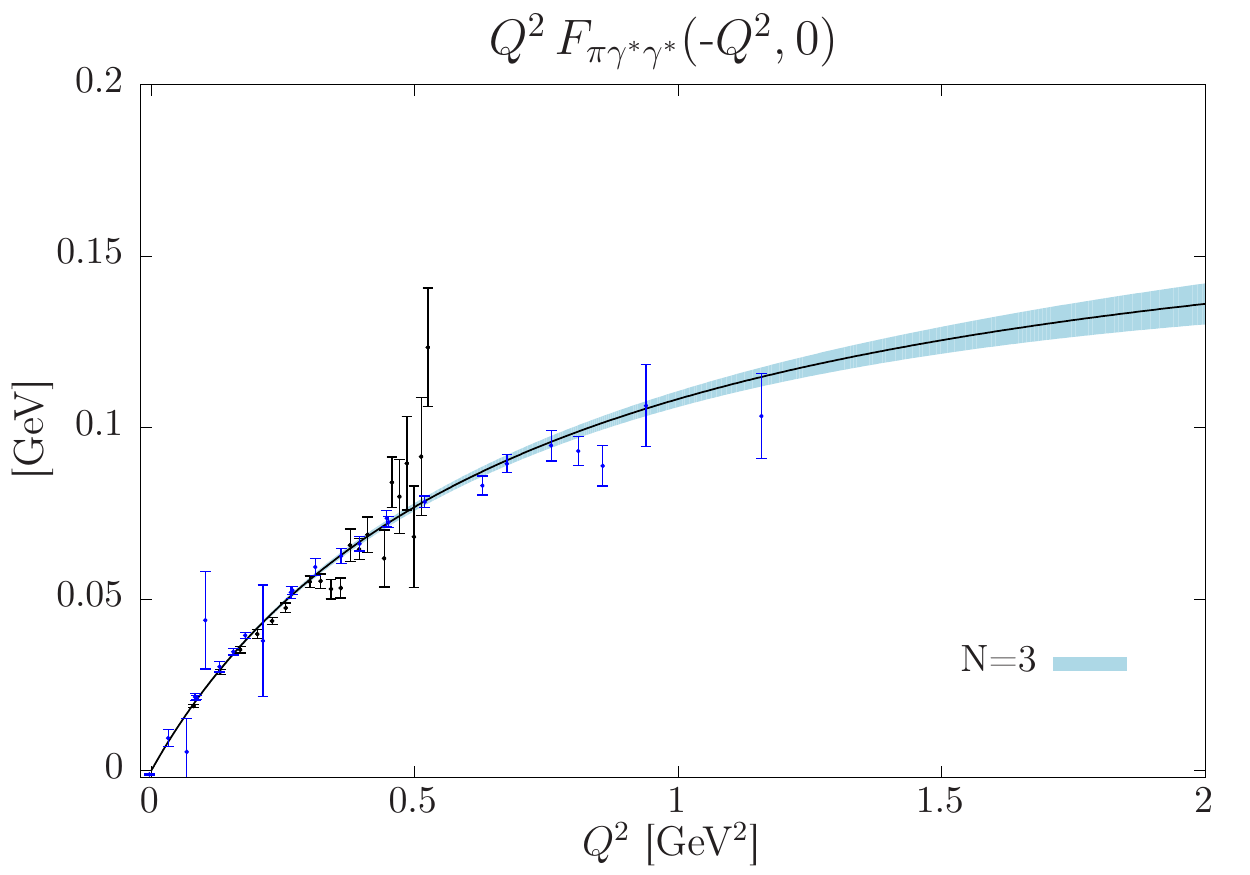}
	\includegraphics*[width=0.49\linewidth]{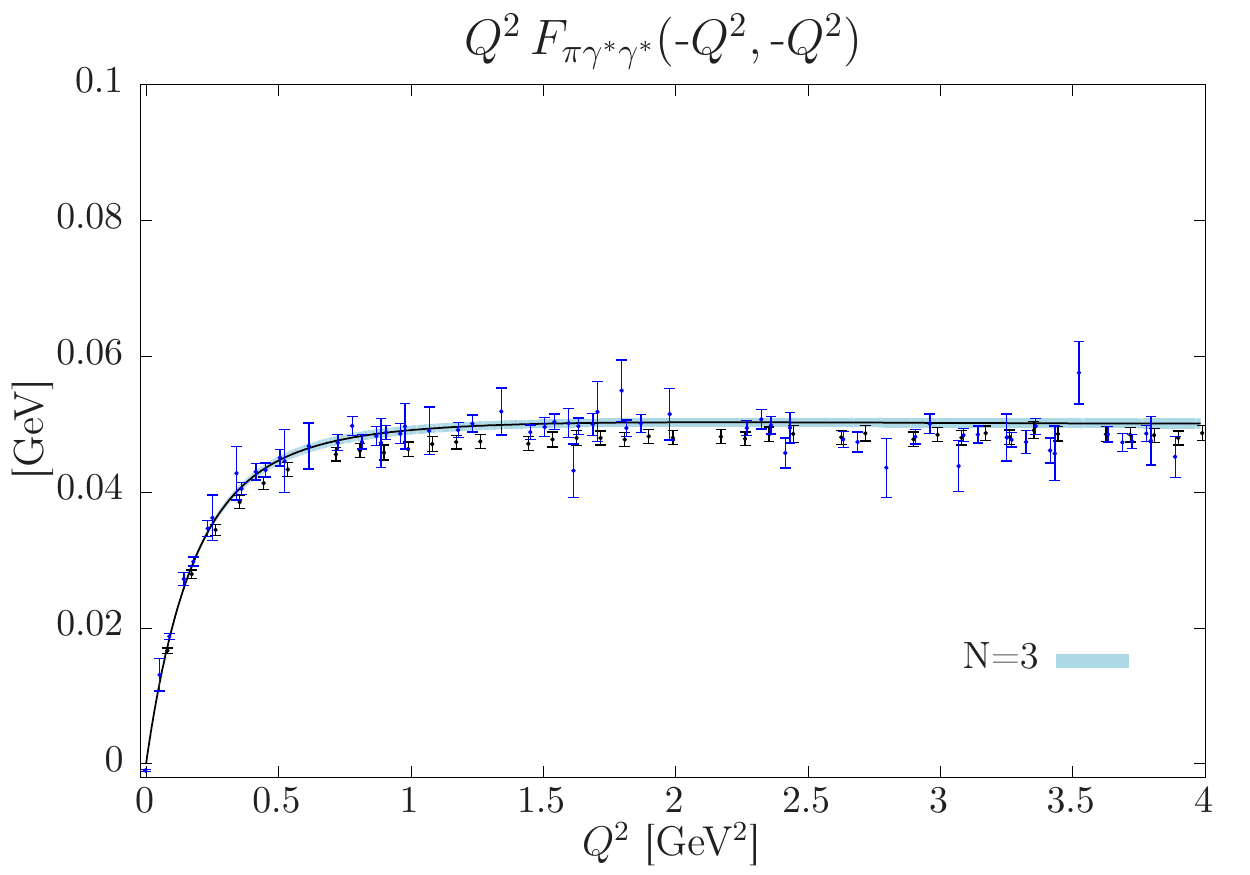}
	
	\caption{Lattice results for the ensemble D200, with a pion mass of 200~MeV, using two local vector currents at the source and at the  sink. Left panel: single virtual, right panel : double virtual. Black points correspond to the results obtained in the pion rest frame while the blue points are obtained in the pion moving frame. Error bands correspond to the global $z$-expansion fitting procedure described in Sec.~\ref{sec:zexp}. Some noisy points, with a small contribution to the fit, are not displayed for clarity.}
	\label{fig:D200}
\end{figure}

The results for the TFF at our lightest pion mass, corresponding to ensemble D200, are plotted in Fig.~\ref{fig:D200} for two different kinematics : the single-virtual case and the double-virtual case with $Q_1^2=Q_2^2$. Black and blue points correspond to the TFF obtained in the pion rest frame and moving frame respectively.

In the pion rest frame, results obtained using $A^{(1)}(\tau)$ are statistically more precise when both photons carry the same virtuality (right panel of Fig.~\ref{fig:D200}). In this case $\omega_1 = m_{\pi}/2$ and the full integrand in Eq.~(\ref{eq:Mlat}) is symmetric. In the single-virtual case, $\omega_1$ takes larger values and the integral probes further the tail of the function $\widetilde{A}^{(1)}(\tau)$, where the noise over signal ratio increases rapidly with $\tau$: the signal is lost for virtualities above $0.5~\GeV^2$. In the moving frame, the situation is similar for $\widetilde{A}^{(1)}(\tau)$ (even if the integrand is not exactly symmetric in the double-virtual case, unless $|\vec{q}_1|^2 \approx |\vec{q}_2|^2$). For $\widetilde{A}^{(2)}(\tau)$, since the sign of the function changes at $\tau=0$, the situation is the opposite: results are more precise in the single-virtual case where there are fewer cancellations between positive and negative contributions and we can reach higher virtualities (see right panel of Fig.~\ref{fig:A}).

We also point out that results obtained in both frames are fully consistent, confirming that sources of breaking the Lorentz invariance
in the lattice calculation are small. In particular, since the two frames are affected by different sources of finite volume effects, it is a
first hint that FSE are small. A more detailed study of FSE is done in Sec.~\ref{sec:fse}.

\subsection{Parametrization of $\FF$ and extrapolation to the physical point \label{sec:zexp}} 

In this section we propose a method to extract the TFF over the whole kinematical range, and at the physical point, based on its analytical properties. We introduce the conformal variables $z_1$ and $z_2$ defined through~\cite{Boyd:1995sq}
\begin{equation}
z_k = \frac{ \sqrt{t_c+Q_k^2} - \sqrt{t_c - t_0} }{ \sqrt{t_c+Q_k^2} + \sqrt{t_c - t_0} } \,, \quad k=1,2 \,,
\label{eq:zvar}
\end{equation}
which map the branch cut, starting at $t_c = 4m_{\pi}^2$, onto the unit circle $|z_k|=1$. Here, $t_0$ is a free parameter representing the virtuality mapping to $z_k=0$. Since the TFF is analytic for $|z_k| < 1$, one can write
\begin{equation}
\FF(-Q_1^2, -Q_2^2)  = \sum_{n,m=0}^{\infty} c_{nm} \, z_1^n \, z_2^m \,,
\label{eq:z_exp}
\end{equation}
where the coefficients $c_{nm} = c_{mn}$ are symmetric due to the Bose symmetry. Since $|z_k|<1$ one can expect a fast convergence of the sum, where only a few terms with $n,m \leq N$ are needed at a given accuracy. 
The optimal choice for $t_0$, which reduces the maximum value of $|z_k|$ in the range $[0, Q_{\max}^2]$, is given by
\begin{equation}
t_0 = t_c \left( 1 - \sqrt{1+ Q_{\max}^2 / t_c } \right) \,.
\end{equation}
This maximum is reached at $z_k=0$ and $z_k=Q^2_{\max}$. Using the values $m_{\pi} = 134.9~\MeV$ and $Q_{\max}^2=4~\GeV^2$, one finds $|z_{\max}| = 0.46$ well below one. The coefficients $c_{nm}$ satisfy the relation
\begin{equation}
\sum_{n,m} |c_{nm}|^2 = \frac{1}{(2i\pi)^2} \oint \frac{ \mathrm{d}z_1}{z_1} \oint \frac{ \mathrm{d}z_2}{z_2} \, | \FF(-Q_1^2, -Q_2^2) |^2  \,,
\label{eq:z_exp_coeff}
\end{equation}
which ensures that the coefficients $c_{nm}$ are not only bounded but also decrease to zero for sufficiently large $n,m$. In practice, the TFF can be multiplied by any analytical function $P(Q_1^2,Q_2^2)$ and the resulting product expanded in powers of the $z_k$. Some choices may improve the convergence rate of the series expansion. At short distances, the behavior of the TFF is predicted by the Brodsky-Lepage behavior in the single-virtual case and by the OPE in the double-virtual case~\cite{BL_3_papers,Nesterenko:1982dn,Novikov:1983jt},
\begin{align}
\FF(-Q^2, 0) &\xrightarrow[Q^2 \to \infty]{} \frac{2 F_{\pi}}{Q^2}  \,, \label{eq:BL} \\
\FF(-Q^2, -Q^2) &\xrightarrow[Q^2 \to \infty]{} \frac{2  F_{\pi}}{3Q^2} \,. \label{eq:OPE} 
\end{align}
Therefore, it is convenient to consider 
\begin{equation}
P(Q_1^2,Q_2^2) = 1 + \frac{Q_1^2 + Q_2^2}{M_V^2} \,,
\label{eq:P_zexp}
\end{equation}
where $M_V = 775~\MeV$ is the vector meson mass. With this choice, the parametrization of the TFF at the finite value of $N$ decreases asymptotically as $1/Q^2$ in all directions in the $(Q_1^2,Q_2^2)$ plane, in accord with the Brodsky-Lepage and the OPE behavior. 
This feature is one major reason for us preferring the $z$-expansion over the LMD+V model~\cite{Knecht:2001xc} used in~\cite{Gerardin:2016cqj}, which e.g. for constant $Q_1^2 \neq 0$ and $Q_2^2 \to \infty$ does not vanish, even though the corresponding integrals for the pion-pole contribution to the hadronic light-by-light scattering in the muon $g-2$ are still convergent with the LMD+V model (see Sec.~\ref{sec:g-2}). However, at the precision level we aim for, the proper high-energy behavior in all directions is important. It should be noted that we do not impose the explicit values of the coefficients on the right-hand side of Eqs.~(\ref{eq:BL}) and (\ref{eq:OPE}), since they receive higher-order corrections in perturbative QCD and at higher twist, see the discussion below. 

\renewcommand{\arraystretch}{1.1}
\begin{table}[b]
\caption{Coefficients of the $z$-expansion, in $\GeV^{-1}$, defined through Eq.~(\ref{eq:z_exp_mod}) in the continuum, at physical quark masses, for different values of $N$. We use $Q_{\max}^2=4~\GeV^2$. The chi-squares per degree of freedom are respectively $\chi^2/\dof = 1.5,\, 1.2,\, 1.1$ and for $N=3$, the associated correlation matrix is given by Eq.~(\ref{eq:cov}).} 
\begin{center}
\begin{tabular}{ccccccccccc}
\hline
$N$	&	$c_{00}$	&	$c_{01}$	&	$c_{11}$	&	$c_{20}$	&	$c_{21}$	&	$c_{22}$	&	$c_{30}$	&	$c_{31}$	&	$c_{32}$	&	$c_{33}$	 \\ 
\hline
1	&	0.2346(65)	&	$-0.0590(39)$	&	\ \ 0.074(19)  \\ 
2	&	0.2350(61)	&	$-0.0651(49)$	&	$-0.284(68)$	&	0.106(33)	&	0.109(46)		&	$-0.29(12)$ 	 \\ 
3	&	0.2345(63)	&	$-0.0746(52)$	&	$-0.338(86)$	&	0.145(43)	&	0.008(127)	&	$-0.92(55)$	&	0.34(10)	&	0.25(27) 	&	$-1.27(79)$	&	1.16(1.40)   \\ 
\hline 
\end{tabular}
\label{tab:z-exp}
\end{center}
\end{table}

\begin{figure}[t!]
	\includegraphics*[width=0.49\linewidth]{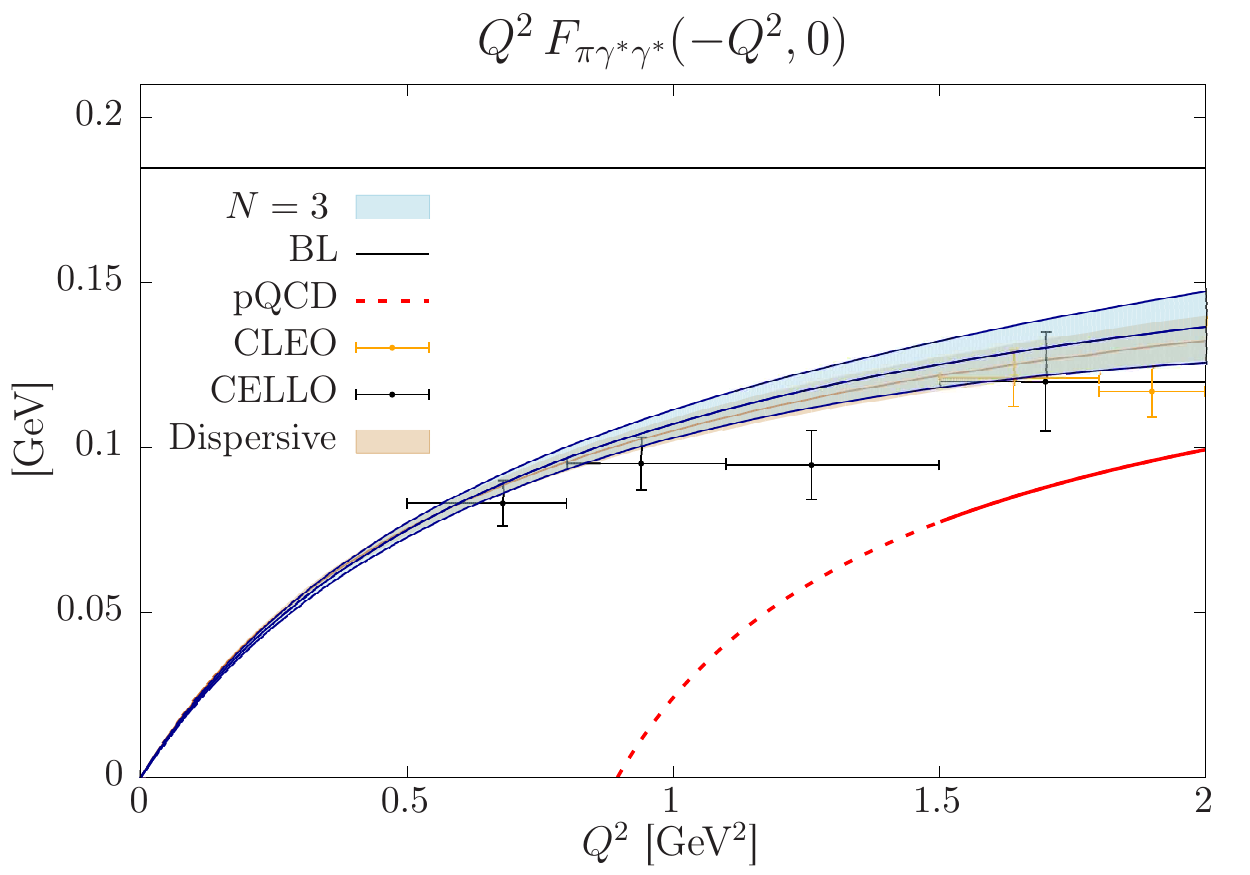}
	\includegraphics*[width=0.49\linewidth]{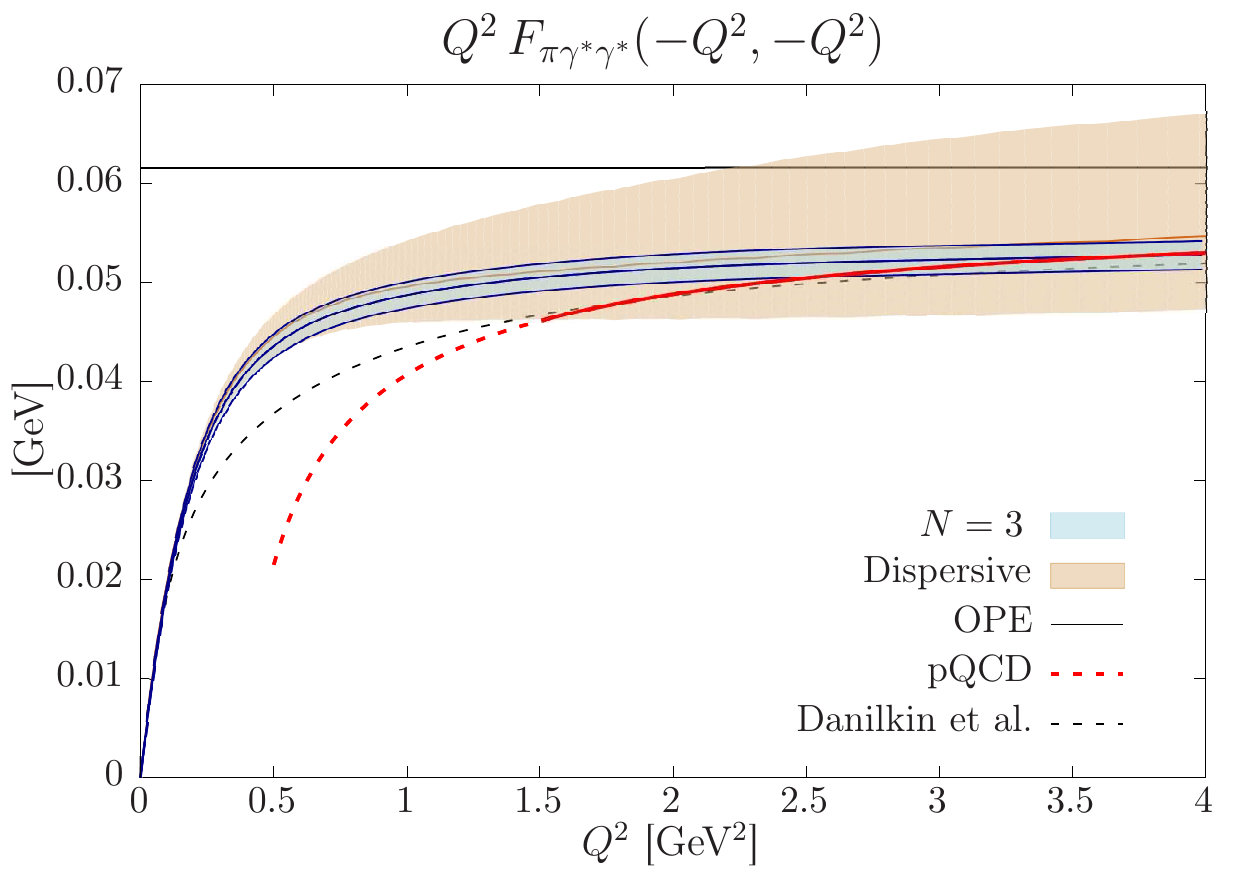}
	
	\caption{Extrapolated TFF at the physical point using the modified $z$-expansion (\ref{eq:z_exp_mod}) with $N=3$. The horizontal black lines correspond  to the Brodsky-Lepage and OPE predictions. The red line corresponds to the asymptotic prediction given by Eq.~(\ref{eq:TFF_asympt}) including higher twists and NLO corrections and assuming an asymptotic DA (see Sec.~\ref{sec:aymptotic}). The result of Ref.~\cite{Hoferichter:2018dmo}, obtained in the dispersive framework, is shown for comparison. The dashed black line in the double-virtual case corresponds to the prediction given by Eq.~(83) in Ref.~\cite{Danilkin:2019mhd}. Experimental data from CELLO and CLEO are displayed in the single-virtual case~\cite{exp}. }
	\label{fig:z-exp}
\end{figure}

The imaginary part of the TFF behaves as $(q^2 - t_c)^{3/2}$ near threshold (P-wave). This property is not fulfilled at finite $N$ but\footnote{The authors thank Martin Hoferichter for pointing out this fact to us.}, as shown in Ref.~\cite{Bourrely:2008za}, this constraint can be implemented by imposing
\begin{equation}
\left[ \frac{  \mathrm{d} \FF  }{ \mathrm{d}z_k } \right]_{z_k=-1} = 0 \,, \quad k=1,2 \,,
\end{equation}
which leads to the modified $z$-expansion
\begin{multline}
P(Q_1^2,Q_2^2) \ \FF(-Q_1^2, -Q_2^2)  = \\  \sum_{n,m=0}^{N} c_{nm} \, \left( z_1^n - (-1)^{N+n+1} \frac{n}{N+1} \, z_1^{N+1} \right) \, \left( z_2^m -  (-1)^{N+m+1} \frac{m}{N+1} \, z_2^{N+1} \right) \,.
\label{eq:z_exp_mod}
\end{multline}
Finally, to take into account the discretization effects and the dependence on the quark masses used in the simulations, the coefficients $c_{nm}$ are allowed to vary linearly with the variable $\widetilde{y} = m_{\pi}^2/(16\pi^2 f_{\pi}^2)$ and quadratically with the lattice spacing,
\begin{equation}
c_{nm}(\widetilde{y}, a) = c_{nm}(\widetilde{y}^{\phys}, 0) + \gamma_{nm} \, (\widetilde{y} - \widetilde{y}_{\phys} )+ \delta^{d}_{nm} \, \left( \frac{a}{a_{\beta=3.55}} \right)^2\,,
\label{eq:pz}
\end{equation}
where $d=1,2$ stands for the two discretizations of the three-point correlation function.
We perform a global fit and the results are summarized in Table~\ref{tab:z-exp}. Due to the large set of data, the correlation matrices are ill-conditioned and we perform uncorrelated fits. The error on the fit parameters is then obtained from a jackknife procedure, using blocking to take into account autocorrelations.
The results for $N=3$ are shown in Fig.~\ref{fig:z-exp}. The chi-square by degree of freedom is $\chi^2/\dof = 1.1$ and  the correlation matrix of the coefficients in the continuum and at the physical pion mass is 
\begin{equation}
\mathrm{cor}(c_{nm}) = \begin{pmatrix}
+1.000 & -0.086 & -0.370 & +0.076 & -0.062 & -0.044 & +0.056 & +0.415 & -0.043 & -0.294 \\
-0.086 & +1.000 & -0.005 & +0.018 & -0.531 & +0.381 & -0.157 & -0.307 & +0.697 & -0.569 \\
-0.370 & -0.005 & +1.000 & -0.890 & +0.458 & +0.441 & -0.756 & -0.468 & +0.227 & +0.154 \\
+0.076 & +0.018 & -0.890 & +1.000 & -0.592 & -0.484 & +0.914 & +0.196 & -0.191 & +0.022 \\
-0.062 & -0.531 & +0.458 & -0.592 & +1.000 & -0.114 & -0.571 & +0.217 & -0.589 & +0.574 \\
-0.044 & +0.381 & +0.441 & -0.484 & -0.114 & +1.000 & -0.519 & -0.705 & +0.638 & -0.328 \\
+0.056 & -0.157 & -0.756 & +0.914 & -0.571 & -0.519 & +1.000 & +0.195 & -0.300 & +0.128 \\
+0.415 & -0.307 & -0.468 & +0.196 & +0.217 & -0.705 & +0.195 & +1.000 & -0.516 & -0.109 \\
-0.043 & +0.697 & +0.227 & -0.191 & -0.589 & +0.638 & -0.300 & -0.516 & +1.000 & -0.758 \\
-0.294 & -0.569 & +0.154 & +0.022 & +0.574 & -0.328 & +0.128 & -0.109 & -0.758 & +1.000 \\
\end{pmatrix} \,.
\label{eq:cov} 
\end{equation}
In Appendix~\ref{app:B}, we study the systematic error associated with the truncation of the sum in Eq.~(\ref{eq:z_exp_mod}) and we conclude that $N=3$ is already sufficient for an accuracy at the percent level. We also checked that using $P(Q_1^2,Q_2^2) = 1$ leads to compatible results in the range where we have lattice data. 
In the single-virtual case, our results are in good agreement with experimental data. In the single-virtual case and in the double-virtual case, we also find good agreement with the results obtained in the dispersive framework~\cite{Hoferichter:2018dmo}; see Fig.~\ref{fig:z-exp}. A more detailed comparison with phenomenology is provided in the next section.

Finally, we remark that a different fit strategy would consist in fitting each ensemble independently and then extrapolating each coefficient $c_{nm}$ to the physical point. First, we found that the global fit procedure is more stable. Second, the continuum and chiral extrapolation of the individual coefficients $c_{nm}$ is nontrivial, as they are correlated. Nonetheless, we provide the $z$-expansion on three individual lattice ensembles in Appendix~\ref{sec:z-exp-indiv} to allow for comparisons prior to the chiral and continuum extrapolation.

\subsection{Systematic errors}

\subsubsection{Hypercubic effects \label{sec:hypercubic} }

\begin{figure}[t!]
	\includegraphics*[width=0.49\linewidth]{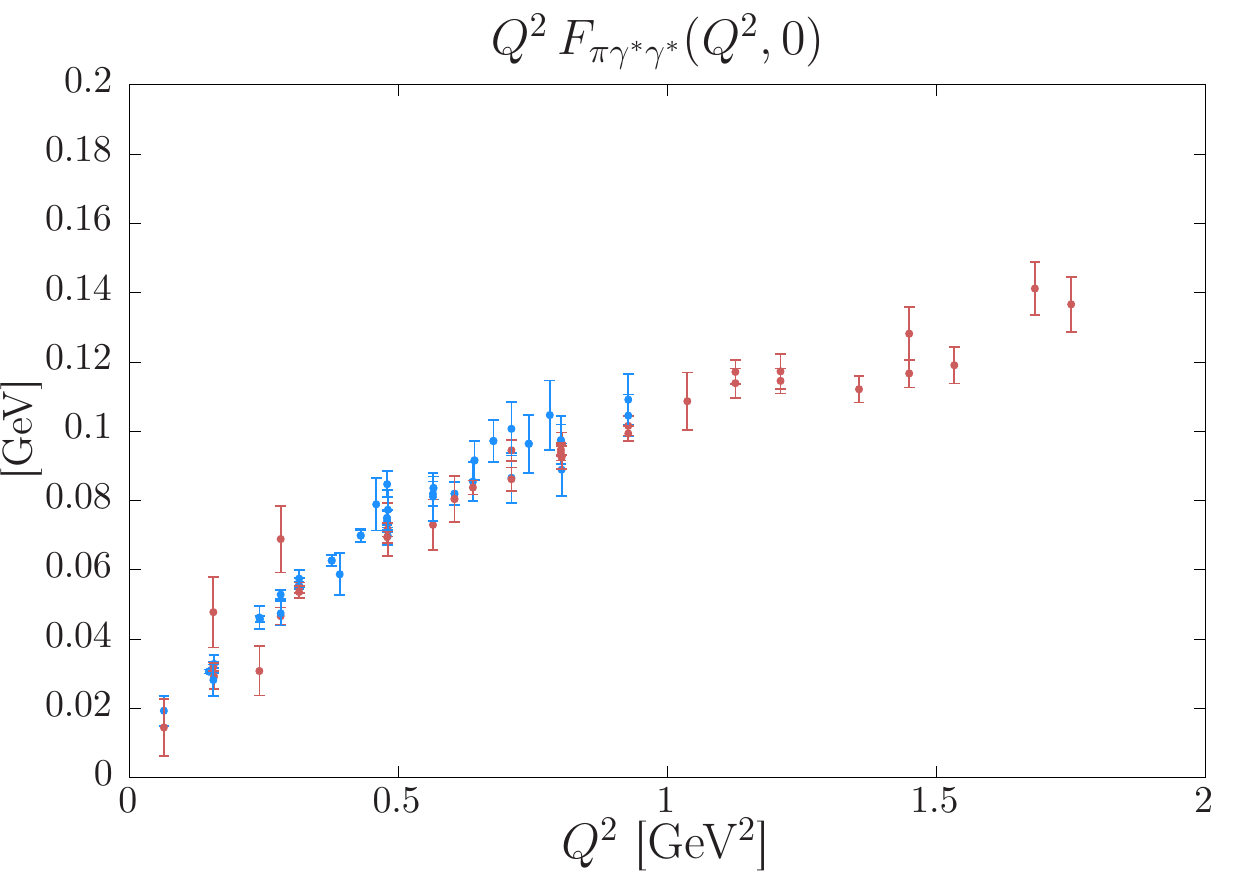}
	\includegraphics*[width=0.49\linewidth]{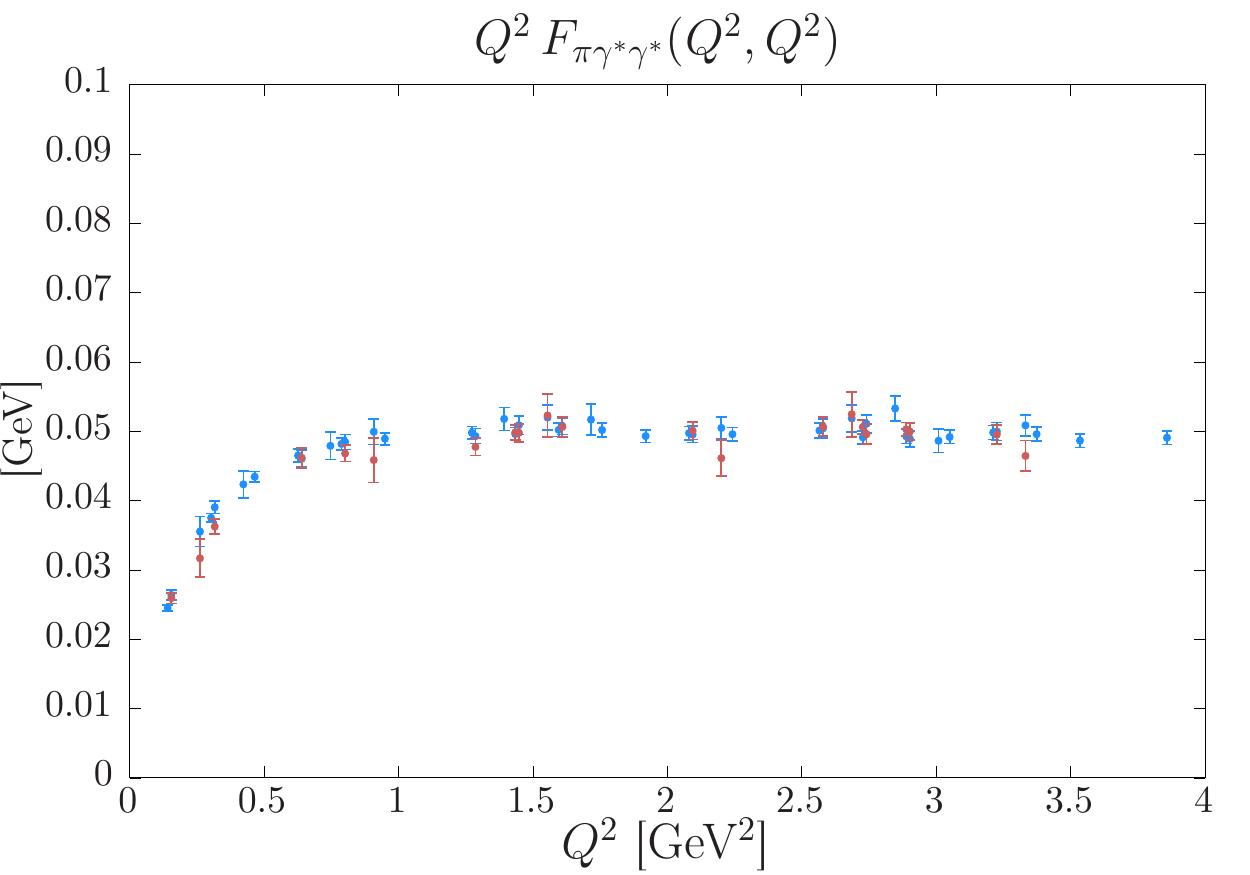}
	
	\caption{Study of hypercubic artifacts for different values of the photon virtualities $Q_1^2$ and $Q_2^2$. Within our statistical accuracy, we do not observe hypercubic effects, even at large virtualities. The results correspond to the ensemble N200. The blue and red  points correspond to the pion rest frame and moving frame respectively.  } 	
	\label{fig:hypercubic}
\end{figure}

At finite lattice spacing, the O(3) rotational symmetry is broken down to the cubic subgroup H(3). Spatial momenta, equivalent in the continuum, but belonging to different H(3) orbits may be affected by different lattice artifacts. This has been observed in previous calculations of form factors (recently in Ref.~\cite{Lubicz:2017syv} for example). As explained in Sec.~\ref{sec:orbs}, in the calculation of the TFF, we have averaged over all equivalent combinations of $(\vec{q}_1^{\ 2},\vec{q}_2^{\ 2})$. In this section, we study the validity of this approach. 

Any polynomial function of the spatial momenta $\vec{q}$, and invariant under H(3), can be expressed in terms of the three invariants
\begin{equation}
q^{[n]} = q_x^n  + q_y^n + q_z^n\,, \quad n=2,4,6 \,.
\end{equation}
Therefore, each H(3) orbit is uniquely characterized by the values of $q_i^{[2]}$, $q_i^{[4]}$ and $q_i^{[6]}$ with $i=1,2$.

In the pion rest frame, both photons have back-to-back spatial momenta and $\vec{q}_1 = - \vec{q}_2$. The first kinematical configuration with two orbits corresponds to $|\vec{q}_1|^2 = 9(\frac{2\pi}{L})^2$  with, 
for example, $\vec{q}_1=\frac{2\pi}{L}(\pm 3,0,0)$ (6 elements) and $\vec{q}_1=\frac{2\pi}{L}(\pm2,\pm2,\pm1)$ (24 elements). 
When the pion has one unit of momentum in the $\hat{z}$-direction, equivalent $(q_1^2,q_2^2)$ can also be obtained from different values of $\vec{q}_1$ not related by H(3) symmetry but by the permutation of $q_1$ and $q_2$. 

In Fig.~\ref{fig:hypercubic}, we show the results for the transition form factor without averaging over equivalent momenta. We observe that, at our level of statistical precision, we are not yet sensitive to hypercubic artifacts.

\subsubsection{Finite-size effects \label{sec:fse} }

\begin{figure}[t!]
	\includegraphics*[width=0.49\linewidth]{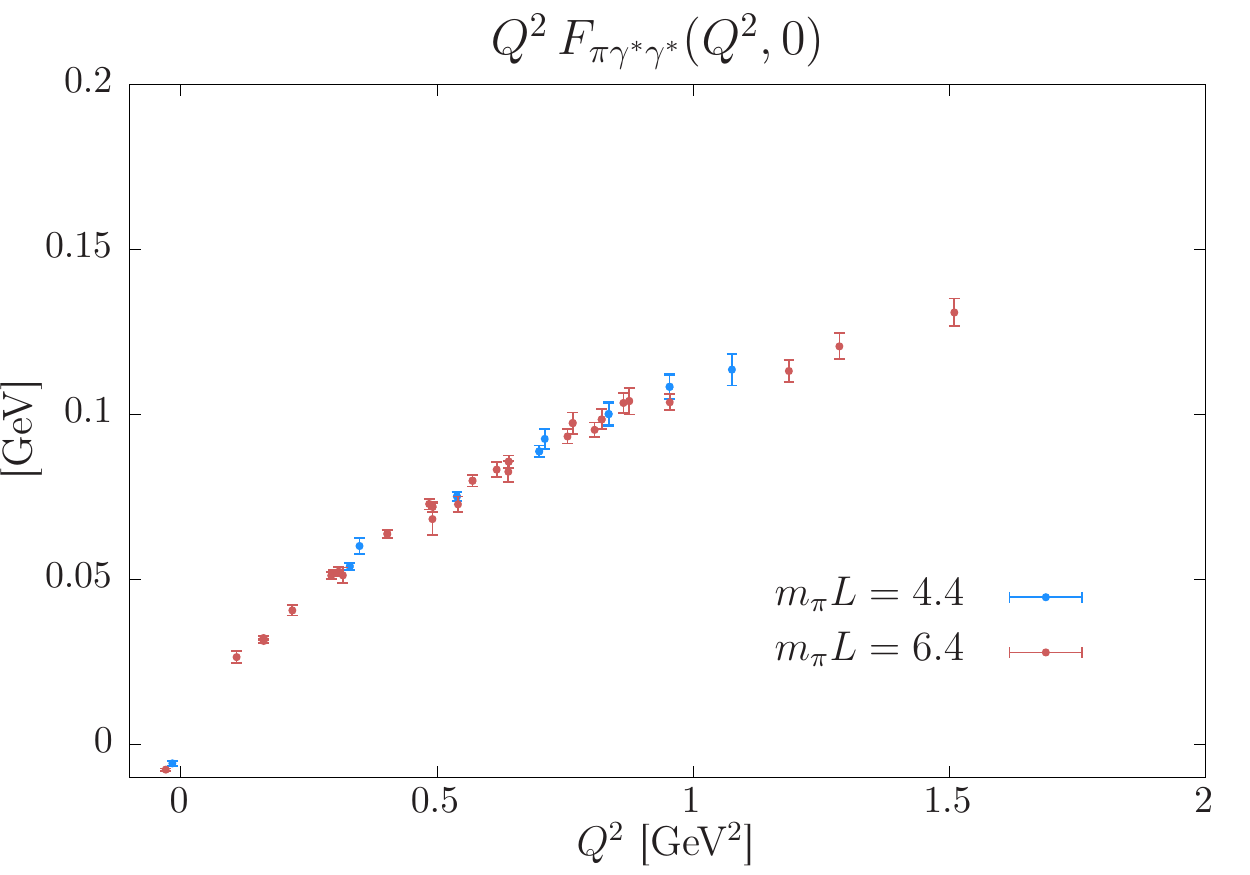}
	\includegraphics*[width=0.49\linewidth]{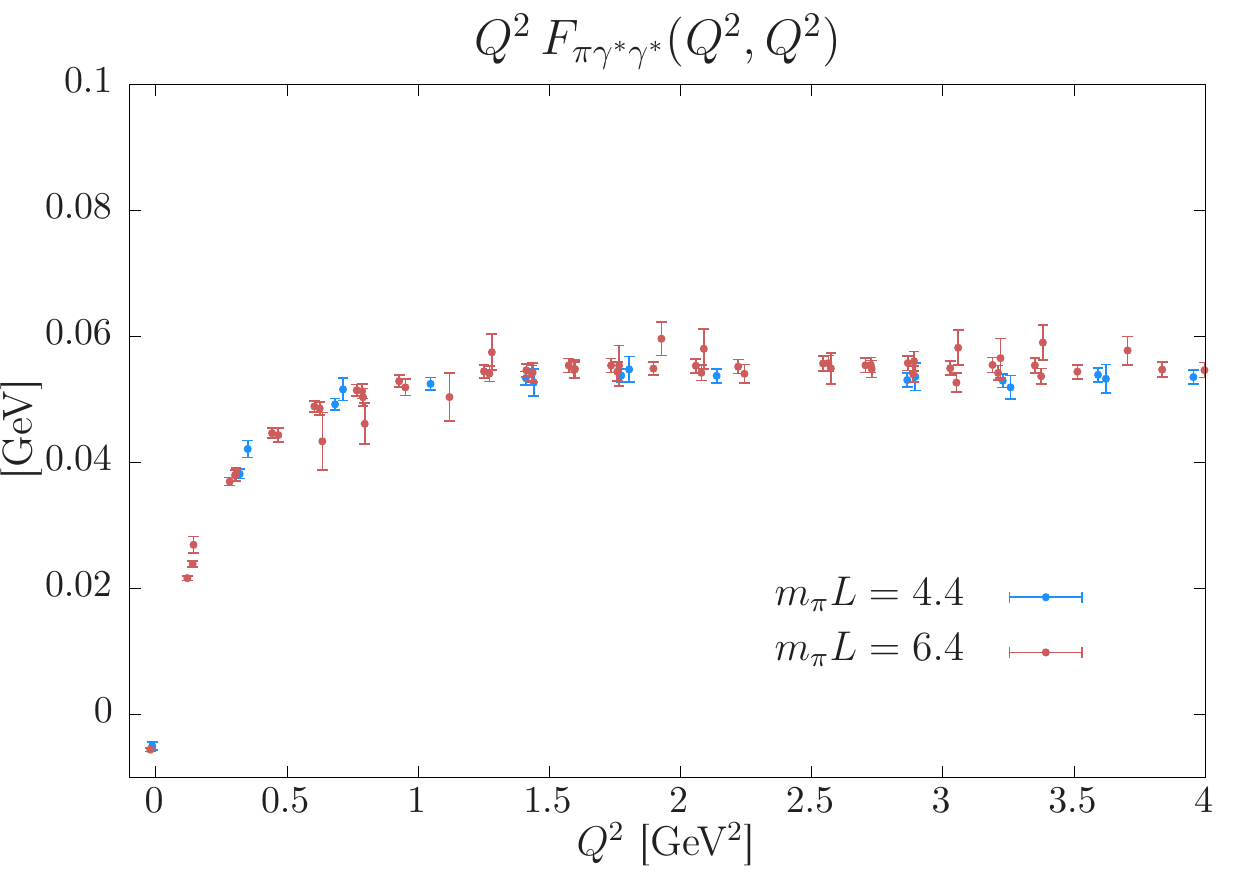}

	\caption{Transition form factor obtained using two different volumes ($L = 2.1~\fm$ and $L = 3.1~\fm$ for H200 and N202 respectively) at the same bare lattice parameters $\kappa$ and $\beta$.}	
	\label{fig:vol_effects}
\end{figure}

Two sets of ensembles (H200/N202 and H105/N101, listed in Table~\ref{tab:sim}) have been generated using the same bare lattice parameters $\kappa$ and $\beta$ but with different volumes ($L/a = 32$ and $48$). The ensembles N202 and N101 correspond to large volumes with $m_{\pi}L \geq 6.5$ where FSE are expected to be negligible. The two other ensembles satisfy $m_{\pi}L \approx 4.0$. The pion decay constant has been computed on all ensembles (see Table~\ref{tab:spectrum}) with a statistical precision below $1.0~\%$ and finite-size effects are sizable: we observe a 2-3~$\sigma$ discrepancy between H200/N202 and FSE corrections using $\chi$PT~\cite{Colangelo:2005gd} failed to explain the difference. However, for H105 and N101, the results are in perfect agreement after FSE corrections.
Since finite-size effects strongly depend on the observable (and on the estimator), the TFF has been computed on these two sets of ensembles. At our level of statistical precision, we do not observe any significant differences between different volumes and the results for H200, compared to N202, are shown in Fig.~\ref{fig:vol_effects}. Since H200 is our smallest ensemble, we neglect finite-size effects in the following and exclude the ensembles H200 from the analysis. To be conservative, we also exclude H105.

\subsubsection{Quark-disconnected contributions}

\begin{figure}[t!]
	\includegraphics*[width=0.47\linewidth]{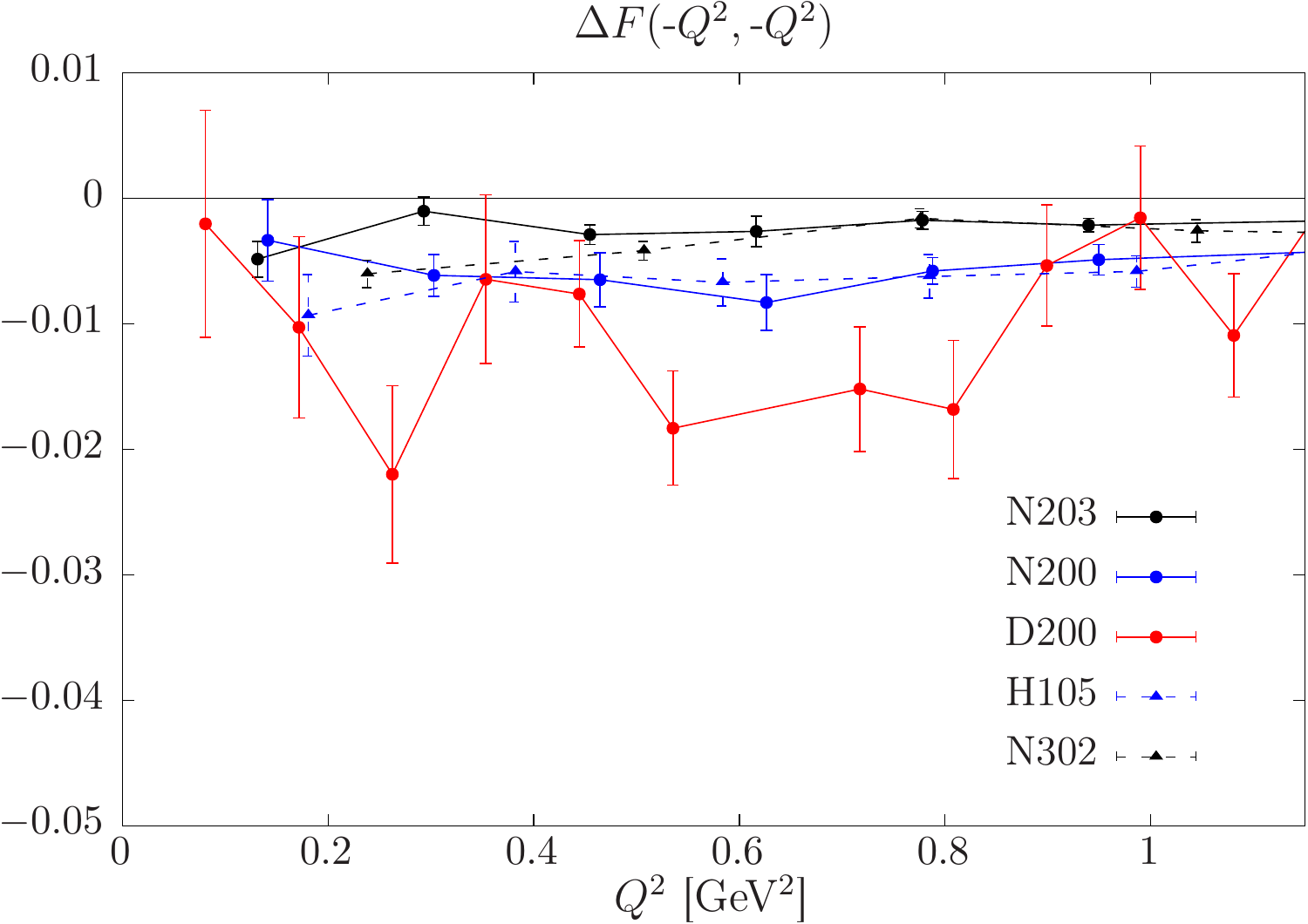}
	\includegraphics*[width=0.49\linewidth]{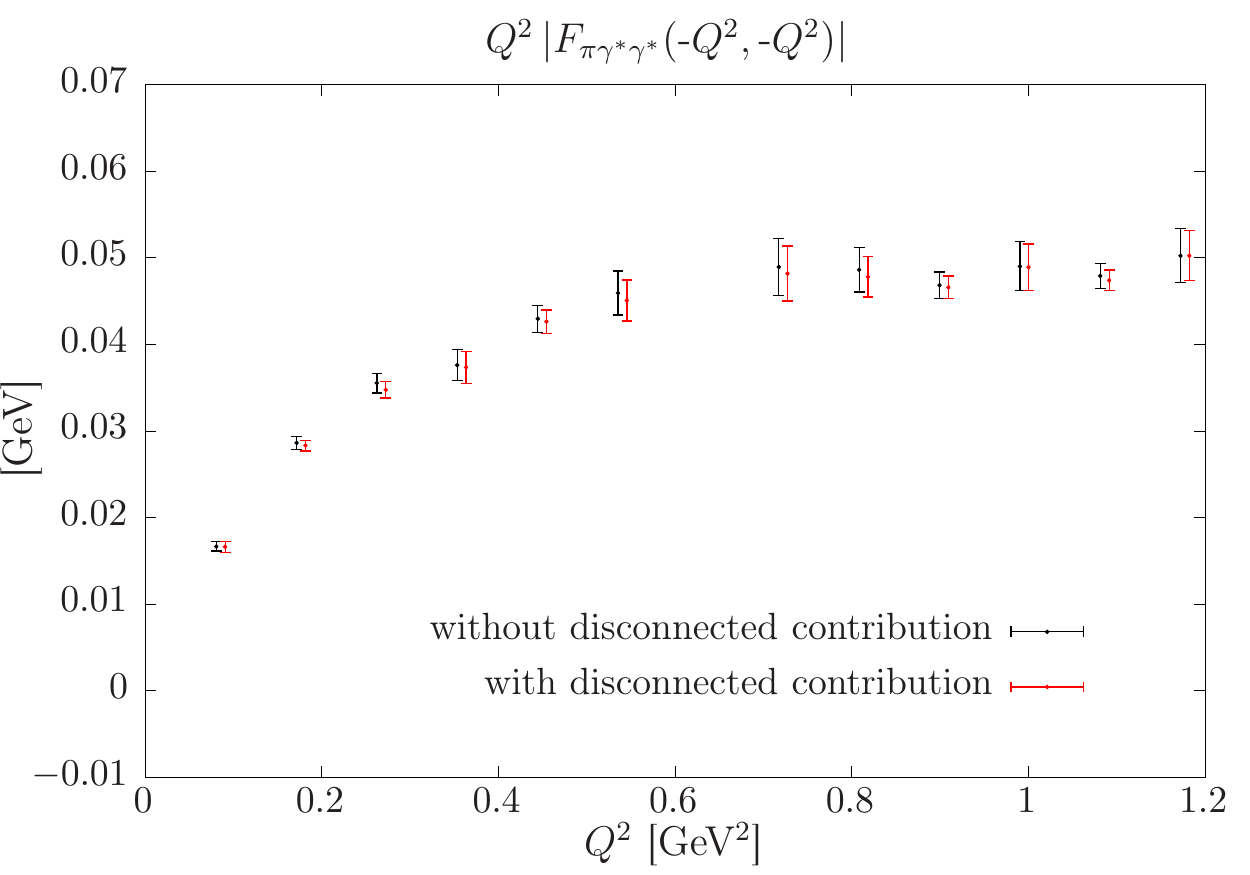}

	\caption{Left panel : $\Delta F(-Q_1^2,-Q_2^2)$ defined through Eq.~(\ref{eq:disc}) for the ensembles H105, N203, N200, D200 and N302. Ensembles at the same lattice spacing $a=0.064~\fm$ are plotted using plain lines whereas ensembles at different lattice spacings are plotted using dashed lines. Right panel : TFF with and without including the disconnected contribution for our lightest pion mass ensemble D200. Red points have been shifted slightly to the right for clarity.}	
	\label{fig:disc}
\end{figure}

The quark-disconnected contributions involve a single quark loop which couples to the vector current. Within the Mainz group, they have been coded~\cite{MainzDisc} and computed as part of various physics projects~\cite{Ce:2018ziv,Djukanovic:2018iir,Gerardin:2018sin} on a number of ensembles (H105, N203, N200, D200 and N302). These contributions vanish exactly for the three ensembles at the heaviest pion mass, where $m_l=m_s$. The single-propagator loops have been computed using two random sources of 512 hierarchical probing
vectors~\cite{Stathopoulos:2013aci} per configuration and for all spatial momenta with $|\vec{q}|^2 \leq 12 \times (\frac{2\pi}{L})^2$. The
two-point correlation functions between a pseudoscalar and a vector current have been computed using stochastic sources with U(1) noise as in Ref.~\cite{Gerardin:2016cqj}.  We define the ratio 
\begin{equation}
\Delta F(-Q_1^2,-Q_2^2) = \frac{ \FF^{\rm disc}(-Q_1^2, -Q_2^2)  }{ \FF^{\rm conn}(-Q_1^2, -Q_2^2) } 
\label{eq:disc}
\end{equation}
where $\FF^{\rm conn}+\FF^{\rm disc}$ and  $\FF^{\rm conn}$ correspond to the TFF obtained with or without including the disconnected contribution. The results are shown in the left panel of Fig.~\ref{fig:disc}. We remark that the tail of the disconnected correlator was treated in the same manner as in the connected correlator. As expected, the disconnected contribution increases as we approach the physical pion mass: the CLS ensembles used in this work were generated using a constant value of the trace of the bare quark masses. Comparing ensembles generated at the same pion mass but with different lattice spacings, we do not observe significant discretization effects. For our ensemble with the lightest pion (D200), the TFF with and without including the disconnected contribution is depicted in the right panel of Fig.~\ref{fig:disc}~: at our level of statistical precision, the disconnected contribution can be neglected. The impact of the disconnected contribution on the pion-pole contribution to the muon $g-2$ is discussed in more detail in Sec~\ref{sec:pion_pole_zexp}.

\section{Phenomenological applications \label{sec:pheno} }

\subsection{Parametrizing and comparing the TFF with phenomenological models \label{subsec:pheno_fits} }

As in Ref.~\cite{Gerardin:2016cqj}, we compare our results with phenomenological models which have been applied to the pion-pole contribution to hadronic light-by-light scattering in the muon $g-2$.
In particular, we consider the VMD model, the LMD model \cite{Moussallam:1994xp,Knecht:1999gb} and the LMD+V model \cite{Knecht:2001xc}. These models are parametrized by
\begin{subequations}
\begin{equation}
\FF^{\VMD}(q_1^2, q_2^2) = \frac{ \alpha M_V^4}{(M_V^2-q_1^2)(M_V^2-q_2^2)} \,, 
\end{equation}
\begin{equation}
 \FF^{\LMD}(q_1^2, q_2^2) = \frac{ \alpha M_V^4+ \beta(q_1^2  + q_2^2)}{(M_V^2-q_1^2)(M_V^2-q_2^2)} \,, 
\end{equation}
\begin{equation}
\FF^{\LMDV}(q_1^2, q_2^2) = \frac{\widetilde{h}_0\, q_1^2 q_2^2 (q_1^2 + q_2^2) + \widetilde{h}_1 (q_1^2+q_2^2)^2  +   \widetilde{h}_2\, q_1^2 q_2^2 + \widetilde{h}_5\, M_{V_1}^2   M_{V_2}^2\,  (q_1^2+q_2^2) +  \alpha\, M_{V_1}^4 M_{V_2}^4}{(M_{V_1}^2-q_1^2)(M_{V_2}^2-q_1^2) (M_{V_1}^2-q_2^2)(M_{V_2}^2-q_2^2)} \,, \label{eq:LMDV}
\end{equation}
\end{subequations}
and we refer the reader to Ref.~\cite{Gerardin:2016cqj} for a more detailed description of the models and phenomenological values of the parameters. We just recall the main properties relevant for this study.
All models satisfy the anomaly constraint at vanishing momenta~\cite{Adler:1969gk,Bell:1969ts} 
\begin{equation}
\FF(0,0) = \frac{1}{4\pi^2F_{\pi}} \,,
\end{equation}
by setting $\alpha = \alpha^{\rm th} = 1/(4\pi^2F_{\pi}) = 0.274~\GeV^{-1}$. 
The VMD model does not fulfill the OPE constraint given by Eq.~(\ref{eq:OPE}) but rather behaves as $\FF^{\VMD}(-Q^2,-Q^2) \sim 1/Q^4$ at large virtualities. The LMD model has the advantage to fulfill the OPE constraint but does not satisfy the Brodsky-Lepage (BL) behavior given by Eq.~(\ref{eq:BL}) since $\FF^{\LMD}(-Q^2,0) \sim \beta/M_V^2$ at large virtualities. Finally, the LMD+V model satisfies the short-distance constraints both for $Q_1^2=Q_2^2\to\infty$ and for $Q_1^2=0$, $Q_2^2\to\infty$ if one sets $\widetilde{h}_1 = 0$.

We have performed global fits using the VMD, LMD and LMD+V Ans\"atze where the fit parameters are, respectively, $(\alpha, M_V)$, $(\alpha, \beta, M_V)$ and $(\alpha, \widetilde{h}_0, \widetilde{h}_2, \widetilde{h}_5, M_{V_1}, M_{V_2})$. As for the $z$-expansion, each fit parameter is assumed to depend linearly on the dimensionless variable $\widetilde{y} = m_{\pi}^2/(16\pi^2 f_{\pi}^2)$ and quadratically on the lattice spacing $a$. The fits are uncorrelated. Since we have computed the TFF using two different discretizations of the vector current, we actually perform a combined fit such that any parameter has the functional form
\begin{equation}
p(a, \widetilde{y}) = p(0, \widetilde{y}_{\phys}) + \gamma_m \, (\widetilde{y} -\widetilde{y}_{\phys} ) 
    + \delta_d \left( \frac{a}{a_{\beta=3.55}} \right)^2 \,, \quad d=1,2 \,.
\end{equation}
 
\begin{figure}[t]
	\includegraphics*[width=0.49\linewidth]{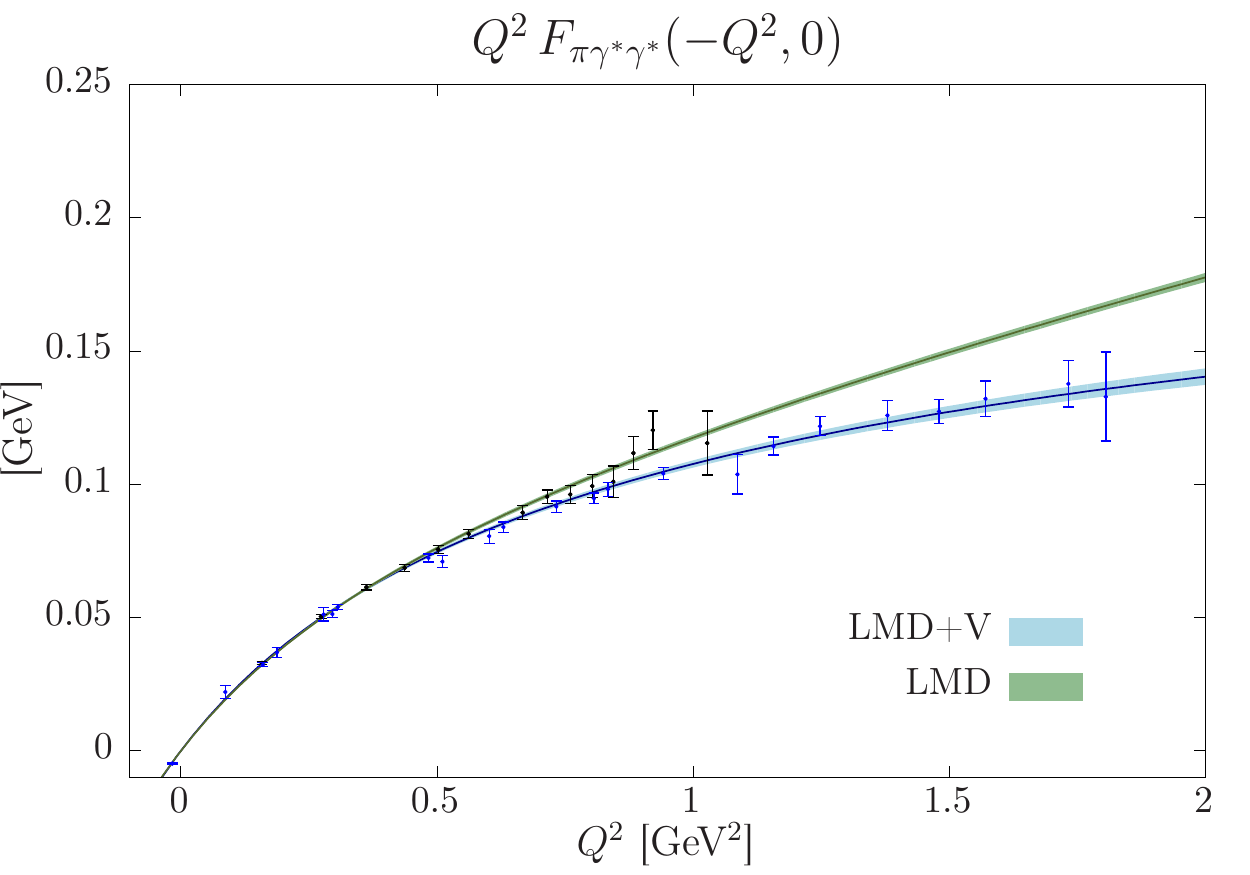}
	\includegraphics*[width=0.49\linewidth]{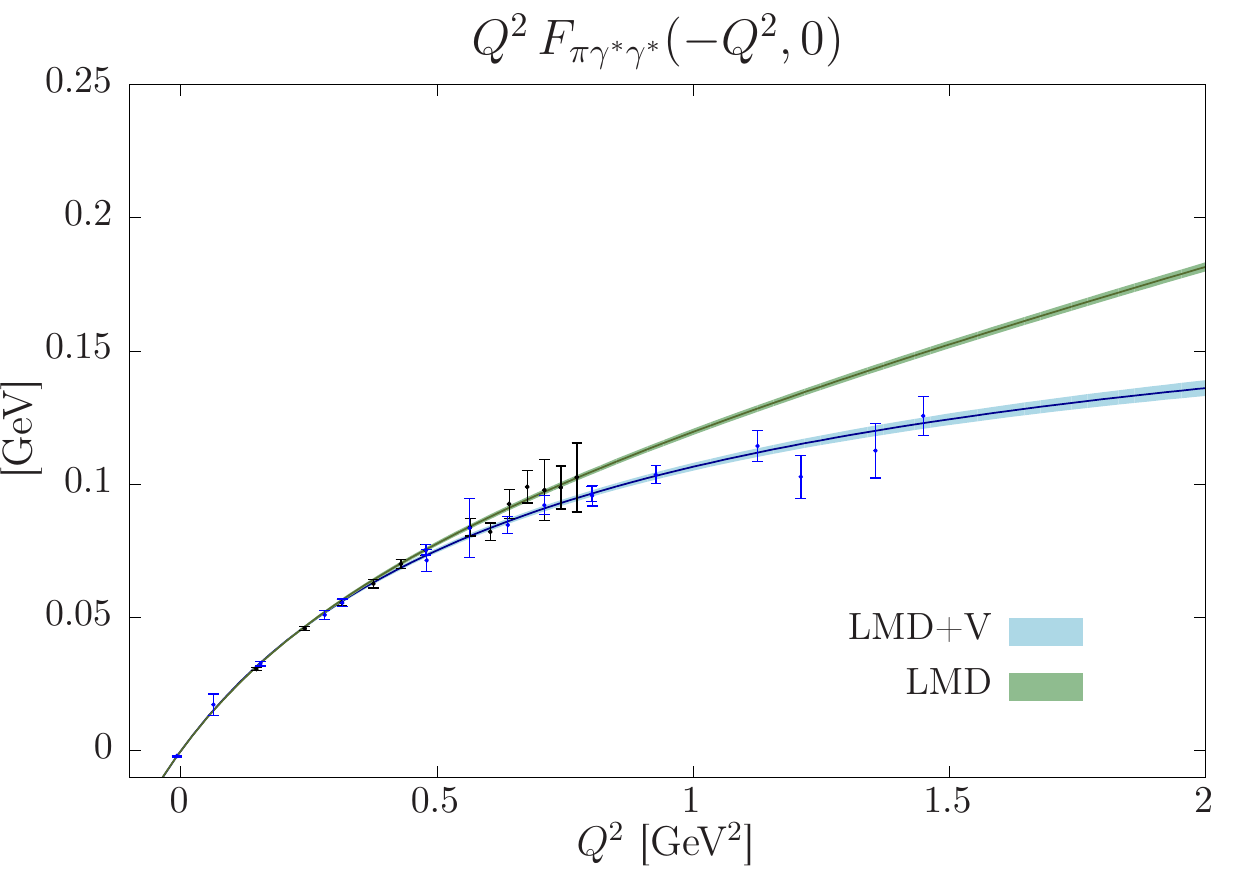}

	\caption{Results of the global LMD and LMD+V fits in the single-virtual case and for the ensembles N203 (left panel) and N200 (right panel). The data strongly favor the LMD+V model, which has the correct asymptotic behavior at these kinematics, over the LMD model.}	
	\label{fig:res_sv}
\end{figure}

\begin{figure}[t]
	\includegraphics*[width=0.49\linewidth]{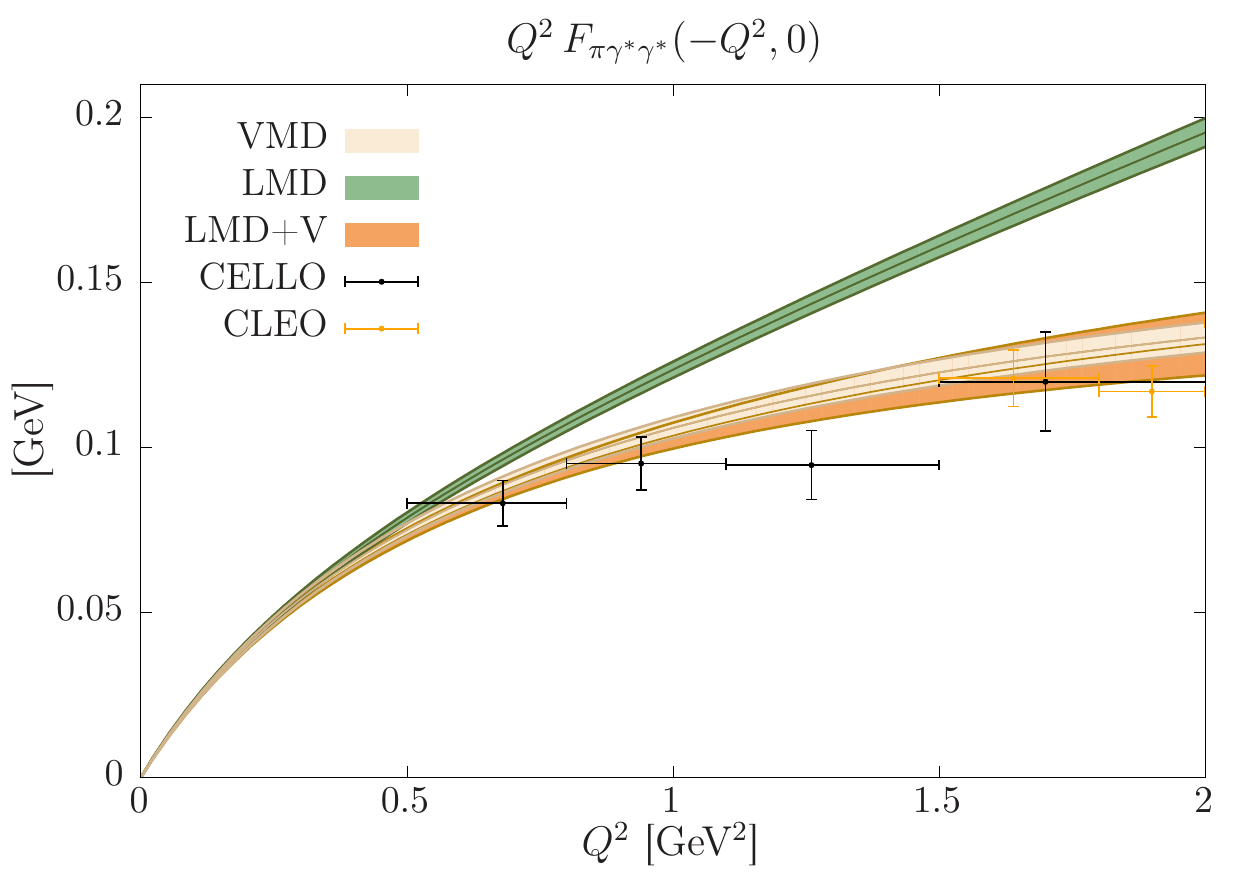}
	\includegraphics*[width=0.49\linewidth]{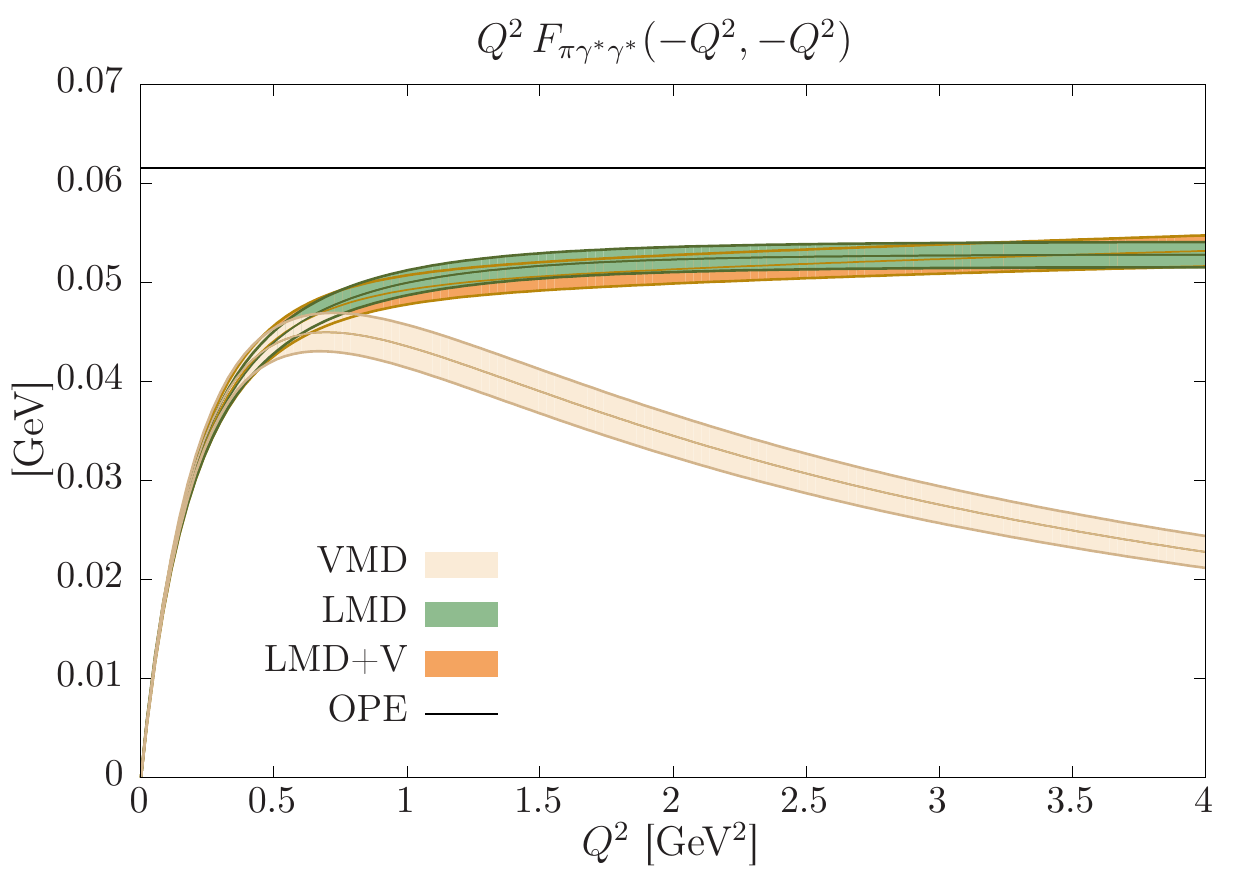}	
	\caption{Transition form factor extrapolated to the physical point using the three phenomenological models described in the main text. 
        In the single-virtual case, we also compare the results with data from the CELLO and CLEO experiments~\cite{exp}.}
	\label{fig:pheno_fits}
\end{figure}

In Ref.~\cite{Gerardin:2016cqj}, we have shown that the VMD model fails to reproduce the lattice data at large $Q^2$ due to the wrong asymptotic behavior in the double-virtual case. Fitting our new data, we obtain $\chi^2/\dof = 4.8$.
However, we obtain a reasonable $\chi^2$ if we restrict the fit to the single-virtual TFF and the result at the physical point reads
\begin{equation}
\alpha^{\VMD} =  0.258(7)~\GeV^{-1} \,, \quad M_V^{\VMD} = 836(18)~\MeV \,, 
\label{eq:resVMD}
\end{equation}
with $\chi^2/\dof = 1.2$. The value of $\alpha^{\VMD}$ is compatible with the theoretical prediction for the anomaly within 2$\sigma$ and $M_V$ is close to the rho meson mass. 
For the LMD model, we fit $\alpha$, $\beta$, and $M_V$ on the full kinematical range. The result of the global fit at the physical point reads
\begin{equation}
\alpha^{\LMD} = 0.270(6)~\GeV^{-1} \,, \quad \beta =  -26.2(0.7)~\MeV \,, \quad M_V^{\LMD} = 656(13)~\MeV \,, 
\label{eq:resLMD}
\end{equation}
with $\chi^2/\dof = 1.5$. The value of $\alpha^{\LMD}$ is compatible with the anomaly constraint with a statistical precision of $2~\%$. However, the value of $\beta$ is lower, in absolute value, than the OPE prediction at leading order $\beta^{\rm OPE} = -F_{\pi}/3 = -30.8~\MeV$. This point will be discussed in Sec.~\ref{sec:aymptotic}: the OPE prediction neglects higher twists and $\mathcal{O}(\alpha_s)$ corrections which are sizable at virtualities accessible on the lattice. 
In Fig.~\ref{fig:res_sv} we show the quality of the fit for the ensembles N203 and N200 in the single-virtual configuration. At large $Q^2$, we observe a significant deviation between the fit and the lattice data which explains the relatively bad $\chi^2$. This feature was not observed in our previous work~\cite{Gerardin:2016cqj} where only virtualities $Q^2 < 0.5~\GeV^2$ were accessible in the single-virtual case.
The fact that the LMD model fails to describe the lattice data in the single-virtual case is not unexpected since the model has the wrong asymptotic behavior. This explains the discrepancy between the TFF extrapolated to the physical point using the LMD model and the experimental data, as shown in Fig.~\ref{fig:pheno_fits}.

Finally, for the LMD+V model we explicitly set $\widetilde{h}_1=0$ to satisfy the OPE constraint. Furthermore, to reduce the number of fit parameters, and inspired by quark models, we assume a constant shift in the spectrum and set $M_{V_2}(\widetilde{y}) = m^{\exp}_{\rho^{\prime}} + M_{V_1}(\widetilde{y}) - m^{\exp}_{\rho}$ with $m^{\exp}_{\rho^{\prime}} = 1.465~\GeV$.  All the remaining parameters are fitted and the result reads
\begin{gather}
\nonumber \alpha^{\LMDV} = 0.261(7)~\GeV^{-1} \,, \quad M_{V_1}^{\LMDV} = 770(92)~\MeV \,,\\ 
\widetilde{h}_0 = -0.030(3)~\GeV \,, \quad \widetilde{h}_2 = 0.277(70)~\GeV^3 \,, \quad \widetilde{h}_5 = -0.187(40)~\GeV \,,
\label{eq:resLMDV1}
\end{gather}
with $\chi^2/\dof = 1.2$. We note that the same fit would be unstable if only lattice data obtained in the pion rest frame were included. 
The coefficient $\widetilde{h}_0$, related to the OPE behavior, is in good agreement with the leading order OPE prediction $\widetilde{h}_0=-F_{\pi}/3$. 
The phenomenological value $\widetilde{h}_2 = 0.327~\GeV^2$~\cite{Melnikov:2003xd} can be fixed by comparing with the higher-twist corrections in the OPE in Eq.~(\ref{eq:OPE}) and our result turns out to be in good agreement as well.
Finally, $\widetilde{h}_5$ is consistent with the phenomenological prediction $\widetilde{h}_5 = -0.166 \pm 0.006~\GeV$ obtained in Ref.~\cite{Knecht:2001xc} by fitting the LMD+V model to the CLEO data in the single-virtual case. At our level of precision, this model provides a good description of our lattice data in the whole kinematical range covered by this study. It is also perfectly compatible with the available experimental data, as can be seen on the left panel of Fig.~\ref{fig:pheno_fits}.

Recently, a new parametrization of the pion TFF over the whole spacelike region has been proposed in~\cite{Danilkin:2019mhd}. The parametrization of the $(Q_1^2,Q_2^2)$ dependence relies on a single parameter, which can be obtained by adjusting the model to the experimental data in the single-virtual configuration. 
The model obeys by construction the leading prediction for the asymptotic behavior at large $Q^2$; however, it does not accurately reproduce our lattice result at intermediate virtualities: the prediction underestimates our result by about 20\% at $Q_1^2=Q_2^2=0.5~\GeV^2$ as can be seen in Fig~\ref{fig:z-exp}. Nonetheless, in the future this model could perhaps be used as the factor $P(Q_1^2,Q_2^2)$ in the expansion (\ref{eq:z_exp_mod}).

\subsection{Normalization of the TFF, the decay width 
  $\Gamma(\pi^0 \to \gamma\gamma)$ and the slope of the form factor \label{sec:norm} } 

To leading order in the fine-structure constant $\alphaQED$, the decay width of $\pi^0 \to \gamma\gamma$ is given by the pion TFF as follows: $\Gamma(\pi^0 \to \gamma\gamma) = (\pi\alphaQED^2m_\pi^3/4) \FF(0,0)^2$. From the measurement $\Gamma(\pi^0 \to \gamma\gamma) = 7.82(22)~{\rm eV}$ by the PrimEx experiment~\cite{Larin:2010kq} we get the TFF at the origin with a precision of 1.4\%, $\alpha^{\rm PrimEx} = 0.276(4)~\GeV^{-1}$, and this is therefore an important benchmark for our lattice calculation. The PrimEx-II experiment aims at reducing the error on the decay width by a factor of two and a preliminary result $\Gamma(\pi^0 \to \gamma\gamma) = 7.80(13)~{\rm eV}$ has been released~\cite{Gasparian:2016oyl,Primex2Mainz}, which leads to $\alpha^{\rm PrimEx-II} = 0.275(2)~\GeV^{-1}$ with 0.8\% precision. A tension exists between the state-of-the-art chiral predictions of the decay width~\cite{pion_decay_NLO, Ananthanarayan:2002kj, pion_decay_updates, Kampf:2009tk} and the new, more precise, preliminary measurement.

In the chiral limit, the normalization of the TFF is exactly predicted by the ABJ chiral anomaly~\cite{Adler:1969gk,Bell:1969ts}, 
\begin{equation}
\FF(0,0) = \frac{1}{4\pi^2 F} \,,
\label{eq:ABJ}
\end{equation}
where $F$ is the pion decay constant in the chiral limit. At finite quark mass, this result receives corrections which can be computed in the framework of chiral perturbation theory ($\chi$PT) or directly using lattice QCD.  The chiral expansion of the pion TFF at vanishing momenta is known up to next-to-next-to-leading (NNLO) order in $\chi$PT~\cite{Kampf:2009tk}. Chiral logarithms are absent at NLO when the result is expressed in terms of the physical value of $F_{\pi}$~\cite{pion_decay_NLO}. They appear only at NNLO but were shown to contribute at the permille level and are negligible at our level of accuracy~\cite{Kampf:2009tk}. This motivates the extrapolation of the normalization of the form factor on the lattice using the \textit{Ansatz}
\begin{equation}
f_{\pi} \, \FF(0,0) = \widetilde{\alpha} + \gamma \, m_{\pi}^2  +
\delta_d \left( \frac{a}{a_{\beta=3.55}} \right)^2 \,,  
\label{eq:KMfit}
\end{equation}
already used in a previous lattice calculation~\cite{Feng:2012ck}. This functional form differs from the one assumed in the previous sections but both parametrizations lead to compatible results at our level of precision. Equation.~(\ref{eq:KMfit}) has the advantage to offer the possibility to extract the low-energy constant (LEC) $C_7^{\mathrm{Wr}}$ in the odd-intrinsic-parity sector of chiral perturbation theory at order $p^6$~\cite{Bijnens:2001bb} via the relation~\cite{Ananthanarayan:2002kj, Kampf:2009tk} 
\begin{equation}
C_7^{\mathrm{Wr}} = - \frac{3}{64} \, \gamma \,, 
\end{equation}
by varying the quark masses and thus $m_\pi$ over a range of values on our lattice ensembles. As explained in Sec.~\ref{sec:ens}, we note that our lattice ensembles lie in the chiral trajectory where the trace of the bare quark matrix is kept constant. If we repeat the analysis done in Sec.~\ref{sec:zexp} but using $f_{\pi} \, \FF(Q_1^2,Q_2^2)$ instead of $\FF(Q_1^2,Q_2^2)$ as our primary observable, and limiting the fit range below 1~GeV, we obtain the result
\begin{equation} \label{eq:results_alpha_C7W}
\alpha = 0.264(8)(4)~\GeV^{-1} \,, 
\quad C_7^{\mathrm{Wr}} = 0.16(18)\times 10^{-3}~\GeV^{-2}  \,,
\end{equation}
where the first error is statistical and the second error is an estimate of the systematic error due to the truncation on the $z$-expansion (see Appendix~\ref{app:B}). The choice $N=1$ is already sufficient to get a $\chi/\dof = 1.1$. The normalization of the TFF is almost unchanged compared to the fit done in Sec.~\ref{sec:zexp}. The result in Table~\ref{tab:z-exp}, with $N=3$, leads to $\alpha = 0.261(13)(2)~\GeV^{-1}$. The value of the normalization $\alpha$ is compatible with the theoretical prediction $\alpha^{\rm th} = 0.274~\GeV^{-1}$, given by Eq.~(\ref{eq:ABJ}) with $F$ replaced by $F_{\pi}$, with a precision of $3.5~\%$. At this level of precision, we are not yet sensitive to the other chiral corrections, not taken into account by this replacement. We note that our choice of $t_0$ in the $z$-expansion is optimal for the pion-pole contribution to $a_\mu^{\rm HLbL}$, discussed in the next subsection, but not for the normalization, as explained in Appendix~\ref{app:B}. The results for $\alpha$ and $C_7^{\mathrm{Wr}}$ obtained from a fit based on the LMD+V model are compatible within error bars with the model-independent results in Eq.~(\ref{eq:results_alpha_C7W}).

The central value of our result for the LEC $C_7^{\mathrm{Wr}}$ in Eq.~(\ref{eq:results_alpha_C7W}) is larger than the bound $|C_7^{\mathrm{Wr}}| < 0.06 \times 10^{-3}~\GeV^{-2}$ used in Ref.~\cite{Kampf:2009tk}. Of course with our current uncertainty we cannot exclude that this LEC vanishes. The bound is essentially based on estimates using a resonance Lagrangian with heavy pseudoscalars and the nonobservation of the decay $\pi(1300) \to \gamma\gamma$~\cite{Moussallam:1994xp, Ananthanarayan:2002kj, pdg:2018}. The LEC $C_7^{\mathrm{Wr}}$ also vanishes exactly~\cite{Knecht:2001xc} in simple resonance Lagrangians with vector mesons only, such as the VMD model for the TFF or the model proposed in Ref.~\cite{Prades:1993ys}, that do not obey short-distance constraints from the OPE. On the other hand, using the LMD model~\cite{Moussallam:1994xp, Knecht:1999gb} or a resonance Lagrangian with vector and heavy pseudoscalar mesons (LMD+P) that obeys these short-distance constraints, one obtains the estimate $C_7^{\mathrm{Wr}} = 0.35(7) \times
10^{-3}~\GeV^{-2}$~\cite{Kampf:2011ty}. The result is actually dominated by the LMD part and the additional contribution from the heavy pseudoscalar is again estimated to be very small. Note, however, that the error might be underestimated. All resonance estimates of LECs are based on the assumption of narrow resonances in large-$N_c$ QCD and carry an intrinsic uncertainty of 20\%-30\%. Under the assumption that the LECs have a natural size of $0.5 \times 10^{-3}~\GeV^{-2}$, one could argue that this implies rather an absolute error of $\pm (0.10-0.15) \times 10^{-3}~\GeV^{-2}$ for all LECs. With the current precision, our value for $C_7^{\mathrm{Wr}}$ in Eq.~(\ref{eq:results_alpha_C7W}) is also compatible with the larger LMD+P estimate.

As mentioned above, the precision obtained for the normalization of the TFF from the lattice cannot compete with the PrimEx or even the PrimEx-II result. However, with our new estimate for $C_7^{\mathrm{Wr}}$ in Eq.~(\ref{eq:results_alpha_C7W}), we can follow the same strategy as in Refs.~\cite{Moussallam:1994xp, Ananthanarayan:2002kj, Kampf:2009tk} and use it together with the experimental decay width $\Gamma(\eta \to \gamma\gamma) = 0.515(18)~\mathrm{keV}$~\cite{pdg:2018} to determine the other LEC $C_8^{\mathrm{Wr}}$ that appears in the expression for the decay $\pi^0 \to \gamma\gamma$ at NLO in $\chi$PT with the result
\begin{equation}
C_8^{\mathrm{Wr}} = 0.56(17) \times 10^{-3}~\GeV^{-2}\, .  
\end{equation}
This result is almost identical to the estimate $C_8^{\mathrm{Wr}} = 0.58(20) \times 10^{-3}~\GeV^{-2}$ from Ref.~\cite{Kampf:2009tk} that was obtained by setting $C_7^{\mathrm{Wr}} = 0$. We took over their estimate of 30\% uncertainty from neglected terms of order $m_s^2$ in the chiral expansion. Using the expression for the decay amplitude $\pi^0 \to \gamma\gamma$ given in Ref.~\cite{Kampf:2009tk}, including strong isospin breaking effects from $m_u \neq m_d$ and electromagnetic corrections, these changes in the LECs lead to the prediction
\begin{equation}
\Gamma(\pi^0 \to \gamma\gamma) = 8.07(10)~\mathrm{eV}\ , 
\end{equation}
confirming the value $\Gamma(\pi^0 \to \gamma\gamma) = 8.09(11)~\mathrm{eV}$ obtained in Ref.~\cite{Kampf:2009tk}. Therefore the tension with the preliminary result of the PrimEx-II experiment~\cite{Primex2Mainz} remains.

Another phenomenologically interesting quantity is the slope of the form factor at the origin. It is defined through
\begin{equation}
b_{\pi^0} = \frac{1}{\FF(0,0)}  \frac{ \mathrm{d}\FF(q^2,0)}{ \mathrm{d} q^2}\Big|_{q^2=0} \,.
\end{equation}
We obtain
\begin{equation}
b_{\pi^0}^{\LMDV} = 1.59(11)~\GeV^2 \,,
\quad b_{\pi^0}^{z-{\rm exp}} =1.57(13)(11)~\GeV^2\,. 
\end{equation}
where the first error is statistical and the second error is the systematic error associated with the $z$-expansion. Both determinations are compatible with each other and with the result obtained in Ref.~\cite{Hoferichter:2018dmo} using a dispersive framework (within our convention, their result reads $b_{\pi^0} = 1.73(5)~\GeV^2$). The model independent determination is affected by a relatively large error which could be reduced in the future by adding large-volume ensembles which probe the low-$Q^2$ region. 

We also try the method of Canterbury approximants proposed in Ref.~\cite{Masjuan:2017tvw} in the context of the pion TFF. It is a generalization of the Pad\'e approximants for bivariate functions and we refer the reader to Ref.~\cite{Masjuan:2015lca} for an introduction to the subject. A general Canterbury approximant has the form 
\begin{equation}
C_M^N(Q_1^2,Q_2^2) = \frac{ \sum_{n,m=0}^{N} a_{nm} Q_1^{2n} Q_2^{2m} }{ \sum_{n,m=0}^{M} b_{nm} Q_1^{2n} Q_2^{2m} } \,,
\end{equation}
with the convention $b_{00} = 1$ and the relations $a_{nm} = a_{mn}$, $b_{nm} = b_{mn}$ reflecting the Bose symmetry of the photons. 
We consider both the diagonal and the subdiagonal sequences of approximants $C_N^N$ and $C_{N+1}^N$. In particular, the sequence $C_{N+1}^N$ automatically satisfies the short-distance behavior $\sim 1/Q^2$ if one sets $b_{NN} = 0$. The lowest order elements for both sequences are 
\begin{gather}
C_1^0(Q_1^2,Q_2^2) = \frac{ a_{00} }{1 + b_{01} (Q_1^2 + Q_2^2) + b_{11} Q_1^2 Q_2^2 } \,, \\[2mm] 
C_1^1(Q_1^2,Q_2^2) = \frac{ a_{00} + a_{01} (Q_1^2 + Q_2^2) + a_{11} Q_1^2 Q_2^2  }{1 + b_{01} (Q_1^2 + Q_2^2) + b_{11} Q_1^2 Q_2^2 } \,. 
\end{gather}
The main drawbacks of this method are, first, the absence of a proof of convergence for the pion TFF and, second, the rapid growth of the number of fit parameters. As a consequence, for large values of $M$, the denominator can become singular at large virtualities (the presence of spurious poles in the spacelike domain) where there are no lattice data. 
In the following, we perform a global fit of the TFF, over the whole kinematic range, assuming a linear dependence of the coefficients $a_{ij}$ and $b_{ij}$ on the dimensionless parameter $\widetilde{y}$ and a quadratic dependence in the lattice spacing $a$. Both sequences already give a good $\chi^2$ for $N=1$. Using higher-order approximants only increases the statistical error or leads to unstable fits. The results for the two sequences are summarized in Table~\ref{tab:canterbury}. We use as our final estimate the average between the approximants $C_2^1$ and $C_1^1$ and we take half the difference between the two sequences as an estimate for the systematic error : $\alpha = 0.261(12)_{\stat}(8)_{\syst}~\GeV^{-1}$. This value is compatible with our determination via the $z$-expansion with a similar statistical error, but a systematic error which is difficult to estimate.

\renewcommand{\arraystretch}{1.2}
\begin{table}[t!]
\caption{Normalization of the transition form factor obtained from the two sequences of Canterbury approximants $C^{N}_{N}$ and $C^{N}_{N+1}$. As before, chi-square corresponds to uncorrelated fits.}  
\begin{center}
	\begin{tabular}{l@{\qquad}c@{\quad}c@{\quad}c}
	\hline
					&	$C^{0}_{1}$	&	$C^{1}_{2}$	&	$C^{1}_{1}$ 		\\
	\hline
$\alpha~[\GeV^{-1}]$		& 	0.325(8)		&	0.253(12)		&	0.268(10)			\\ 
	\hline 
$\chi^2/\dof$   			& 	4.1			&	1.2			&	1.3				\\
	\hline 
	\end{tabular}
\end{center}
\label{tab:canterbury}
\end{table}

\subsection{Asymptotic behavior of the pion transition form factor \label{sec:aymptotic}}

When at least one of the photon virtualities is large, and assuming (collinear) factorization, the TFF can be written as a convolution integral~\cite{BL_3_papers}
\begin{align}
\FF(q_1^2,q_2^2) &= \frac{2F_{\pi} }{3}  \int_0^1 \mathrm{d}x \, T_H(x,Q_1^2,Q_2^2;\mu^2) \, \varphi_{\pi}(x;\mu^2) \,,
\label{eq:DIS}
\end{align}
where $T_H(x,Q_1^2,Q_2^2,\mu^2)$ is a hard scattering kernel, calculable in perturbative QCD, and $\varphi_{\pi}(x;\mu^2)$ is the nonperturbative pion distribution amplitude (DA) of twist two (normalized to one). The latter is scale dependent but otherwise universal and therefore plays a major role in the study of hard exclusive processes.
The factorization and renormalization scales are chosen to be equal $\mu_R=\mu_F=\mu$ and of order $\overline{Q}^2 = Q_1^2 + Q_2^2$. Here, $x$ and $1-x$ are, respectively, the quark and antiquark longitudinal momentum fractions of the meson’s total momentum. Neglecting isospin breaking effects, the pion distribution amplitude is symmetric under the interchange of $x$ and $\overline{x} = 1-x$ such that $\varphi_{\pi}(x) =  \varphi_{\pi}(1-x)$. At leading twist and leading order in perturbative QCD, the hard scattering kernel is given by~\cite{BL_3_papers}
\begin{equation}
T_H^{\rm LO}(x,Q^2) = \frac{1}{x Q_1^2 + (1-x) Q_2^2} = \frac{2}{\overline{Q}^2 + \omega (1- 2x)\overline{Q}^2} \,,
\end{equation}
where we have introduced the asymmetry parameter $\omega$ defined through 
\begin{equation}
\omega = \frac{Q_2^2-Q_1^2}{Q_2^2+Q_1^2} \,, \quad Q_1^2 = \frac{1-\omega}{2} \, \overline{Q}^2 \,, \quad Q_2^2 = \frac{1+\omega}{2} \, \overline{Q}^2 \,.
\end{equation}
In this section, the parameter $\omega$ should not be confused with the parameter $\omega_1$ introduced in Eq.~(\ref{eq:Mlat}).
At next-to-leading order in perturbative QCD, the hard scattering kernel has been computed for all values of $\omega$ in the $\overline{{\rm MS}}$ scheme in Refs.~\cite{delAguila:1981nk,Braaten:1982yp,Kadantseva:1985kb} and the result is 
\begin{equation}
T_H^{\rm NLO}(x,Q^2) = \frac{1}{\overline{Q}^2 + \omega (1- 2x)\overline{Q}^2} \left[ 1 + C_F \frac{\alpha_s(\mu)}{2\pi} t( \overline{x} , \overline{w}) \right]+ (x \leftrightarrow 1-x) \,,
\end{equation}
where an explicit expression for $t(\overline{x}, \overline{w})$ with $\overline{w} = (1-\omega)/2$ is given by Eq.~(5.2) of Ref.~\cite{Braaten:1982yp}. Then, using $C_F=4/3$ and the symmetry properties of the pion DA, Eq.~(\ref{eq:DIS}) reduces to
\begin{align}
\FF(q_1^2,q_2^2) = \frac{4 F_{\pi} }{3\overline{Q}^2}\int_0^1 \frac{ \varphi_{\pi}(x) }{ 1 + \omega(1-2x) } \, \left[ 1 + \frac{2 \alpha_s(\mu)}{3\pi} t( \overline{x},w) \right]  \mathrm{d}x \ + \ \mathcal{O}\left( \frac{1}{\overline{Q}^4}, \frac{\alpha_s^2}{\overline{Q}^2} \right) \,.
\label{eq:TFF_asympt}
\end{align}
The asymptotic expression of the twist two pion DA, $\varphi_{\pi}^{\rm as}(x) = 6x(1-x)$, leads to the OPE and BL asymptotic predictions for the double and single off-shell form factors respectively \cite{BL_3_papers, Efremov:1979qk}. However, this asymptotic result is expected to hold only at very large virtualities. 
Higher twist corrections, discussed in Refs.~\cite{Novikov:1983jt,Khodjamirian:1997tk}, are given by 
\begin{equation}
- \frac{2 F_{\pi}}{3} \frac{80}{9} \frac{\delta^2}{\overline{Q}^4} \left( \frac{ -2 \omega^3 + 3\omega + 3 (\omega^2-1) \mathrm{tanh}^{-1}(\omega) }{\omega^5} \right)  \,,
\end{equation}
where the parameter $\delta^2 = 0.20(2)~\GeV^2$ has been evaluated in Ref.~\cite{Novikov:1983jt} using QCD sum rules and more recently in Ref.~\cite{Bali:2018spj} using lattice QCD.

\subsubsection{Double-virtual form factor : Symmetric case} 

When both photons share the same virtualities ($\omega=0$) one can see from Eq.~(\ref{eq:TFF_asympt}) that the asymptotic value of the TFF is not sensitive to the shape of the pion DA (one can show that $t(\overline{x},1/2) = -3/2$ is also independent of the quark longitudinal momentum fraction $x$) and the result reads
\begin{equation}
\FF(-Q^2,-Q^2) = \frac{2 F_{\pi} }{3Q^2} \left[ 1 - \frac{\alpha_s(Q)}{\pi} - \frac{8}{9} \frac{\delta^2}{Q^2}  \right]  + \mathcal{O}\left( \alpha_s^2(Q), \frac{1}{Q^4} \right) \,.
\label{eq:OPE_ht}
\end{equation}
Checking the validity of Eq.~(\ref{eq:OPE_ht}) at asymptotically large $Q^2$ provides a strong test of perturbative QCD. 
Here we will ask whether the given functional form describes our lattice data at moderately large $Q^2$.

\subsubsection{Double-virtual form factor : General case} 

When both photons share different virtualities, the situation is more difficult because the result depends on the precise shape of the pion DA, which is largely unknown, although a recent lattice calculation of the two lowest Gegenbauer moments has been carried out on a similar set of lattice ensembles in Ref.~\cite{Bali:2019dqc}. However, as pointed out in Ref.~\cite{Diehl:2001dg}, this dependence is small for small values of the asymmetry parameter $\omega$. At leading twist, we have 
\begin{align}
\nonumber \FF(-Q_1^2,-Q_2^2) = \frac{4 F_{\pi} }{3\overline{Q}^2} & \left[ 1 - \frac{\alpha_s(Q)}{\pi} + \frac{1}{5} \omega^2 \left( 1 - \frac{5}{3} \frac{\alpha_s(Q)}{\pi} \right)  \right. \\
& \left. +\frac{12}{35} \omega^2 a_2(\mu) \left( 1 + \frac{5}{12} \frac{\alpha_s(Q)}{\pi} \left\{ 1 - \frac{10}{3} \ln \frac{\overline{Q}^2 }{2\mu_F^2}  \right\} \right) \right] + \mathcal{O}(\omega^4, \alpha_s^2) \,,
\end{align}
where $a_2$ is the second coefficient in the expansion of the pion distribution amplitude in terms of Gegenbauer polynomials. For $Q_2^2 = 2 Q_1^2$, where $\omega = 1/3$, corrections to the asymptotic DA are of the order of $2\%$. We can therefore fit our lattice data in a much wider range without being sensitive to the actual shape of the pion DA.

\subsubsection{Fits} 

\renewcommand{\arraystretch}{1.4}
\begin{table}[t!]
\caption{Value of the higher twist coefficient $\delta^2$ as a function of $Q^2_{\rm min}$.} 
\begin{center}
	\begin{tabular}{l@{\qquad}c@{\quad}c@{\quad}c@{\quad}c@{\quad}c}
	\hline
	$Q^2_{\rm min}~[\GeV^2]$	&	1.5		&	1.75		&	2.0		&	2.25 	\\
	\hline
	$\delta^2~[\GeV^2]$		& 	0.13(5)	&	0.14(6)	&	0.15(6)	&	0.15(7)	\\ 
	\hline 
	\end{tabular}
\end{center}
\label{tab:OPE}
\end{table}

We perform a global fit using Eq.~(\ref{eq:OPE_ht}) where $\delta^2$ is considered as a free fit parameter and is allowed to vary linearly with $\widetilde{y}$ and $a^2$. We restrict the fit to virtualities with $\omega < 1/3$ and $Q^2>Q^2_{\rm min}$ and use 
the four-loop running strong coupling in the $\overline{\rm MS}$ scheme~\cite{vanRitbergen:1997va}.
Of course, one might question the applicability of perturbative QCD down to such low values of  $|Q_{\rm min}| = 1.2~\GeV$, even in Euclidean space, see Ref.~\cite{Eidelman:1998vc} for an analysis of nonperturbative effects in the Adler function. 
The results at the physical point and for different values of $Q^2_{\rm min}$ are listed in Table~\ref{tab:OPE} and we quote as our final estimate the value at $Q^2_{\rm min} = 2~\GeV^2$ 
\begin{equation}
\delta^2 = 0.15(6)~\GeV^2 \,.
\end{equation}
This result is compatible with the sum rule determination~\cite{Novikov:1983jt} and the lattice determination~\cite{Bali:2018spj} although with larger error. 
 
\subsection{Determination of the pion-pole contribution to HLbL scattering	 in the $(g-2)_{\mu}$}
\label{sec:g-2}

The hadronic light-by-light scattering contribution to the anomalous magnetic moment of the muon is one of two dominant sources of uncertainty along with the hadronic vacuum polarization. For the HLbL scattering, a model independent dispersive approach has been proposed recently~\cite{HLbL_DR} and the dominant contribution is expected to be the pion-pole contribution which requires the pion TFF as input. Until recently, most estimates were based on model calculations where errors are difficult to estimate. In our previous work~\cite{Gerardin:2016cqj}, we provided the first lattice QCD calculation of the pion-pole contribution, with a statistical precision of about 12\%. Recently, a data-driven determination based on the calculation of the TFF using dispersion theory was published~\cite{Hoferichter:2018dmo}. The result is compatible with our previous determination but with reduced uncertainty. In this section, we propose to use our results to improve our estimate of the pion-pole contribution from lattice QCD.

Following~\cite{Jegerlehner:2009ry}, the pion-pole contribution to the hadronic light-by-light scattering in the muon $(g-2)$ can be expressed as a three-dimensional integral involving two model-independent weight functions $w_1$ and $w_2$ and the product of a single-virtual TFF times a double-virtual TFF for arbitrary spacelike virtualities,
\begin{multline}
a_\mu^{\mathrm{HLbL}; \pi^0}  =  \left( \frac{\alphaQED}{\pi} \right)^3 \int_0^\infty \!\!\!dQ_1 \!\!\int_0^\infty \!\!\!dQ_2 \!\! \int_{-1}^{1} \!\!d\tau \,   \Big( w_1(Q_1,Q_2,\tau) \, \FF(-Q_1^2, -(Q_1 + Q_2)^2) \, \FF(-Q_2^2,0)  \\ 
 + w_2(Q_1,Q_2,\tau) \, \FF(-Q_1^2, -Q_2^2) \, \FF(-(Q_1+Q_2)^2,0) \Big) \,. 
\label{contribution_2}
\end{multline}  
The integrals run over the lengths $Q_i = |(Q_i )_{\mu} |$, $i = 1, 2$ of the two Euclidean four-momentum vectors and the angle $\theta$
between them, $Q_1 \cdot Q_2 = Q_1Q_2 \cos \theta$, where we defined $\tau = \cos \theta$. The analytical expressions for the dimensionless weight functions $w_i(Q_1,Q_2,\tau)$, $i = 1,2$ can be found in Ref.~\cite{Jegerlehner:2009ry}. In particular, it was shown in Ref.~\cite{Nyffeler:2016gnb} that the relevant integration range involves spacelike virtualities below $2~\GeV^2$, precisely the kinematical region where we have lattice data.

\subsubsection{Phenomenological models} 

A first estimate of the pion-pole contribution is obtained using the LMD+V model, which provides a good description of our lattice data. The parameters at the physical point are determined from the global fit procedure described in Sec.~\ref{subsec:pheno_fits} and we obtain
\begin{align} \label{eq:amulmdv}
\amuLMDV &= (58.6 \pm 2.7) \times 10^{-11},
\end{align} 
where the error is statistical but includes the error from the continuum and chiral extrapolations.
It can be compared with our previous estimate, $\amuLMDV = (65.0 \pm 8.3) \times 10^{-11}$, obtained with two dynamical quarks~\cite{Gerardin:2016cqj}. 
The slightly lower central value in Eq.~(\ref{eq:amulmdv}) arises due to the slightly lower values of the parameters $\alpha$ and $\widetilde{h}_2$ emerging from the fit, but the results are in agreement within the quoted uncertainties.
We also checked that computing $\amuLMDV$ on each ensemble separately and then, in a second step, extrapolating the results to the continuum limit and at the physical pion mass using a fit linear in $\widetilde{y}$ and $a^2$ gives a similar result.
In this case, we would obtain $(58.4 \pm 2.4) \times 10^{-11}$.
Since we are using a phenomenological model to describe the lattice data, the result could be biased and the error underestimated. Moreover, for the LMD+V model, we do not fit all the model parameters but made some assumptions on the second vector-resonance mass $M_{V_2}$ to stabilize the fit.

\subsubsection{Canterbury approximants} 

A second possibility is to use the results obtained from the method of the Canterbury approximants presented in Sec.~\ref{sec:norm}. The short-distance constraints in Eqs~(\ref{eq:BL}) and (\ref{eq:OPE}) are explicitly implemented using the sequence with $M=N+1$ with the additional constraint $b_{NN} = 0$. Using the approximant $(N, M)=(1,2)$, which provides a good description of our lattice data, we find at the physical point
\begin{equation}
\amu = (58.3 \pm 4.2) \times 10^{-11} \,. 
\end{equation}
The next approximant leads to unstable fits. 
We note that this result is compatible with the LMD+V model determinations, which suggests that the model dependence of the central value is small. As for the LMD+V model, it is, however, difficult to assess a systematic error, especially because we are limited to rather low values of $M$. 
       
\subsubsection{Final result : $z$-expansion} 
\label{sec:pion_pole_zexp}
      
Finally, the $z$-expansion provides a systematically improvable determination. The choice for the function $P(Q_1^2, Q_2^2)$ in Eq.~(\ref{eq:P_zexp}) guarantees the $1/Q^2$ falloff of the TFF in all directions, and, using the results of Sec.~\ref{sec:zexp} with $N=3$, we quote as our final result 
\begin{equation}
a_{\mu}^{\mathrm{HLbL}; \pi^0} = (59.7 \pm 3.4 \pm 0.9 \pm 0.5) \times 10^{-11} =  (59.7 \pm 3.6)\times 10^{-11}\,, 
\label{eq:amu_zexp}
\end{equation} 
where the first error is statistical, the second is the systematic error inferred from the study in Appendix~\ref{app:B} and the third error comes from the disconnected contribution. In particular, the first error includes the error on the lattice spacing, the renormalization of the vector currents and the extrapolation to the physical point. In the last step of Eq.~(\ref{eq:amu_zexp}), we have added the errors in quadrature.
Since the TFF has units of $\GeV^{-1}$, the relative scale-setting uncertainty on the pion-pole contribution is 2 times the relative error on the lattice spacing expressed in fm.
The scale is known with a precision of $1\%$ which translates in an uncertainty of about 2\% on $\amu$. The uncertainty on the renormalization of the vector current is negligible. Therefore, the statistical precision of the correlation functions and the extrapolation to the physical point are the dominant sources of uncertainties in our calculation. They could be improved in a future calculation, in particular by including a lattice ensemble at the physical pion mass in the analysis.
In contrast to phenomenological analyses, our result for $a_{\mu}^{\mathrm{HLbL}; \pi^0}$  is significantly more accurate than our determination of $\FF(0,0)$.

\begin{figure}[t]
	\includegraphics*[width=0.49\linewidth]{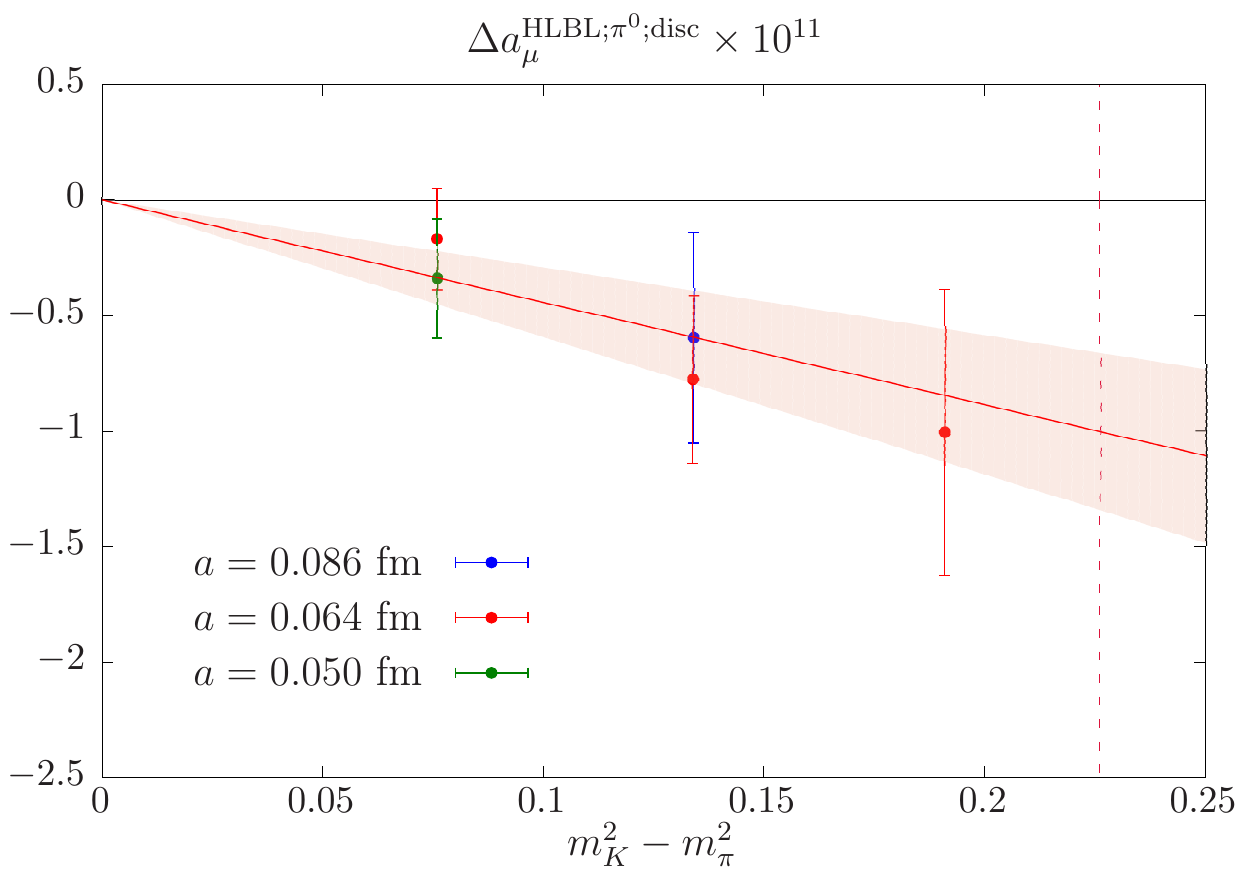}
	\includegraphics*[width=0.49\linewidth]{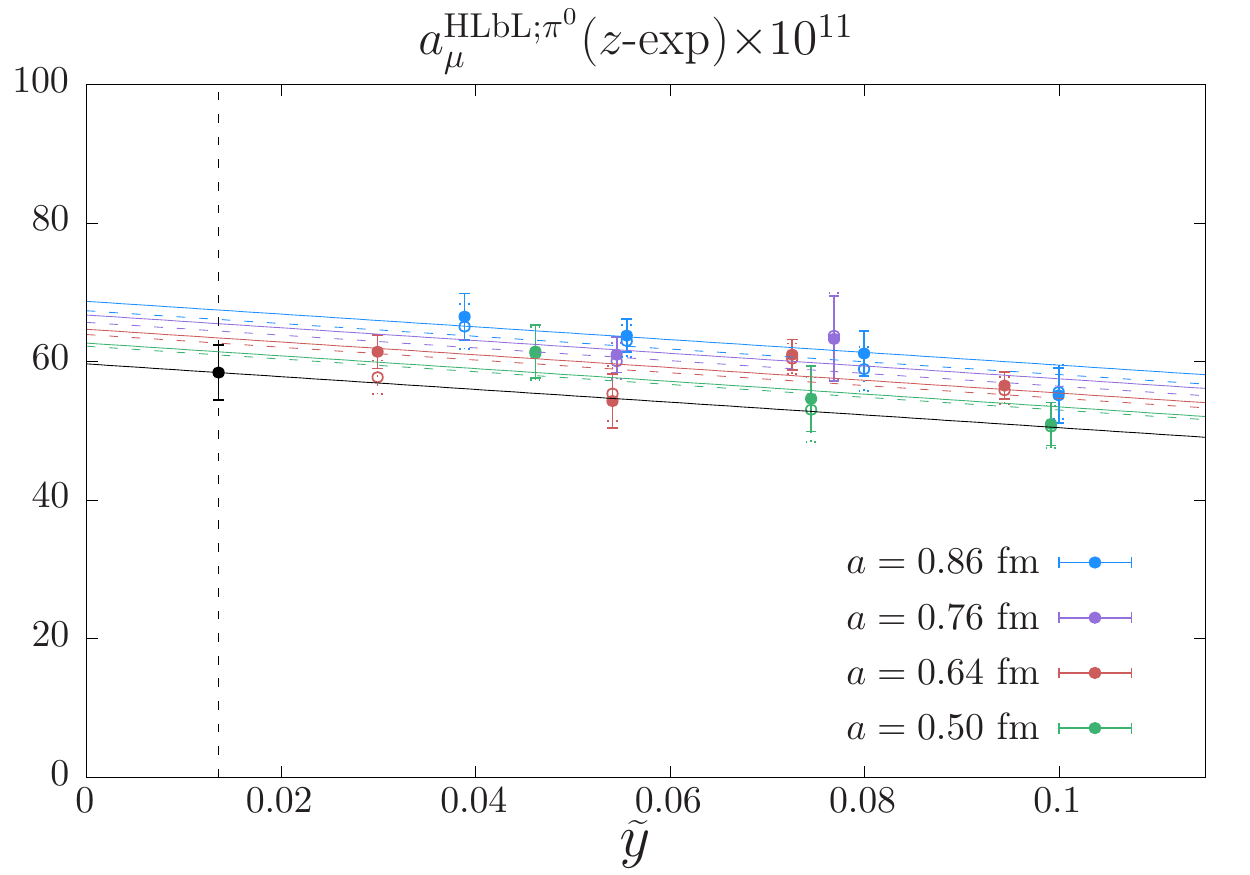}
	
	\caption{Left panel: Difference in $\amu$ with or without including the disconnected diagrams. The red line corresponds to a linear fit and the vertical dashed line to the physical point. Right panel: Continuum and chiral extrapolation of $\amu$ using the local $z$-expansion performed on each ensemble.}
	\label{fig:amu}
\end{figure}

In the left panel of Fig.~\ref{fig:amu}, we show the difference $\Delta a_{\mu}^{\mathrm{HLbL};\pi^0;\mathrm{disc}}$ between the results obtained with and without including the disconnected contribution. At the SU(3)$_{\rm flavor}$ symmetric point, the difference vanishes exactly, while it turns negative as the pion mass approaches its physical value. At a constant value of the trace of the quark-mass matrix, the disconnected contribution is proportional to $m_s-m_l$ close 
to the SU(3)$_{\rm flavor}$ symmetric point. In practice, we parametrize it as being linear in  $m_K^2-m_{\pi}^2$ and obtain $\Delta a_{\mu}^{\mathrm{HLbL};\pi^0;\mathrm{disc}} = -1.0(0.3) \times 10^{-11}$ at the physical point. To be conservative, and because our data do not allow us to perform a precise continuum extrapolation, we associate 50\% uncertainty to this contribution.

Similar to the case of the LMD+V model, we could perform the $z$-expansion on each ensemble separately and then extrapolate $\amu$ to the physical point. We would obtain $(58.5 \pm 4.0) \times 10^{-11}$, compatible with the results from the global fit. We note that the continuum and chiral extrapolations are very mild, as can be seen in the right panel of Fig.~\ref{fig:amu}. We prefer the global fit method which reduces the number of fit parameters and relies on the chiral extrapolation of the pion TFF itself instead of the pion-pole contribution. 

Our final value given by Eq.~(\ref{eq:amu_zexp}) is compatible with our previous determination~\cite{Gerardin:2016cqj} using two dynamical quarks but with an improved accuracy. It is also in good agreement with the data-driven determination recently published in Ref.~\cite{Hoferichter:2018dmo} and based on dispersion theory, with a similar precision to results published in Ref.~\cite{Masjuan:2017tvw}, based on a fit to experimental data using Canterbury approximants.

\subsubsection{Combination of lattice and experimental data}

The experimental precision on the normalization of the TFF, dominated by the PrimEx experiment~\cite{Larin:2010kq}, is still better than our lattice determination. It is thus possible to reduce the error on $\amu$ by exploiting this experimental measurement. We thus performed a constrained linear fit: to propagate the error on the normalization of the TFF, the constraint is imposed on each jackknife sample with a Gaussian distribution reproducing the PrimEx result. We obtain
\begin{equation}
a_{\mu}^{\mathrm{HLbL}; \pi^0} = (62.3 \pm 2.0 \pm 0.9 \pm 0.5) \times 10^{-11} = (62.3\pm 2.3)\times 10^{-11}\,.
\label{eq:amu_zexp_exp}
\end{equation} 
Two comments on these results. First, this value is slightly higher than in Eq.~(\ref{eq:amu_zexp}). This can be explained by our lower value for the normalization of the TFF. Second, the statistical error has been reduced by a factor of 1.7 which can be attributed to the small experimental error on the decay width, emphasizing the importance of precise lattice data at low virtualities.
Thus, to further improve the lattice determination, it would be interesting to add large volume ensembles, which more tightly constrain the normalization of the TFF. The CLS ensemble E250~\cite{Mohler:2017wnb}, at the physical mass and with a volume of 6.2~fm would be valuable.

\section{Conclusion} 
\label{sec:ccl}

We have computed the neutral pion transition form factor with $N_f=2+1$ dynamical quarks at the physical point. The results are summarized in Table~\ref{tab:z-exp} where we provide the coefficients of the modified $z$-expansion and the associated correlation matrix. Our results are in good agreement with experimental data from CELLO and CLEO in the single-virtual case. In the double-virtual case, where no experimental data exist yet, we can compare our result with the recent dispersive analysis~\cite{Hoferichter:2018dmo}. Finally, when both virtual photons carry large virtualities, our result reproduces the asymptotic prediction from perturbative QCD once one includes the first-order $\alpha_s$ correction and higher-twist corrections.

In Sec.~\ref{sec:pheno}, we first compared our result with popular phenomenological models, often used to estimate the pion-pole contribution to hadronic light-by-light scattering in the muon $g-2$. The LMD+V model, which satisfies both the OPE and the Brodsky-Lepage constraints at short distances, provides a good parametrization over the whole kinematic range covered by our lattice data. The fit parameters turn out to be close to their phenomenological values obtained in Ref.~\cite{Nyffeler:2009tw}. However, our lattice data show significant deviations from the VMD or LMD models.  Second, we have extracted the normalization of the TFF using either the phenomenological models or the more systematic parametrization based on the $z$-expansion. We reproduce the experimental result with a precision of 3.5\% : $\alpha = 0.264(9)~\GeV^{-1}$. This is an important benchmark of our calculation. The precision is not yet competitive with the experimental estimate from the PrimEx Collaboration~\cite{Larin:2010kq}, but we were able to extract the LEC appearing in the odd-intrinsic-parity sector of $\chi$PT at order $p^6$, $C_7^{\mathrm{Wr}} = 0.16(18)\times 10^{-3}~\GeV^{-2}$. This value lies in-between some recent estimates based on various resonance Lagrangians that do or do not fulfill short-distance constraints from the OPE. Ref.~\cite{Kampf:2009tk} revisited the analysis of pion decay $\pi^0 \to \gamma\gamma$ at NLO in $\chi$PT and assumed that this LEC vanishes. It turns out that with our new result from the lattice, the other relevant LEC at NLO does not change much compared to the analysis in Ref.~\cite{Kampf:2009tk}. We get $C_8^{\mathrm{Wr}} = 0.56(17)\times 10^{-3}~\GeV^{-2}$ and essentially reproduce with $\Gamma(\pi^0 \to \gamma\gamma) = 8.07(10)~\mathrm{eV}$ the result given in that reference, using similar input from experiment and theory.

Finally, our model-independent lattice estimate for the pion-pole contribution to hadronic light-by-light scattering in $(g-2)_\mu$, given by Eq.~(\ref{eq:amu_zexp}), reads $\amu=(59.7 \pm 3.6) \times 10^{-11}$ and corresponds to a precision of 6\%. This is our main result. The precision can be further improved by imposing the experimental constraint on the normalization of the TFF from the PrimEx experiment, and we obtain in this way the lattice and data-driven estimate $\amu=(62.3 \pm 2.3) \times 10^{-11}$ with a precision of 4\%. This precision has already reached a sufficient level in view of the forthcoming experiment at Fermilab, for which a precision of $10\%$ on the theory estimate of the full HLbL contribution is desired.

In the future, we plan to use our result to estimate the dominant finite-size effects in the full lattice calculation of the HLbL contribution to the muon $g-2$. Since the QCD four-point correlation function will be computed on the same set of lattice ensembles, we should be able to constrain the tail of the integrand at large distances and thus reduce the statistical error.
We also plan to include another lattice ensemble, with physical pion mass and large volume, to constrain even better the chiral and continuum extrapolations. The large volume should help us reach smaller virtualities and will constrain the normalization of the TFF even better.
Finally, besides the dominant pion-pole contribution, it would be interesting to have a first principle estimate for the $\eta$ and $\eta^{\prime}$ pseudoscalar-pole contributions.
According to model calculations~\cite{Nyffeler:2016gnb}, the size of these contributions is of the order of 20\% of the pion-pole contribution and they are therefore not negligible. The same lattice methodology can be used to extract the relevant pseudoscalar transition form factors, even though disconnected diagrams will play a more important role. In that case, the weight functions appearing in Eq.~(\ref{contribution_2}) are peaked at slightly larger virtualities. However, since a precision of 20 or 30\% should suffice, a lattice calculation should be feasible. Such a study would have a high impact since only sparse experimental data are available in the double-virtual case for spacelike momenta~\cite{BaBar:2018zpn}.


\begin{acknowledgments}
We thank Martin Hoferichter and Bastian Kubis for helpful discussions.
We are grateful for the access to the ensembles used here, made available to us through CLS, as well as to the samples of quark-disconnected loops generated by Konstantin Ottnad on a subset of these ensembles.
The quark-connected correlation functions were computed on the platforms ``Clover'' at the Helmholtz-Institut Mainz and ``Mogon~II'' at Johannes Gutenberg University Mainz. 
The CLS lattice ensembles and the quark-disconnected loops used here were partly generated through computing time provided by the Gauss Centre for Supercomputing (GCS) through the John von Neumann Institute for Computing (NIC) on the GCS share of the supercomputer JUQUEEN at J\"ulich Supercomputing Centre (JSC).
This work was partly supported by the European Research Council (ERC) under the European Union's Horizon 2020 research and innovation programme through Grant Agreement No.\ 771971-SIMDAMA, as well as  by the Cluster of Excellence \emph{Precision
Physics, Fundamental Interactions, and Structure of Matter} (PRISMA$^+$ EXC 2118/1) funded by the German Research Foundation (DFG) within the German Excellence Strategy (Project ID 39083149), as well as by the SFB 1044 funded by the DFG. 
A.N.\ is grateful to the Albert Einstein Center for Fundamental Physics (AEC) at the University of Bern for hospitality and support during the completion of this work.  
\end{acknowledgments}

\appendix

\section{Analytic expressions of $\widetilde{A}_{\mu\nu}^{\rm VMD}(\tau)$ and $\widetilde{A}_{\mu\nu}^{\rm LMD}(\tau)$ \label{app:A} } 

In this appendix, we provide analytical expressions for the function $\widetilde{A}_{\mu\nu}^{\rm LMD}(\tau)$, introduced in Eq.~(\ref{eq:Mlat}), 
for a general pion momentum $\vec p$, assuming an LMD transition form factor~\cite{Moussallam:1994xp,Knecht:1999gb}. 
The VMD case is simply obtained by setting $\beta=0$ 
in the equations below. Using the LMD model and Eq.~(\ref{eq:decomp}), we obtain
\begin{align}
\widetilde{A}_{\mu\nu}(\tau)  = \frac{ Z_{\pi}  }{ 4 \pi E_{\pi} } \int_{-\infty}^{\infty} \, \mathrm{d} \widetilde{\omega} \, 
\left(P^E_{\mu\nu} \widetilde{\omega} + Q^E_{\mu\nu} \right)\, \frac{ \alpha \, M_V^4 + \beta \, (q_1^2 + q_2^2) }{\left( M_V^2+|\vec{q}_1|^2+\widetilde{\omega}^2 \right) \left( M_V^2+|\vec{q}_2|^2-(E_{\pi}-i\widetilde{\omega})^2 \right)}\,  e^{-i \widetilde{w} \tau} \,,
\end{align}
where $P^{E}_{\mu\nu} = i \epsilon_{\mu\nu0i} p^i$ and $Q^{E}_{\mu\nu} = \epsilon_{\mu\nu i 0}\, E_{\pi}\, q_1^i\, - i \epsilon_{\mu\nu i j}\, q_1^i \, p^j$ are independent of $\widetilde{\omega}$. The integrand has four distinct simple poles 
\begin{align}
\widetilde{\omega}_1^{(\pm)} = \pm i\sqrt{M_V^2+|\vec{q}_1|^2} = \pm i k_1 \quad , \quad \widetilde{\omega}_2^{(\pm)} = - i\left( E_{\pi}  \mp \sqrt{M_V^2+|\vec{q}_2|^2}  \right) = -i(E_{\pi} \mp k_2)  \,,
\end{align}
where we used the notations $k_1 = \sqrt{M_V^2 + |\vec{q}_1|^2}$ and  $k_2 = \sqrt{M_V^2 + |\vec{q}_2|^2}$. Therefore 
\begin{align}
\widetilde{A}_{\mu\nu}^{\LMD}(\tau) = \frac{Z_{\pi} }{ 4 \pi E_{\pi}} \int_{-\infty}^{\infty} \, \mathrm{d} \widetilde{\omega} \, \left(P^E_{\mu\nu} \widetilde{\omega} + Q^E_{\mu\nu} \right)  \, \frac{ \alpha \, M_V^4 + \beta \, (q_1^2 + q_2^2) }{ \left( \widetilde{\omega} - \widetilde{\omega}_1^{(+)} \right) \left(\widetilde{\omega} - \widetilde{\omega}_1^{(-)} \right)   \left( \widetilde{\omega} - \widetilde{\omega}_2^{(+)} \right) \left(\widetilde{\omega} - \widetilde{\omega}_2^{(-)} \right) } \, e^{- i\widetilde{\omega} \tau} \,.
\end{align}

\underline{Case $\tau > 0$} 

\begin{multline}
\label{eq:LMDtaupos}
\widetilde{A}_{\mu\nu}^{\LMD}(\tau) = - \frac{Z_{\pi}}{4 E_{\pi}} \left[  \left( P^E_{\mu\nu} \widetilde{\omega}_1^{(-)} + Q^E_{\mu\nu} \right)  \frac{\alpha M_V^4 + \beta(2M_V^2 + E_{\pi}^2 - 2E_{\pi}k_1 + |\vec{q}_1|^2 - |\vec{q}_2|^2) }{ k_1(E_{\pi} - k_1 + k_2)(E_{\pi} - k_1 - k_2) }  e^{-k_1 \tau} + \right.  \\
 \left. \left( P^E_{\mu\nu} \widetilde{\omega}_2^{(-)} + Q^E_{\mu\nu} \right)  \frac{\alpha M_V^4 + \beta(2M_V^2 + E_{\pi}^2 + 2E_{\pi}k_2 - |\vec{q}_1|^2 + |\vec{q}_2|^2) }{ k_2(E_{\pi} - k_1 + k_2)(E_{\pi} + k_1 + k_2) }  e^{-(E_{\pi} + k_2) \tau} \right] \,.
\end{multline}

\underline{Case $\tau < 0$} 

\begin{multline}
\label{eq:LMDtauneg}
\widetilde{A}_{\mu\nu}^{\LMD}(\tau) = - \frac{Z_{\pi}}{4 E_{\pi}} \left[  \left( P^E_{\mu\nu} \widetilde{\omega}_1^{(+)} + Q^E_{\mu\nu} \right)  \frac{\alpha M_V^4 + \beta(2M_V^2 + E_{\pi}^2 + 2E_{\pi}k_1 + |\vec{q}_1|^2 - |\vec{q}_2|^2) }{ k_1(E_{\pi} + k_1 - k_2)(E_{\pi} + k_1 + k_2) }  e^{k_1 \tau} + \right.  \\
 \left. \left( P^E_{\mu\nu} \widetilde{\omega}_2^{(+)} + Q^E_{\mu\nu} \right)  \frac{\alpha M_V^4 + \beta(2M_V^2 + E_{\pi}^2 - 2E_{\pi}k_2 - |\vec{q}_1|^2 + |\vec{q}_2|^2) }{ k_2(E_{\pi} + k_1 - k_2)(E_{\pi} - k_1 - k_2) }  e^{+(k_2 - E_{\pi}) \tau} \right] \,.
\end{multline}
The expression for $\widetilde{A}^{(1),\LMD}(\tau)$ is then obtained using Eq.~(\ref{eq:decomp}). The function $\widetilde{A}_{\mu\nu}^{\LMD}(\tau)$ is continuous for all values of $\tau$ when $P^{E}_{\mu\nu}=0$ but is discontinuous at $\tau=0$ when $P^{E}_{\mu\nu} \neq 0$. However, the function has finite limits when $\tau \to 0^{\pm}$ respectively. A similar method can easily be applied to the LMD+V model.

Finally, we remark that Eqs.\ (\ref{eq:LMDtaupos}) and (\ref{eq:LMDtauneg}) indicate the type of intermediate states that contribute to the Euclidean three-point function $C_{\mu\nu}^{(3)}$ defined in Eq. (\ref{eq:C3}), according to the LMD model. Taking into account
$\vec p=\vec q_1+\vec q_2$ and the relation (\ref{eq:Amunu}) between $\widetilde{A}_{\mu\nu}(\tau)$ and $C_{\mu\nu}^{(3)}$, we find that for $\tau>0$, either a vector meson with momentum $\vec q_1$ or a two-particle state consisting of a pion with momentum $\vec p$ and a vector meson with momentum $-\vec q_2$ is propagating between the two vector currents; for $\tau<0$, either a vector meson with momentum $\vec q_2$ or a two-particle state consisting of a pion with momentum $\vec p$ and a vector meson with momentum $-\vec q_1$ is propagating.

\section{$z$-expansion : systematic error \label{app:B} } 

\begin{figure}[t!]
	\label{fig:z-exp-test-fit}
	\includegraphics*[width=0.49\linewidth]{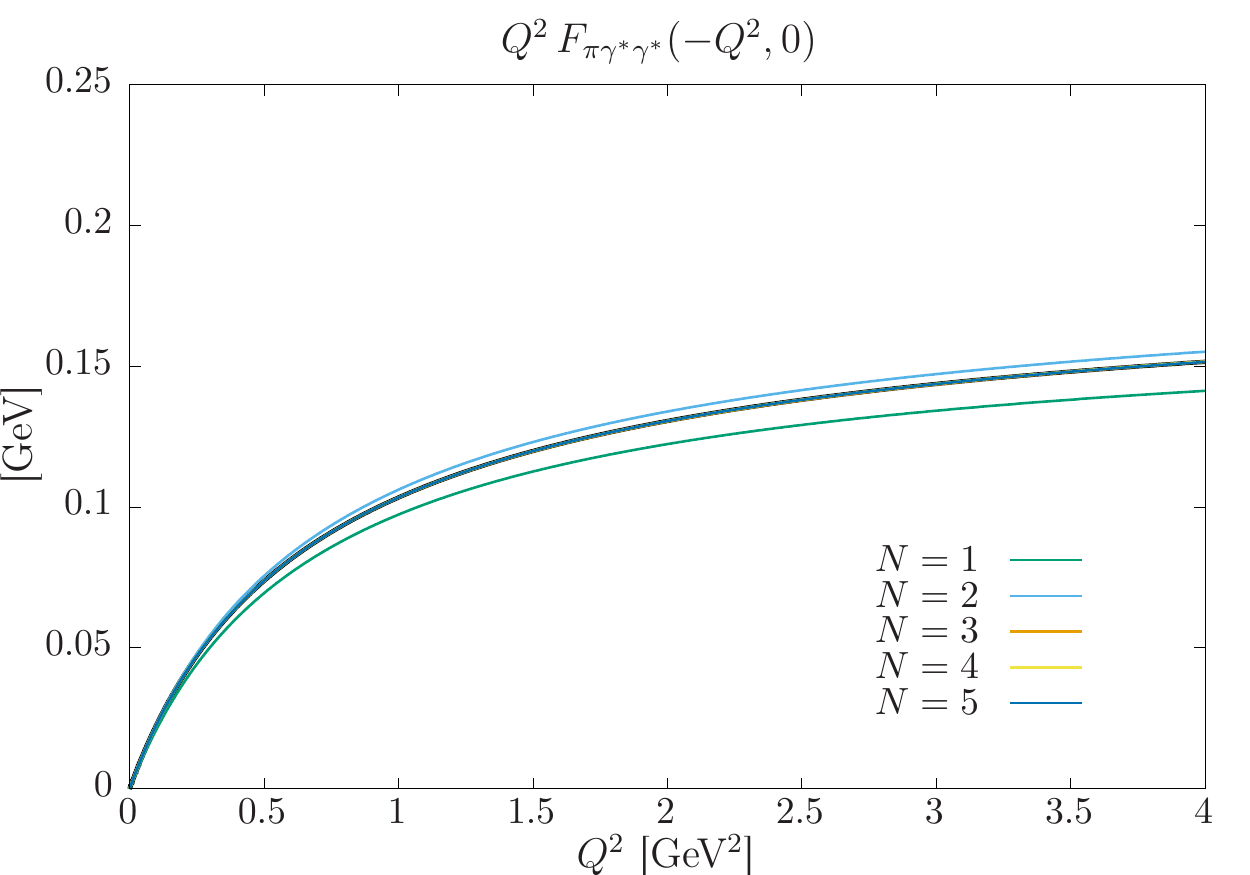}
	\includegraphics*[width=0.49\linewidth]{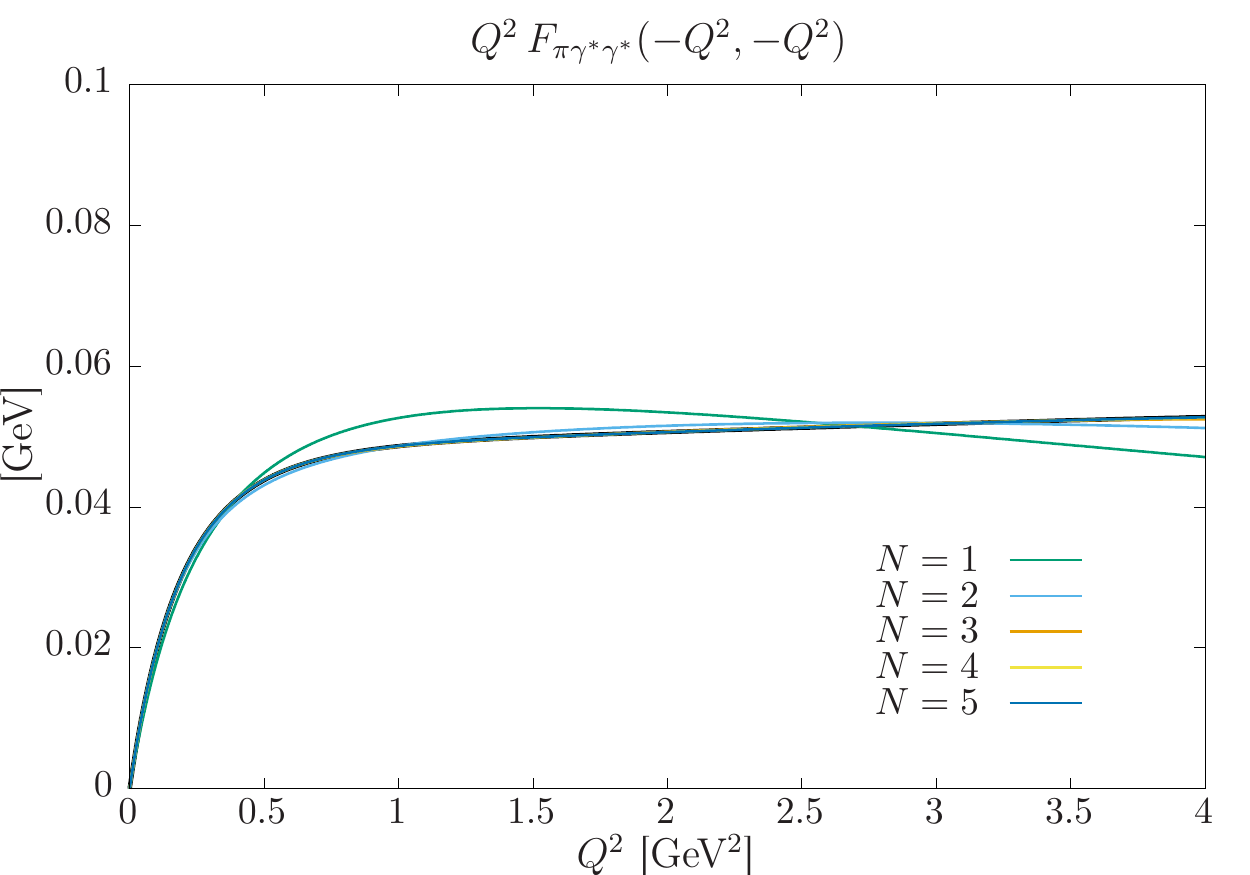}
	\caption{Fits of the LMD+V TFF using the modified $z$-expansion in the range 0 - 4~$\GeV^2$ and for different values of $N$. The LMD+V TFF is defined using Eq.~(\ref{eq:LMDV}) and the fit parameters in Eq.~(\ref{eq:resLMDV1}). \label{fig:LMDVz}}
\end{figure}

The transition form factor admits a modified double $z$-expansion given by Eq.~(\ref{eq:z_exp_mod}) where the conformal variables $z_1$ and $z_2$ are given by
\begin{equation}
z_k = \frac{ \sqrt{t_c + Q_k^2}-\sqrt{t_c-t_0} }{ \sqrt{t_c + Q_k^2}+\sqrt{t_c-t_0} }  \,, \quad t_0 = t_c \left( 1 - \sqrt{1+\frac{Q_{\max}^2}{t_c} } \right) \,, \quad t_c = 4 m_{\pi}^2 \,.
\end{equation}
As explained in the main text, $t_0$ is the optimal choice which reduces the maximum value of $|z_k|$ in the range $[0, Q_{\max}^2]$. 
Using $m_{\pi} = 134.9~\MeV$ one finds
\begin{subequations}
\begin{align}
|z_{\max}|_{Q_{\max}^2=2~\GeV^2} &=  0.40 \,, \\ 
|z_{\max}|_{Q_{\max}^2=3~\GeV^2} &=  0.44 \,, \\ 
|z_{\max}|_{Q_{\max}^2=4~\GeV^2} &=  0.46 \,, 
\end{align}
\end{subequations}
well below one. The expansion parameter $z_k$ vanishes for $Q_k^2 = -t_0$ (with $Q_{\max}^2=4~\GeV^2$, it corresponds to $t_0 = -0.47~\GeV^2$) and is maximum at $z_k=0$ and $z_k=Q^2_{\max}$. Finally, the virtualities can be expressed in terms of the variables $z_k$
\begin{equation}
Q_k^2 = \left( \frac{1+z_k}{1-z_k} \right)^2 (t_c-t_0) - t_c \,.
\end{equation}
In Sec.~\ref{sec:zexp}, the coefficients of the double $z$-expansion were obtained by fitting the lattice data for a given value of~$N$. The latter should be large enough to minimize the systematic error coming from the truncation of the sum but its value is also limited by the statistical precision of the data. We would like to estimate the systematic error induced by this truncation. Therefore, as a test, we fit the LMD+V TFF, defined using Eq.~(\ref{eq:LMDV}) and the fit parameters in Eq.~(\ref{eq:resLMDV1}),
with the $z$-expansion given by Eq.~(\ref{eq:z_exp_mod}). The model satisfies both short-distance constraints given by Eqs.~(\ref{eq:BL}) and (\ref{eq:OPE}). 
For this purpose, we use a regular grid with $0\leq Q_k^2 \leq Q_{\max}^2$ and a step size $\delta Q_k^2 = 0.025~\GeV^2$. 
In Table~\ref{tab:z-exp-LMDV}, and for different values of $N$ and $Q_{\max}^2$, we provide the results of the fit for the normalization of the TFF $\alpha$, the slope at the origin $b_{\pi}$ and the pion-pole contribution $\amu$ which can be compared to the exact known results. We also provide the maximum deviation between the exact TFF and the fit. The level of agreement is illustrated in Fig.~\ref{fig:LMDVz}.

\renewcommand{\arraystretch}{1.2}
\begin{table}[t]
\caption{Results of the fits of the LMD+V model using the modified double $z$-expansion. The model is defined using Eq.~(\ref{eq:LMDV}) and the fit parameters in Eq.~(\ref{eq:resLMDV1}). The exact values are $\alpha=0.264~\GeV^{-1}$, $b_{\pi}=1.62~\GeV^{-2}$ and $\amu = 59.2 \times 10^{-11}$. Results for $\amu$ are given in units of $10^{-11}$. The last column, $d_{\rm max}$, corresponds to the maximum deviation in percent between the exact TFF and the fit.} 
\begin{center}
\begin{tabular}{l|ccc|ccc|cccr}
\hline
	&	\multicolumn{3}{c|}{$Q^2_{\max}= 1~\GeV^2$} 	&	\multicolumn{3}{c|}{$Q^2_{\max}= 2~\GeV^2$} 	&	\multicolumn{4}{c}{$Q^2_{\max}= 4~\GeV^2$}  \\ 
$N$	&	$\alpha~[\GeV^{-1}]$	&	$b_{\pi}~[\GeV^{-2}]$	&	$\amu$	&	$\alpha~[\GeV^{-1}]$	&	$b_{\pi}~[\GeV^{-2}]$	&	$\amu$	&	$\alpha~[\GeV^{-1}]$	&	$b_{\pi}~[\GeV^{-2}]$	&	$\amu$ 	&	$d_{\rm max}$\\
\hline
$1$	& 0.260 &	1.51 & 55.4  	&	0.253 & 1.51 & 54.4		& 0.232 & 1.43 & 49.1 &	12.2\% \\
$2$	& 0.265 &	1.63 & 57.0  	&	0.264 & 1.55 & 58.2 		& 0.262 & 1.40 & 58.4 &	3.5\% \\
$3$	& 0.264 &	1.62 & 57.9  	&	0.264 & 1.65 & 57.9		& 0.265 & 1.71 & 58.3 &	0.6\% \\
$4$	& 0.264 &	1.62 & 57.8  	&	0.264 & 1.61 & 58.3		& 0.264 & 1.60 & 59.2 &	0.2\% \\
$5$	& 0.264 &	1.62 & 58.4  	&	0.264 & 1.63 & 59.4 		& 0.264 & 1.62 & 59.8 &	0.06\% \\
\hline 
\end{tabular}
\label{tab:z-exp-LMDV}
\end{center}
\end{table}

We conclude that using $N=3$ and $Q^2_{\max} = 4~\GeV^2$ is already sufficient to get a precision below 1\% for the TFF in the range $[0, Q^2_{\max}]$. The normalization of the TFF and the pion-pole contribution, $\amu$, are also recovered within a precision below 1\%. The slope parameter is obtained with a precision of 6\%.
We point out that, using the optimal value $t_0$, the parameters $z_k$ reach their maximal value at $Q_k^2=0$, precisely where the normalization of the TFF is obtained. Therefore, fitting the data in a wide range of virtualities is not the best method to determine the normalization of the TFF. However, the minimal value of $z_k$ is obtained at $Q^2_k = -t_0 \approx 0.5~\GeV^2$, an optimal choice for the pion-pole contribution since the integrand is peaked around this value. 

\section{Results of the $z$-expansion for some individual lattice ensembles\label{sec:z-exp-indiv}} 

In Table~\ref{tab:z-exp-indiv}, we provide the coefficients of the $z$-expansion for three ensembles used in this work: N101, N200 and D200. The correlation matrices are respectively given by Eqs~(\ref{eq:covN101}), (\ref{eq:covN200}) and (\ref{eq:covD200}). The ensembles N200 and D200 have the same lattice spacing but different pion masses whereas N200 and N101 have similar pion masses but different lattice spacings. We stress that our main result in Sec.~\ref{sec:zexp} is based on a global fit, where all the ensembles are fitted simultaneously using the ansatz (\ref{eq:pz}).
 
\renewcommand{\arraystretch}{1.2}
\begin{table}[h!]
\caption{Coefficients of the $z$-expansion, in $\GeV^{-1}$, defined in Eq.~(\ref{eq:z_exp_mod}) and obtained from a fit on a single ensemble using local vector currents at the source and at the sink. The vector meson masses used in Eq.~(\ref{eq:P_zexp}) are given in Table~\ref{tab:sim}.} 
\begin{center}
\begin{tabular}{lccccc}
\hline
Id	&	$c_{00}$	&	$c_{01}$	&	$c_{11}$	&	$c_{20}$	&	$c_{21}$	 \\ 
\hline
N101	&	0.2456(46)	&	$-0.0755(61)$	&	$-0.265(99)$	&	0.065(54)	&	\ \ 0.02(14)	 \\ 
N200	&	0.2301(44)	&	$-0.0675(43)$	&	$-0.280(64)$	&	0.071(32)	&	\ \ 0.09(11)	 \\ 
D200	&	0.2484(48)	&	$-0.0728(51)$	&	$-0.399(94)$	&	0.148(47)	&	$-0.02(14)$	 \\ 
\hline 
\end{tabular}
\vskip 0.1in
\begin{tabular}{lccccc}
\hline
Id	&	$c_{22}$	&	$c_{30}$	&	$c_{31}$	&	$c_{32}$	&	$c_{33}$	 \\ 
\hline
N101	&	$-0.038(85)$	&	0.143(125)	&	0.12(41) 	&	$-0.35(62)$	&	$-0.18(91)$   \\ 
N200	&	$-0.250(58)$	&	0.079(95) \ \	&	0.26(32) 	&	$-0.16(76)$	&	$-0.90(91)$   \\ 
D200	&	$-2.44(1.15)$	&	0.283(122)	&	1.14(55) 	&	$-1.64(79)$	&	$-0.15(89)$   \\ 
\hline 
\end{tabular}
\label{tab:z-exp-indiv}
\end{center}
\end{table}
 
\begin{equation}
\mathrm{cor}(c_{nm}) = \begin{pmatrix}
+1.000 & -0.338 & +0.126 & -0.242 & +0.188 & +0.054 & -0.227 & +0.049 & +0.162 & -0.299 \\ 
-0.338 & +1.000 & -0.051 & +0.096 & -0.310 & +0.371 & -0.113 & -0.406 & +0.508 & +0.016 \\
+0.126 & -0.051 & +1.000 & -0.946 & +0.757 & +0.550 & -0.852 & -0.470 & +0.230 & +0.274 \\
-0.242 & +0.096 & -0.946 & +1.000 & -0.854 & -0.517 & +0.932 & +0.331 & -0.179 & -0.140 \\
+0.188 & -0.310 & +0.757 & -0.854 & +1.000 & +0.243 & -0.832 & -0.034 & -0.272 & +0.317 \\
+0.054 & +0.371 & +0.550 & -0.517 & +0.243 & +1.000 & -0.577 & -0.934 & +0.672 & +0.285 \\
-0.227 & -0.113 & -0.852 & +0.932 & -0.832 & -0.577 & +1.000 & +0.365 & -0.268 & -0.116 \\
+0.049 & -0.406 & -0.470 & +0.331 & -0.034 & -0.934 & +0.365 & +1.000 & -0.679 & -0.399 \\
+0.162 & +0.508 & +0.230 & -0.179 & -0.272 & +0.672 & -0.268 & -0.679 & +1.000 & -0.369 \\
-0.299 & +0.016 & +0.274 & -0.140 & +0.317 & +0.285 & -0.116 & -0.399 & -0.369 & +1.000 \\
\end{pmatrix} \,.
\label{eq:covN101} 
\end{equation}

\begin{equation}
\mathrm{cor}(c_{nm}) = \begin{pmatrix}
+1.000 & -0.237 & -0.409 & +0.084 & +0.140 & -0.113 & +0.053 & +0.388 & -0.405 & +0.171 \\
-0.237 & +1.000 & +0.152 & -0.171 & -0.095 & +0.267 & -0.394 & -0.211 & +0.414 & -0.299 \\
-0.409 & +0.152 & +1.000 & -0.812 & +0.466 & +0.471 & -0.688 & -0.510 & +0.331 & +0.077 \\
+0.084 & -0.171 & -0.812 & +1.000 & -0.790 & -0.419 & +0.882 & +0.117 & +0.044 & -0.218 \\
+0.140 & -0.095 & +0.466 & -0.790 & +1.000 & +0.096 & -0.779 & +0.270 & -0.487 & +0.512 \\
-0.113 & +0.267 & +0.471 & -0.419 & +0.096 & +1.000 & -0.402 & -0.808 & +0.435 & +0.002 \\
+0.053 & -0.394 & -0.688 & +0.882 & -0.779 & -0.402 & +1.000 & +0.113 & -0.093 & -0.069 \\
+0.388 & -0.211 & -0.510 & +0.117 & +0.270 & -0.808 & +0.113 & +1.000 & -0.734 & +0.189 \\
-0.405 & +0.414 & +0.331 & +0.044 & -0.487 & +0.435 & -0.093 & -0.734 & +1.000 & -0.769 \\
+0.171 & -0.299 & +0.077 & -0.218 & +0.512 & +0.002 & -0.069 & +0.189 & -0.769 & +1.000 \\
\end{pmatrix} \,.
\label{eq:covN200} 
\end{equation}
 
\begin{equation}
\mathrm{cor}(c_{nm}) = \begin{pmatrix}
+1.000 & -0.147 & -0.109 & -0.006 & -0.101 & +0.139 & +0.018 & -0.079 & +0.178 & -0.246 \\ 
-0.147 & +1.000 & +0.097 & -0.040 & -0.257 & +0.411 & -0.185 & -0.441 & +0.544 & -0.082 \\
-0.109 & +0.097 & +1.000 & -0.936 & +0.676 & +0.456 & -0.812 & -0.440 & +0.210 & +0.418 \\
-0.006 & -0.040 & -0.936 & +1.000 & -0.746 & -0.434 & +0.890 & +0.322 & -0.177 & -0.241 \\
-0.101 & -0.257 & +0.676 & -0.746 & +1.000 & +0.058 & -0.811 & +0.041 & -0.332 & +0.529 \\
+0.139 & +0.411 & +0.456 & -0.434 & +0.058 & +1.000 & -0.490 & -0.954 & +0.839 & +0.032 \\
+0.018 & -0.185 & -0.812 & +0.890 & -0.811 & -0.490 & +1.000 & +0.361 & -0.229 & -0.253 \\
-0.079 & -0.441 & -0.440 & +0.322 & +0.041 & -0.954 & +0.361 & +1.000 & -0.834 & -0.166 \\
+0.178 & +0.544 & +0.210 & -0.177 & -0.332 & +0.839 & -0.229 & -0.834 & +1.000 & -0.370 \\
-0.246 & -0.082 & +0.418 & -0.241 & +0.529 & +0.032 & -0.253 & -0.166 & -0.370 & +1.000 \\
\end{pmatrix} \,.
\label{eq:covD200} 
\end{equation}


\end{document}